\documentclass[amsmath,superscriptaddress,showpacs,aps,prb,twocolumn]{revtex4-1}
\UseRawInputEncoding
\usepackage[colorlinks,linkcolor=blue,anchorcolor=blue,citecolor=blue,urlcolor=blue]{hyperref}
\usepackage{amsmath}
\usepackage{amssymb}
\usepackage{graphicx}
\usepackage{color}
\usepackage{bm}
\usepackage{subfigure}
\pdfoutput=1

\begin{document}
\title{Superfluid-Mott insulator quantum phase transition in a cavity optomagnonic system }
\author{ Qian Cao}
\affiliation{Lanzhou Center for Theoretical Physics, Key Laboratory of Theoretical Physics of Gansu Province, LanZhou University, LanZhou, Gansu, $730000$, China}

\author{Lei Tan}
\email{tanlei@lzu.edu.cn}
\affiliation{Lanzhou Center for Theoretical Physics, Key Laboratory of Theoretical Physics of Gansu Province, LanZhou University, LanZhou, Gansu, $730000$, China}
\affiliation{Key Laboratory for Magnetism and Magnetic Materials of the Ministry of Education, Lanzhou University, Lanzhou $730000$, China}

\author{Wu-Ming Liu}
\affiliation{Beijing National Laboratory for Condensed Matter Physics, Institute of Physics, Chinese Academy of Sciences, Beijing $100190$, China}

\begin{abstract}
The emerging hybrid cavity optomagnonic system is a very promising quantum information processing platform for its strong or ultrastrong photon-magnon interaction on the scale of micrometers in the experiment. In this paper, the superfluid-Mott insulator quantum phase transition in a two-dimensional cavity optomagnonic array system has been studied based on this characteristic. The analytical solution of the critical hopping rate is obtained by the mean field approach, second perturbation theory and Landau second order phase transition theory. The numerical results show that the increasing coupling strength and the positive detunings of the photon and the magnon favor the coherence and then the stable areas of Mott lobes are compressed correspondingly. Moreover, the analytical results agree with the numerical ones when the total excitation number is lower. Finally, an effective repulsive potential is constructed to exhibit the corresponding mechanism. The results obtained here provide an experimentally feasible scheme for characterizing the quantum phase transitions in a cavity optomagnonic array system, which will offer valuable insight for quantum simulations.
\end{abstract}

\maketitle
\section{INTRODUCTION}
Quantum simulation provides a useful tool for solving many problems such as quantum phase transition, quantum magnetism, and high-temperature superconductivity[\onlinecite{Georgescu}]. Quantum phase transition of an interaction system composed of multiple particles are widely investigated, such as heavy fermions in Kondo lattices[\onlinecite{ohneysen}], ultracold atoms in optical lattices [\onlinecite{EhudAltman, Jiu-RongHan, Orth}], and the ensemble of two-level systems interacting with a bosonic field (i.e., Dicke model)[\onlinecite{Dicke}]. Especially, the superfluid-Mott insulator quantum phase transition of bosons, which forms one of the paradigm examples of a quantum phase transition, was first studied in the Bose Hubbard (BH) model due to the competition of the on-site interaction and the hopping term theoretically and experimentally[\onlinecite{Fisher, Jaksch, Sachdev, Greiner}]. Given the precise control of coupling strengths, the properties of scalability and individual accessibility of coupled cavities[\onlinecite{Sherson, Gheri, DiCarlo, Olmschenk}], the Jaynes$-$Cummings Hubbard (JCH) model, which describes the dynamics of the coupled-cavity arrays with each embedded within a two-level atom has attracted tremendous attentions[\onlinecite{Tahan, Hartmann, Angelakis}] in recent years. Based on the JCH model and extended JCH model, the superfluid-Mott insulator quantum phase transition of light reminiscent of the ones of atoms in the BH model are extensively simulated[\onlinecite{lin2021mott, song2021realizing, zhu2021quantum, Chen_2021, huang2020observation, Kamide, gomes2012mott, Soykal, Zheng}]. More importantly, the quantum  phase  transition of light depends crucially on the intrinsic atom-photon interaction in the JCH model, where the atom-photon coupling leads to the formation of repelled collective polaritonic excitations, and this on-site repulsive potential compete with the hopping of photons between neighbouring cavities. On the other hand, the quantum phase transition of light has great potential application for a new source of quantum-correlated photons[\onlinecite{Noh_2016, Georgescu, Schmidt, Houck, Hartmannn,Kamide}].

Hereafter, with the fabrication of optomechanical cavity systems at the desired frequency accuracy and the coupling strength, the cavity optomechanical array system provides another experimentally feasible avenue to simulate the superfluid-Mott insulator quantum phase transition[\onlinecite{Mari, Mian, Peano}]. Compared with coupled cavity arrays system, the phonon-mediated cavity field and the two-level system form polaritons by coupling, which provide an effective on-site repulsion, then the system can also simulate the superfluid-Mott insulator quantum phase transition. Interestingly, the enhanced phonon-photon coupling favors the coherence of the system[\onlinecite{JinLouMa}].

In analogy to the cavity optomechanical system, cavity optomagnonic system, a new class of hybrid quantum systems based on collective spin excitations in ferromagnetic materials[\onlinecite{Dany}], has received increasing attention in recent years, which provides a new and promising platform for studying macroscopic quantum effects[\onlinecite{Dany,Zhao,Osada,Gao,Haigh,Xufeng,Zhang,Tabuchi,Goryachev}]. The collective spin excitation in ferromagnetic crystals is called a magnon, which can interact coherently with microwaves and optical photons as well as phonons via magnetic dipole, magneto-optical, and magnetostrictive interactions, respectively[\onlinecite{Kittel, Sinha, Shen, Demokritov}]. Experimentally, yttrium iron garnet (YIG) spheres are characterized with high collective spin excitation density, low dissipation, great frequency tunability and the longer coherence time, which are widely used in the study of magnon-photon coupling due to its strong and even ultrastrong couplings[\onlinecite{Hillebrands}, \onlinecite{Tan}, \onlinecite{Dany}]. In addition, the strong coupling between cavity photon and magnon has been observed at both low and high temperature experimentally[\onlinecite{Xiao}]. Based on these features of cavity optomagnonic system, many intriguing phenomena have been explored, such as magnon dark modes and gradient memory[\onlinecite{ZhangXufeng}], coherent and dissipative magnon-photon interaction[\onlinecite{Lachance, Hisatomi, Viola, Harder, Grigoryan, Wang, Yu, Yi-Pu}], the high-order sideband generation[\onlinecite{wen}], the self-sustained pscillations and chaos[\onlinecite{Kusminskiy, Yong, Xu}], non-Hermitian physics[\onlinecite{Harder}, \onlinecite{Dengke}, \onlinecite{Qiang}], entanglement[\onlinecite{Deyi, ShiYao, Yang, Huatang, Li_2019}], magnon-induced nearly perfect absorption[\onlinecite{Rao}], magnon Fock state[\onlinecite{Bittencourt}],  magnon squeezing[\onlinecite{Shi-Yao}] and so on. Recently, several novel progresses associated with  the photon blockade in cavity optomagnonic system are also investigated[\onlinecite{Chengsong}, \onlinecite{Fei}, \onlinecite{Fen}]. We note that the interplay of the photon blockade and photon hopping is not considered. In view of the unique advantages of magnons, it is very interesting to further explore the superfluid-Mott insulator quantum phase transition of light in the hybrid macroscopic quantum interface of atoms, photons and magnons.

In this work, we will investigate whether there is a superfluid-Mott insulator quantum phase transition in a cavity optomagnonic array system. Compared with the JCH model, three new degrees of freedom were added by the inclusion of YIG spheres, and the effects of these three new degrees on the quantum phase transition were inveatigated. Firstly, the analytical solutions for low excitation number are obtained based on the mean field approximation, the second order perturbation theory and the Landau phase transition theory. Then, the phase diagrams are discussed numerically using the mean-field theory. Finally, the effective repulsive potential is presented to show the mechanism of the superfluid-Mott insulator transition.

This paper is organized as follows: in Sec. \ref{II}, we describe a cavity optomagnonic array system used for studying the superfluid-Mott insulator quantum phase transition of light. Sec. \ref{III} is devoted to discussing the analytical solutions of this system, and the numerical solutions are given in Sec. \ref{IV}. Conclusions are made in Sec. \ref{V}.
\section{MODEL AND HAMILTONIAN}\label{II}
\begin{figure}[!htbp]
\includegraphics[width=8.0cm,height=4cm]{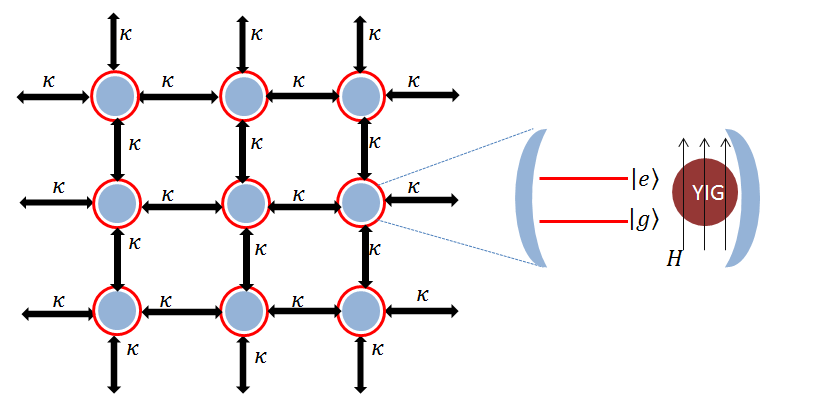}
\caption{Schematic diagram of a two-dimensional hybrid coupled optomagnonic cavity array setup. Each optomagnonic cavity contains a  two-level atom and a YIG sphere is placed near the maximum magnetic field of the cavity mode and in a uniform bias magnetic field, which establishes the magnon-photon coupling.}
\label{fig1}
\end{figure}

Consider a cavity optomagnonic system composed of a 2D array of identical coupled optomagnonic cavities, with each cavity containing a two-level atom (TLA) interacting with the photon mode (see Fig. \ref{fig1}). The total Hamiltonian ($\hbar=1$) of the system can be written as
\begin{footnotesize}
\begin{flalign}
\hat{\mathcal{H}}_T=\sum _i \hat{H}_i^{cpm}-\sum _{i,j}  \kappa _{ij} \hat{a}_i^{\dagger }\hat{a}_j-\sum _i \mu _i \hat{N}_i \label{2.1}
\end{flalign}
\end{footnotesize}

\begin{footnotesize}
\begin{align}
&\hat{H}_i^{cpm}=\omega _c \hat{a}_i^{\dagger }\hat{a}_i+\omega _a\hat{\sigma}_i^{\dagger }\hat{\sigma} _i+\omega _m \hat{m}_i^{\dagger }\hat{m}_i\notag\\&\quad +g_a\left(\hat{\sigma} _i\hat{a}_i^{\dagger }+\hat{\sigma} _i^{\dagger }\hat{a}_i\right)+G_m\left(\hat{m}_i \hat{a}_i^{\dagger }+\hat{m}_i^{\dagger }\hat{a}_i\right) \label{2.2}
\end{align}
\end{footnotesize}
where $i, j$ are the indexes for the individual optomagnonic cavity and range over all nearest neighbors sites, the subscript $T$ is the abbreviation of the total. Here, the cavity optomagnonic system is denoted as the superscript $cpm$ for convenience. $\hat{a}_i^{\dagger }(\hat{a}_i)$ and $\hat{m}_i^{\dagger }(\hat{m}_i)$ are the photonic and magnonic creation (annihilation) operators, respectively. $\hat{\sigma} _i^{\dagger }(\hat{\sigma} _i)$ are the atomic raising and lowering operators, respectively. $\hat{\mathcal{N}}_i=\sum_i\hat{N}_i=\sum_i(\hat{a}_i^{\dagger }\hat{a}_i+\hat{m}_i^{\dagger }\hat{m}_i+\hat{\sigma} _i^{\dagger }\hat{\sigma} _i)$ is the total polariton number operator[\onlinecite{0Quantum}, \onlinecite{Hohenadler}]. $N_i$ is the total number of photonic, magnonic and atomic excitations of the $i$th site in the cavity optomagnonic system. From the commutation relationship between the total polariton number operator $\hat{\mathcal{N}}_i$ and the Hamiltonian $\hat{\mathcal{H}}_T$, one can find that $\hat{\mathcal{N}}_i$ is a conserved quantity. It is feasible to describe this model in the grand-canonical ensemble and the chemical potential is $\mu_i$, which is the Lagrange multiplier in the grand-canonical ensemble  ensuring the conservation of the total excitation number in the phase transition between the Mott insulator and superfluid phases. $G_m$ $(g_a)$ represents the coupling strength of cavity mode and magnon (atom), respectively. $\omega_a$, $\omega_c$ and $\omega_m$ are the frequencies of cavity photon, atom and magnon respectively, where $\omega_m=\gamma H$, $\gamma$ is gyromagnetic ratio and $H$ is the modulated magnetic field, which can be given by the
Holstein-Primakoff (H-P) transformation[\onlinecite{Dany}, \onlinecite{Holstein}, \onlinecite{Huebl}]. We introduce the detuning between the atom and the cavity, $\Delta_a=\omega_a-\omega_c$, and the detuning between the magnon mode and the cavity mode is $\Delta_m=\omega_m-\omega_c$. In addition, the second term of Eq. \eqref{2.1} denotes the photon hopping between the nearest-neighbor cavities with the hopping rate $\kappa_{ij}$. We assume the hopping rate of photons $\kappa_{ij}=\kappa$ between adjacent sites $i$ and $j$. The chemical potential $\mu_i=\mu$ is the same for all optomagnonic cavities for simplicity.

We utilize the mean field approximation method to study the superfluid-Mott insulator quantum phase transition of light. In the mean field approximation, we introduce the superfluid order parameter $\psi\equiv\left\langle a_i\right\rangle$ to study the quantum phase transition. Generally, $\psi$ is a complex number, but its phase factor can be gauged away without affecting the Hamiltonian. Thus, $\psi$ can be taken to be real in the present system. When $\psi=0$, the system is in the Mott insulator phase. Otherwise, the system is in the superfluid phase. The boundary between $\psi=0$ and $\psi\neq0$ phases defines a quantum phase transition in the system. The final form Eq.\eqref{2.3} of the Hamiltonian is obtained by using the decoupling approximation, where $z=4$ is the number of nearest neighbours.

\begin{footnotesize}
\begin{align}
\mathcal{H}_T=\sum _i \left[H_i^{cpm}-\text{z$\kappa \psi $} \left(\hat{a}_i{}^{\dagger }+\hat{a}_i\right)-\text{$\mu N_i$}+\text{z$\kappa \psi $}^2\right] \label{2.3}
\end{align}
\end{footnotesize}

Therefore, we choose the eigenstates as the bare states of the cavity optomagnonic system, which is composed of the direct product of the cavity photon states, magnon states and atom  states, i.e., $|$ photon$(n)$, magnon$(m)$, atom$(e,g) \rangle$. The dimension of the subspace is $(2N+1)\times(2N+1)$. So, we choose a complete set of basis vectors $|N-m-1,m,e \rangle$, $|0,N-1,e \rangle$, $|N-m,m,g \rangle$, $|1,N-1,g \rangle$, $|0,N,g \rangle$ to give the matrix form of the Hamiltonian Eq. \eqref{2.4} after the mean field approximation with $n$ running $0,1,2,$ to $N$, while $m$ takes $0, 1, 2, ..., N-2$. Therefore, the matrix dimension of $H_{(N)}^{MF}$ can be defined $(2N+1)\times(2N+1)$, and likewise, the matrix dimension of $H_{(N)}^{hop}$ is defined as $(2N+1)\times(2(N+1)+1)$. The matrix expression can be obtained as follows. The superscript $MF$ and $hop$ are the abbreviation of the Mean field approximation and the hopping term respectively.

\begin{footnotesize}
\begin{align}
\mathcal{H}^{MF}&=\text{$z\kappa \psi $}^2I+\notag\\&\quad
\left(
\begin{array}{ccccccc}
 H_{(0)}^{MF} & H_{(0)}^{hop} & 0 & 0 &    &    & 0 \\
 H_{(0)}^{hop}{}^T & H_{(1)}^{MF} & H_{(1)}^{hop} & 0 &    &    & 0 \\
 0 & H_{(1)}^{hop}{}^T & H_{(2)}^{MF} & H_{(2)}^{hop} &    &    & 0 \\
 0 & 0 & H_{(2)}^{hop}{}^T & H_{(3)}^{MF} &    &    & 0 \\
    &    &    &    & \ddots & \ddots &   \\
    &    &    &    & \ddots & \ddots &   \\
 0 & 0 & 0 & 0 &    &    & H_{(N)}^{MF} \\
\end{array}
\right)
\label{2.4}
\end{align}
\end{footnotesize}

\begin{widetext}
\begin{footnotesize}
\begin{flalign}
H^{MF}_{(N)}=\left(
\begin{array}{cc}
 H_2 &
\begin{array}{ccccc}
 \sqrt{N} g_a & 0 &    & 0 & 0 \\
 0 & \sqrt{N-1} g_a &    & 0 & 0 \\
 0 & 0 & \ddots & 0 & 0 \\
 0 & 0 &    & g_a & 0 \\
\end{array}
 \\
\begin{array}{cccc}
 \sqrt{N} g_a & 0 &    & 0 \\
 0 & \sqrt{N-1} g_a &    & 0 \\
    &    & \ddots &    \\
 0 & 0 &    & g_a \\
 0 & 0 &    & 0 \\
\end{array}
 & H_1 \\
\end{array}
\right)_{(2 N+1)\times(2 N+1)}
\label{2.5}
\end{flalign}
\end{footnotesize}

The specific forms of $H_1$ and $H_2$ are in the Appendix A.
\begin{footnotesize}
\begin{flalign}
H^{hop}_{(N)}=\left(
\begin{array}{ccccccccc}
 -\sqrt{N-1} \text{$z$$\kappa \psi $} & 0 &    & 0 & 0 & 0 &    & 0 & 0 \\
 0 & -\sqrt{N-2} \text{$z$$\kappa \psi $} &    & 0 & 0 & 0 &    & 0 & 0 \\
    &    & \ddots &    &    &    &    &    &    \\
 0 & 0 &    & 0 & -\sqrt{N} \text{$z$$\kappa \psi $} & 0 &    & 0 & 0 \\
 0 & 0 &    & 0 & 0 & -\sqrt{N-1} \text{$z$$\kappa \psi $} &    & 0 & 0 \\
    &    &    &    &    &    & \ddots &    &    \\
 0 & 0 &    & 0 & 0 & 0 &    & -\text{$z$$\kappa \psi $} & 0 \\
\end{array}
\right)_{(2 N+1)\times(2 N+3)}
\label{2.6}
\end{flalign}
\end{footnotesize}
\end{widetext}

For example, when the total excitation number is $N=0$ ($N=1$), the basis vectors are selected as $|0,0,g \rangle$ ( $|1,0,g \rangle$, $|0,1,g \rangle$, $|0,0,e \rangle$ ). And the matrix of $H^{MF}_{(0)}=(0)$ for $N=0$. For $N=1$, the matrix dimension of $H^{MF}_{(1)}$ should be substituted by
\begin{footnotesize}
\begin{flalign}
 H^{MF}_{(1)}=\left(
\begin{array}{ccc}
 \omega _a-\mu  & g_a & 0 \\
 g_a & \omega _c-\mu  & G_m \\
 0 & G_m & \omega _m-\mu  \\
\end{array}
\right)
\label{2.7}
\end{flalign}
\end{footnotesize}

Introducing the detuning between the photon frequency and the two-level transition frequency $\Delta_a$ (magnon frequency $\Delta_m$), the resulting Hamiltonian is
\begin{footnotesize}
\begin{flalign}
 H^{MF'}_{(1)}=\left(
\begin{array}{ccc}
 \Delta _a-\mu  & g_a & 0 \\
 g_a & \omega _c-\mu  & G_m \\
 0 & G_m & \Delta _m-\mu  \\
\end{array}
\right)
\label{2.8}
\end{flalign}
\end{footnotesize}
the eigenvalues are given ($\Delta_a=\Delta_m=\Delta$)
\begin{footnotesize}
\begin{align}
E_{1,0}^{'}=\Delta-\mu
\end{align}
\end{footnotesize}
\begin{footnotesize}
\begin{align}
E_{1,-}^{'}=\frac{1}{2} \left(\Delta-2\mu+\omega_c -\sqrt{\Delta ^2-2 \Delta  \omega_c +4 g_a^2+4 G_m^2+\omega_c ^2} \right)
\end{align}
\end{footnotesize}
\begin{footnotesize}
\begin{align}
E_{1,-}^{'}=\frac{1}{2} \left(\Delta-2\mu+\omega_c +\sqrt{\Delta ^2-2 \Delta  \omega_c +4 g_a^2+4 G_m^2+\omega_c ^2} \right)
\end{align}
\end{footnotesize}
The splitting between states with the same excitation number of a polariton is given by

\begin{footnotesize}
\begin{align}
\delta_{E}=E^{'}_{1,-}-E^{'}_{1,+}=\sqrt{\Delta ^2-2 \Delta  \omega_c +4 g_a^2+4 G_m^2+\omega_c ^2}  \label{1}
\end{align}
\end{footnotesize}
Note that, the splitting $\delta_E$ does not only depend on the detunings of the photon-atom and the photon-magnon but also depend on their coupling strengths. That is to say, in a cavity optomagnonic system, the strong photon-magnon and photon-atom coupling are all involved in the polariton mapping[\onlinecite{0Quantum}]. Meantime, three new degrees of freedom are added in the new model compared with JCH model, which are the excitation number $m$ of the magnon, the coupling strength $G_m$ and the detuning $\Delta_m$ between the cavity field and the magnon. From the above discussion, we can determine that the excitation number of the magnon $(m)$ is constrained by the total excitation number $(N)$. In the following section, we will first deduce the analytical expressions for the order parameter and the critical hopping rate based on the second-order perturbation method and the Landau theory for the continuous phase transitions.
\section{ANALYTICAL SOLUTIONS}\label{III}
To get the simple analytical expression, we assume that the cavity frequency, atomic frequency, and magnon frequency are the same, i.e. $\omega_m=\omega_c=\omega_a\equiv\omega$. Considering Eq. \eqref{2.5} for lower excitations without the hopping term, the eigenvalues are given in Appendix B.

Then we need to think about the eigenstates that correspond to each eigenvalue. Here, we take $E_{1,-}$, $E_{2,-}$ as examples to obtain the analytical solutions of the order parameter for the quantum phase transition. The expressions of the corresponding eigenstate are
\begin{footnotesize}
\begin{equation}
\begin{aligned}
\phi_{1}\equiv\frac{1}{\sqrt{B_{1}}}\left(|0,1,g \rangle+a_1|1,0,g \rangle+d_1 |0,0,e \rangle\right) \label{4}
 \end{aligned}
\end{equation}
\end{footnotesize}
\begin{footnotesize}
\begin{equation}
\begin{aligned}
\phi_{2}\equiv\frac{1}{\sqrt{B_{2}}}\left({|0,2,g \rangle+a|1,0,e \rangle+b|0,1,e \rangle+c|2,0,g \rangle+d|1,1,g \rangle}\right) \label{3.8}
 \end{aligned}
\end{equation}
\end{footnotesize}
The ground state is $\phi_{0}=|0,0,g \rangle$, and the parameters involved in Eqs. \eqref{4}-\eqref{3.8} are detailed in Eqs. \eqref{3.0}-\eqref{3.9}.

The second order perturbation theory is a commonly used method to study the analytical solution of the superfluid-Mott insulator quantum phase transition[\onlinecite{Tahan}, \onlinecite{Oosten}]. Therefore, we take the hopping term $H^{hop}$ as the perturbation term and calculate the analytical solution of the system. One can get the second-order corrections to the energy and the normalized eigenstaes are shown in Eqs. \eqref{3.14}-\eqref{3.17}.

Then, according to the second-order perturbation theory, we can write the approximative wave function as $\Phi=\frac{1}{\sqrt{N_t}}(\phi _{1}^{(0)}+\phi _{1}^{(1)})$, where $N_t=1+\frac{\left| \left\langle \phi _{2}^{(0)}\left|\hat{a}^{\dagger }\right|\phi _{1}^{(0)}\right\rangle \right|^2}{(E_{1,-}^{(0)}-E_{2,-}^{(0)})^2}+\frac{\left| \left\langle \phi _{0}^{(0)}\left|\hat{a}\right|\phi _{1}^{(0)}\right\rangle \right|^2}{(E_{1,-}^{(0)}-E_{0,0}^{(0)})^2}$ is a normalization coefficient.

Based on the definition formula of the order parameter $\psi$  and the wave function $\Phi$, the analytical formula of the order parameter $\psi$ can be given
\begin{footnotesize}
\begin{equation}
\psi=\sqrt{\frac{\frac{-z\kappa\left|\left\langle \phi _{2}^{(0)}\left|\hat{a}^{\dagger }\right|\phi _{1}^{(0)}\right\rangle\right|^2}{E_{1,-}^{(0)}-E_{2,-}^{(0)}}-\frac{z\kappa\left|\left\langle \phi _{0}^{(0)}\left|\hat{a}\right|\phi _{1}^{(0)}\right\rangle\right|^2}{E_{1,-}^{(0)}-E_{0,0}^{(0)}}-1}{\frac{\left|z\kappa\left\langle \phi _{2}^{(0)}\left|\hat{a}^{\dagger }\right|\phi _{1}^{(0)}\right\rangle\right|^2}{(E_{1,-}^{(0)}-E_{2,-}^{(0)})^2}+\frac{\left|z\kappa\left\langle \phi _{0}^{(0)}\left|\hat{a}\right|\phi _{1}^{(0)}\right\rangle\right|^2}{(E_{1,-}^{(0)}-E_{0,0}^{(0)})^2}}} \label{3.18}
\end{equation}
\end{footnotesize}
Furthermore, the expansion of the energy in power series in $\psi$ can be given by
\begin{footnotesize}
\begin{equation}
E_{1,-}=E_{1,-}{}^{(0)}+E_{1,-}{}^{(2)}+\text{z$\kappa \psi $}^2+O(\psi ^4)
\end{equation}
\end{footnotesize}
$E_{1,-}^{(2)}$ is given by Eq. \eqref{3.4}.

Hereafter, according to Landau's second-order phase transition theory, the phase boundary of the Mott insulator phase and the superfluid phase can be determined when the coefficient of the square term $\psi$ is zero[\onlinecite{Sethna}, \onlinecite{huang}]. Then, the critical hopping rate $\kappa_c$ can be acquired. The system hold the Mott insulator state when $\kappa<\kappa_c$, and in other cases, the system is in a superfluid state. Based on these expressions, the boundaries between the superfluid phase and the Mott insulator phase with the different coupling strength $G_m$ are shown in Figs. \ref{fig5}(e)-\ref{fig5}(f), and the specific analysis will be discussed in the next Section.

\begin{footnotesize}
\begin{displaymath}
z\kappa_c = \left\{ \begin{array}{ll}
   \frac{B_1E_{1,-}}{a_{1}^2} & \textrm{$N=0$;}\\
   \frac{-\left(E_{1,-}^{(0)}-E_{2,-}^{(0)}\right) E_{1,-}^{(0)}}{\left| \left\langle \phi _{2}^{(0)}\left|\hat{a}^{\dagger }\right|\phi _{1}^{(0)}\right\rangle \right|^2E_{2,-}^{(0)}+\left| \left\langle \phi _{0}^{(0)}\left|\hat{a}\right|\phi _{1}^{(0)}\right\rangle \right|^2(E_{1,-}^{(0)}-E_{2,-}^{(0)})} & \textrm{$N=1$;}
    \end{array} \right. \ \tag{17}
    \label{12}
\end{displaymath}
\end{footnotesize}


Up till now, the order parameter $\psi$ and the critical hopping rate are obtained based on a mean-field theory in order to analyze the quantum phase transition of the system. In what follows, we will use a numerical method to discuss the behaviors of the superfluid-Mott insulator quantum phase transition and compare with the analytical results mentioned above against the controlling parameters of our model.
\section{MOTT-SUPERFLUID TRANSITION}\label{IV}

\begin{figure}[!htbp]
\includegraphics[width=4.2cm,height=3.8cm]{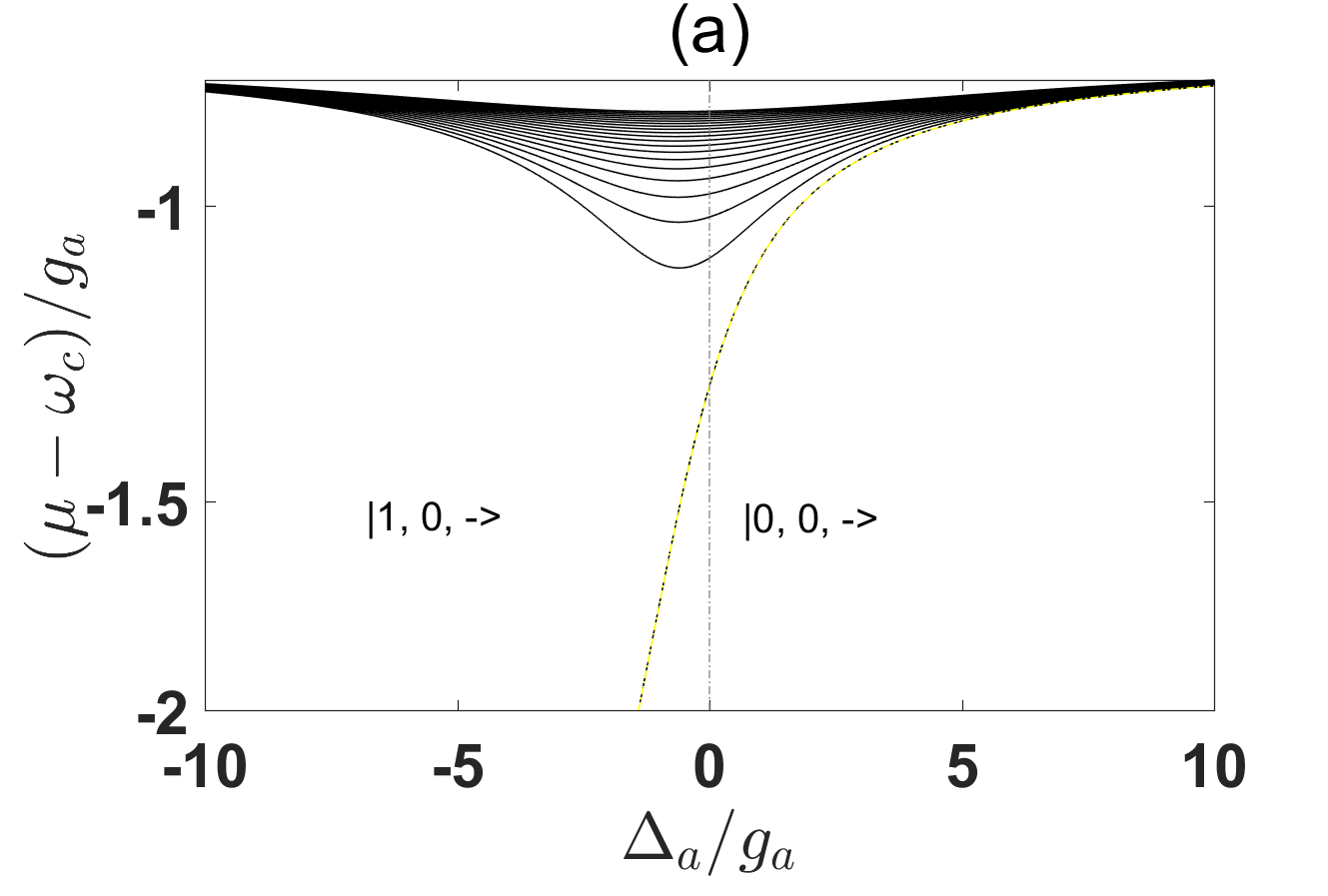}
\includegraphics[width=4.2cm,height=3.8cm]{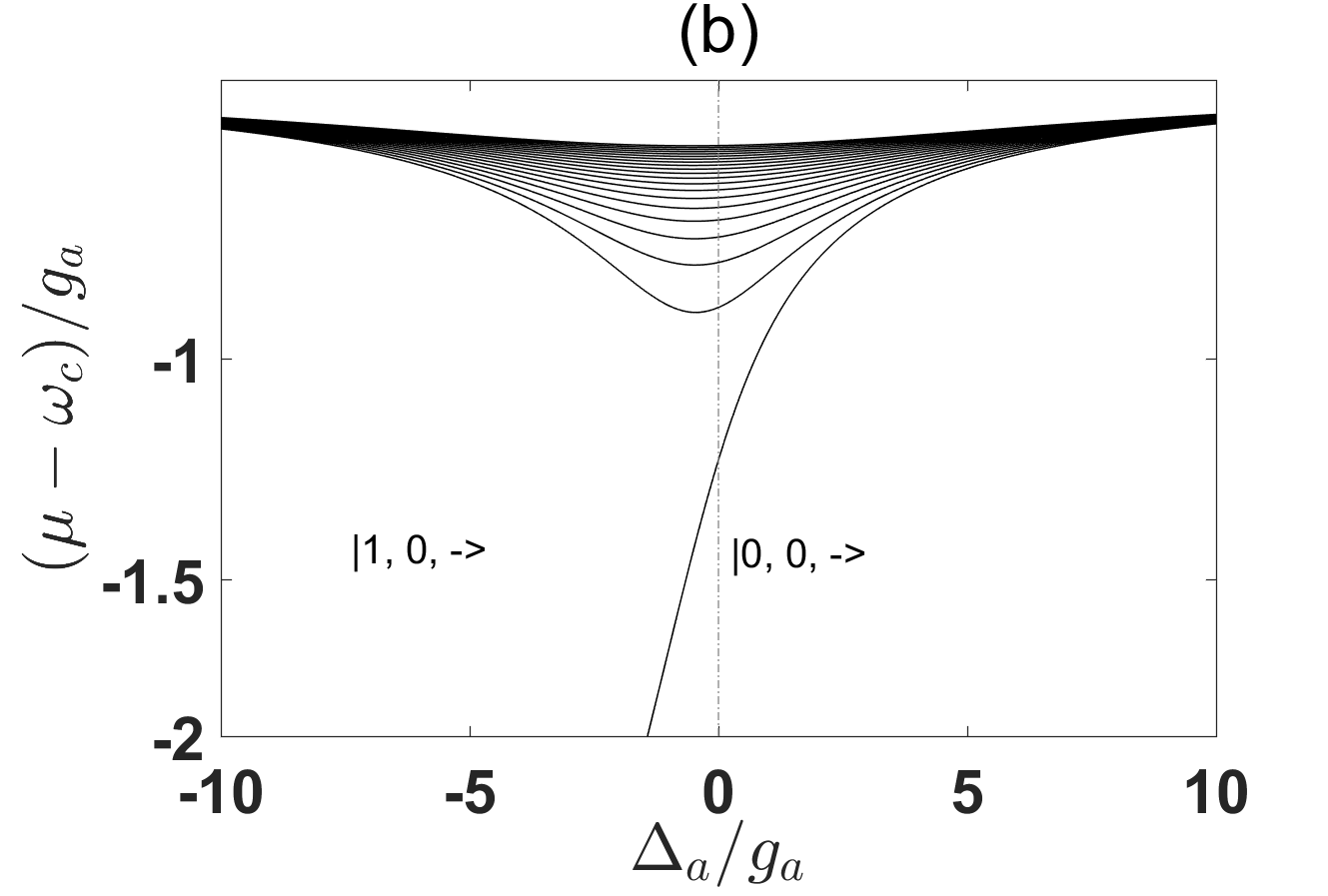}
\includegraphics[width=4.2cm,height=3.8cm]{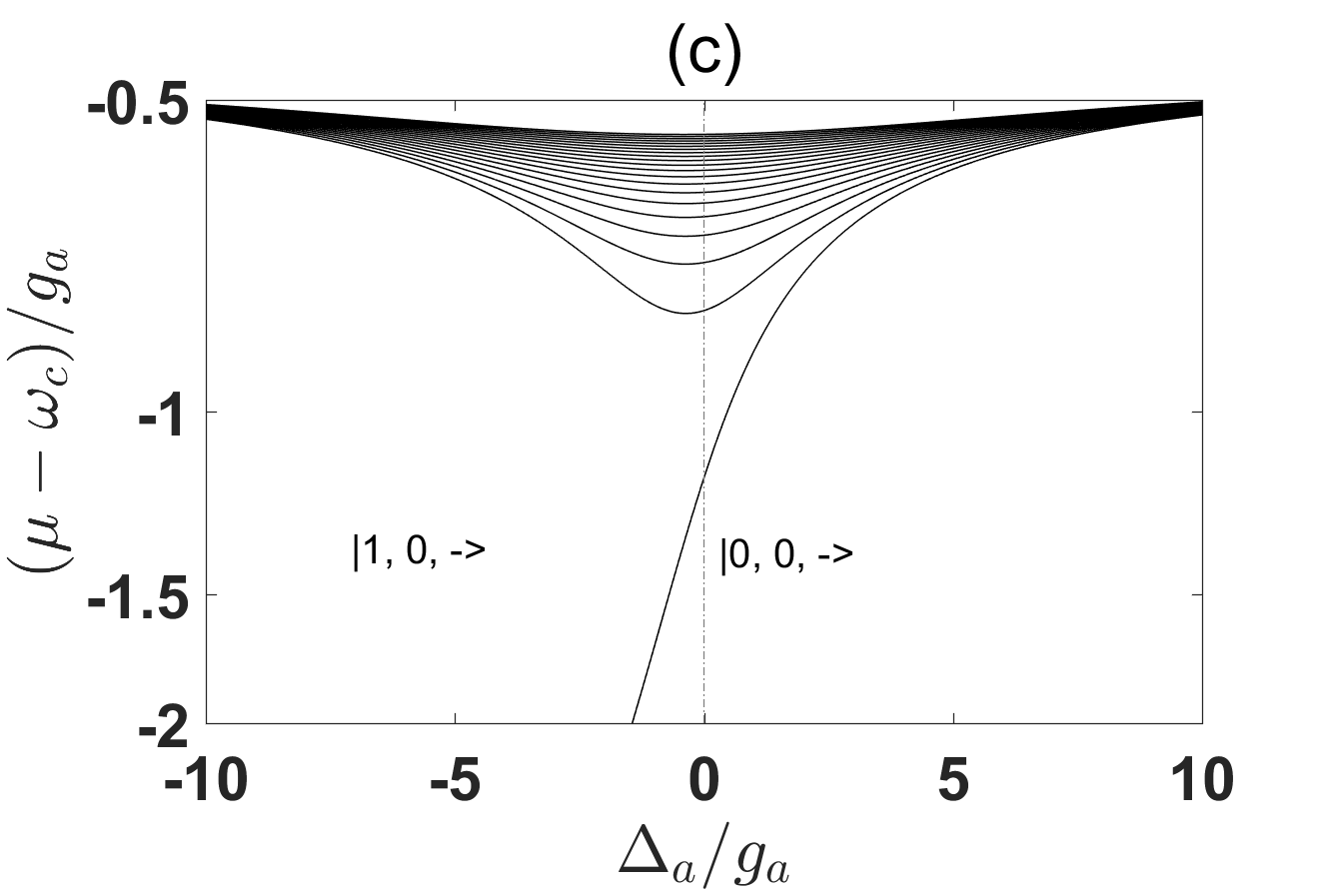}
\includegraphics[width=4.2cm,height=3.8cm]{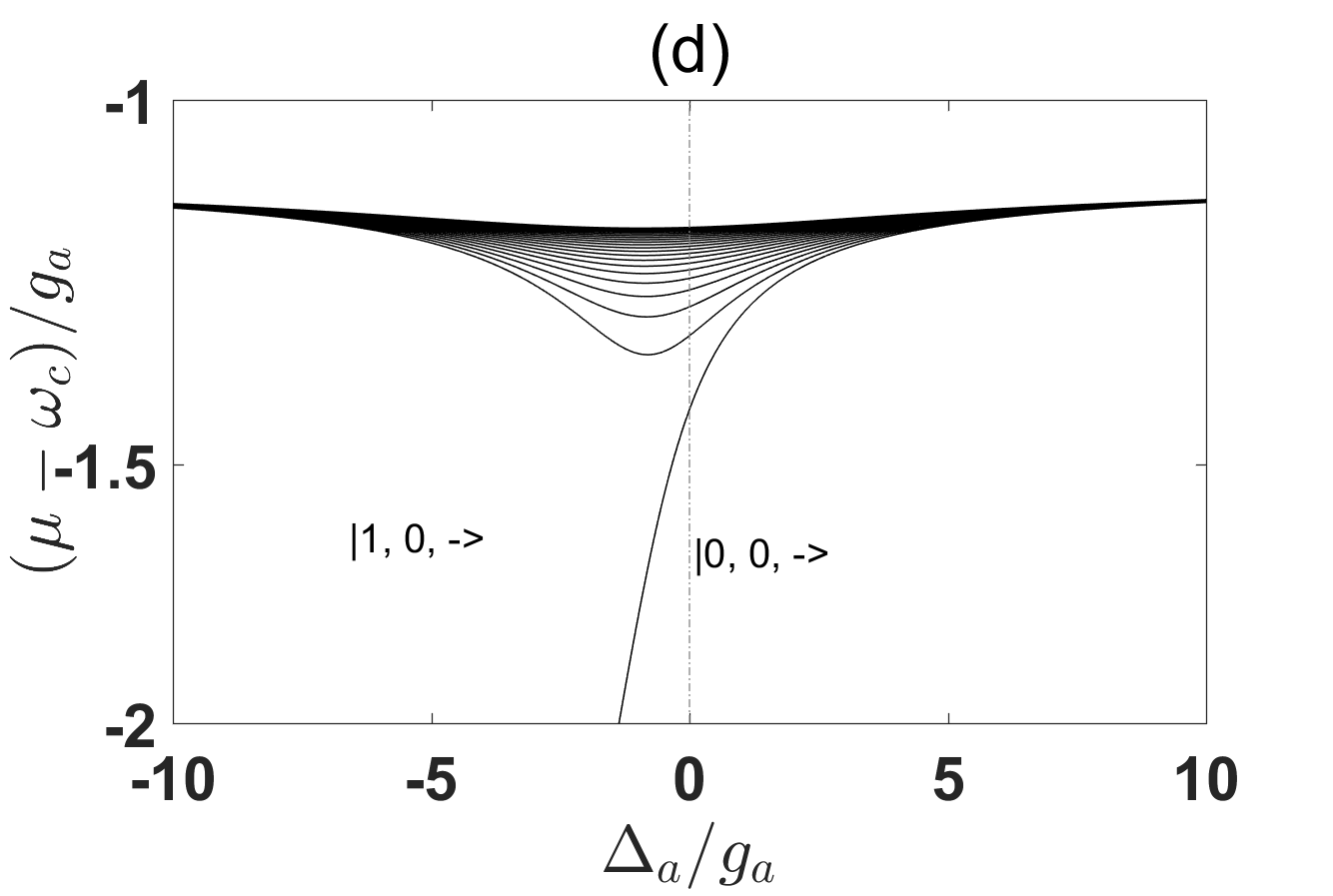}
\includegraphics[width=4.2cm,height=3.8cm]{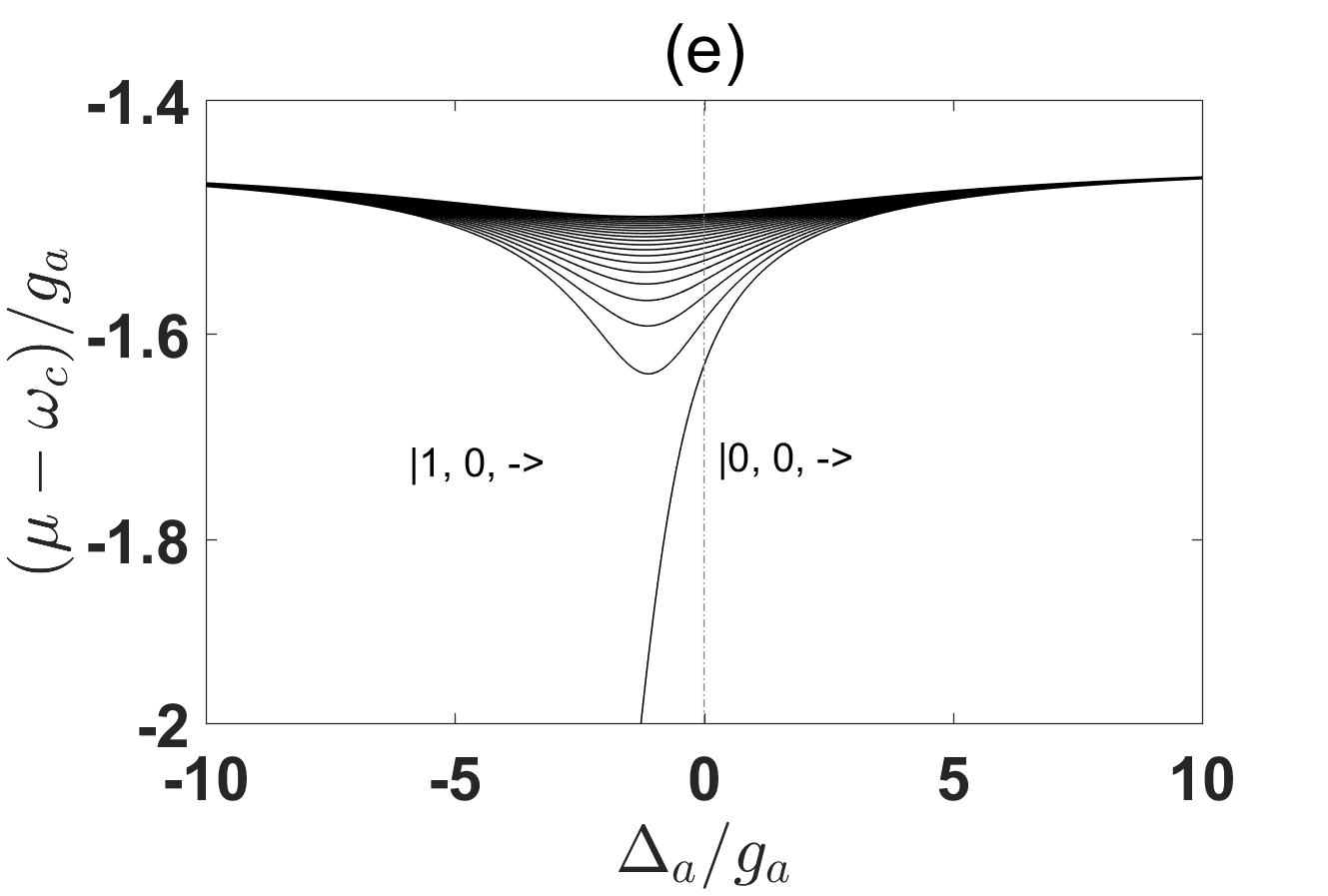}
\includegraphics[width=4.2cm,height=3.8cm]{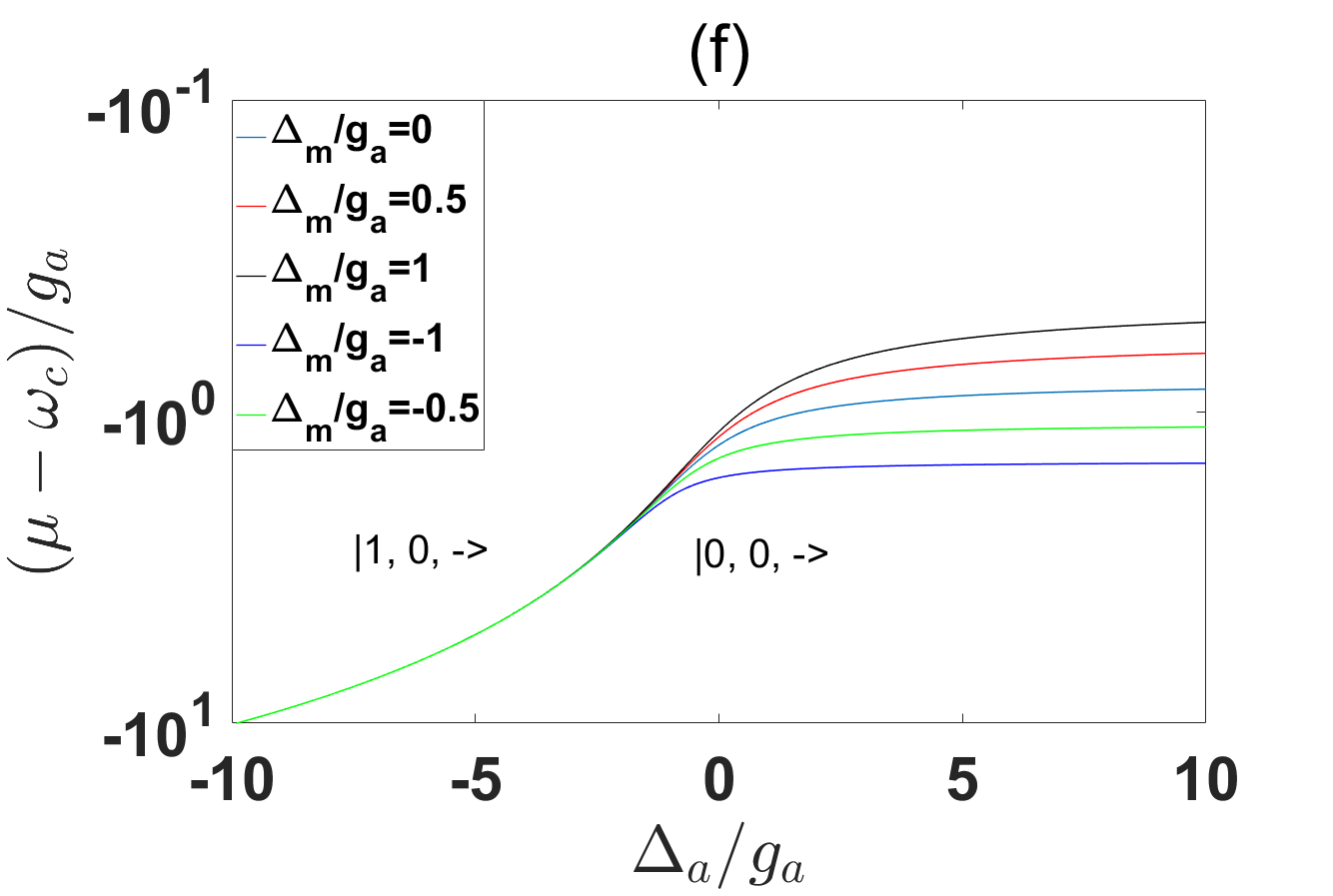}
\caption{Boundaries between different Mott lobes as a function of $\mu$ and $\Delta_a$ when the hopping rate approaches zero for different $\Delta_m$ with $G_m/g_a=0.8$. (a) $\Delta_m=0$, The yellow dashed line is the analytical result between states $|0, 0, -\rangle$ and $|1, 0, -\rangle$. (b) $\Delta_m=0.5$, (c) $\Delta_m=1$, (d) $\Delta_m=-0.5$, (e) $\Delta_m=-1$. And (f) exhibits the boundary between states $|0, 0, -\rangle$ and $|1, 0, -\rangle$ for different $\Delta_m$.}
\label{fig2}
\end{figure}

We first investigate the critical chemical potential as a function of $\Delta_a$, which is usually defined as $E_{N+1,-}-\mu(n+1)=E_{N,-}-\mu n$. Figure \ref{fig2} exhibits the change of boundaries between different Mott lobes for various detuning $\Delta_m$. Meanwhile, the analytic solution between states $|0, 0, -\rangle$ and $|1, 0, -\rangle$ is also shown in Fig. \ref{fig2}(a). Obviously, the analytic solutions are conformed with the nuermical solution on the lower excitation number. Figures \ref{fig2}(a)-\ref{fig2}(e) also show the Mott lobes with different $\Delta_m$, which exhibit that the Mott lobes are smaller and closer with $|\Delta_a/g_a|$ increasing. That's means that the regions of stability become observably smaller with the excition number and the detuning $|\Delta_a/g_a|$ inreasing and it's easy to notice that the states $|0, 0, -\rangle$ and $|1, 0, -\rangle$ are the most stable ones in Figs. \ref{fig2}(a)-\ref{fig2}(e). The phase boundary between the lowest and the second lowest states with different $\Delta_m$ is shown in Fig. \ref{fig2}(f), which illustrates that the stable region of state $|0, 0, -\rangle$ becomes large with $\Delta_m$ increasing. In addition, it can be found that the Mott lobes are  asymmetric with respect to $\Delta_a$, which is different compared with JCH model[\onlinecite{Tahan}]. This means that the phase boundaries at higher excitation are also asymmetric with respect to $\Delta_a$. Compared Figs. \ref{fig2}(b)-\ref{fig2}(c) with Figs. \ref{fig2}(d)-\ref{fig2}(e), it can also be found that the stable regions for a  negative detuning is smaller than ones of the positive detuning.

\begin{figure}[!htbp]
\includegraphics[width=4.2cm,height=3.8cm]{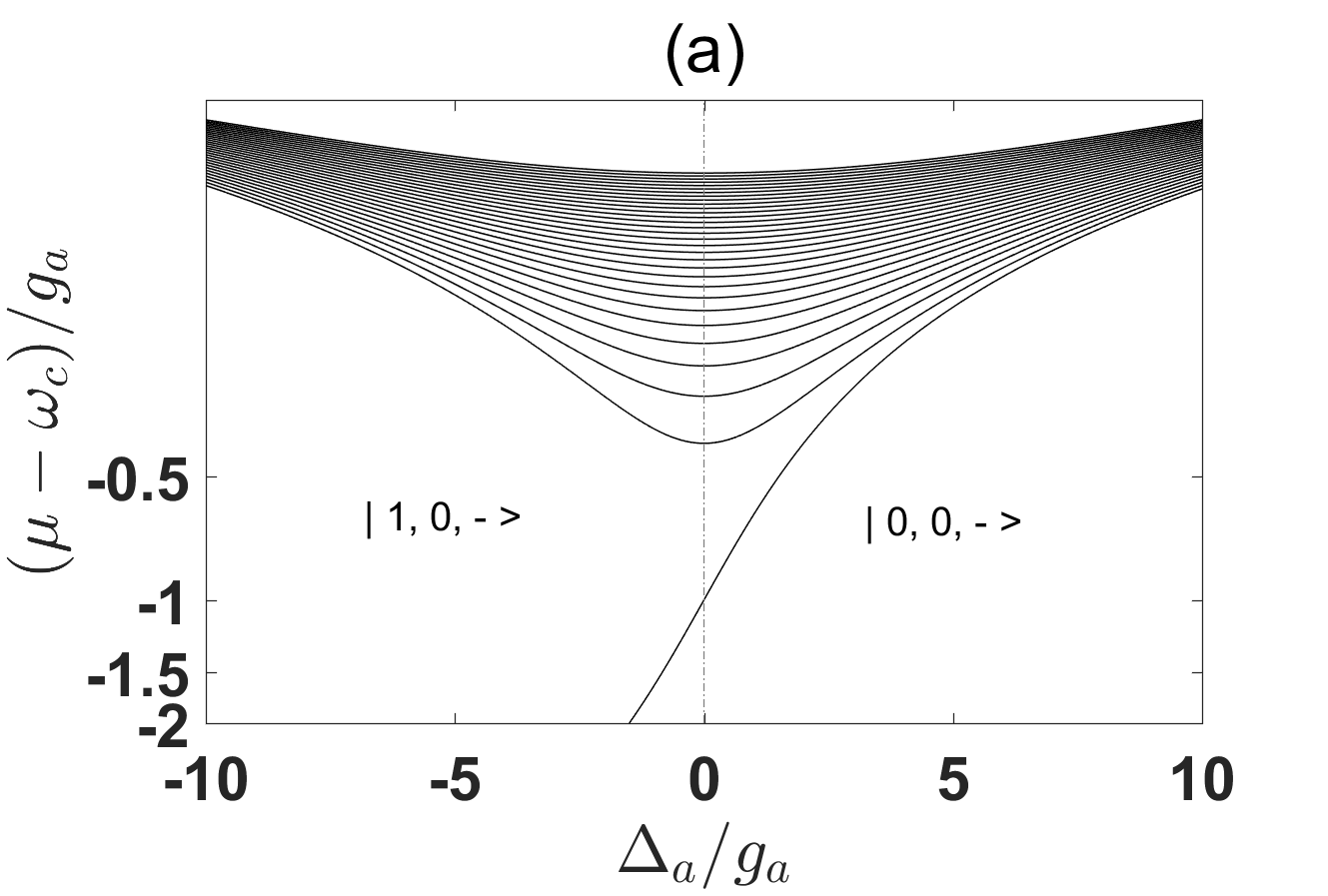}

\includegraphics[width=4.2cm,height=3.8cm]{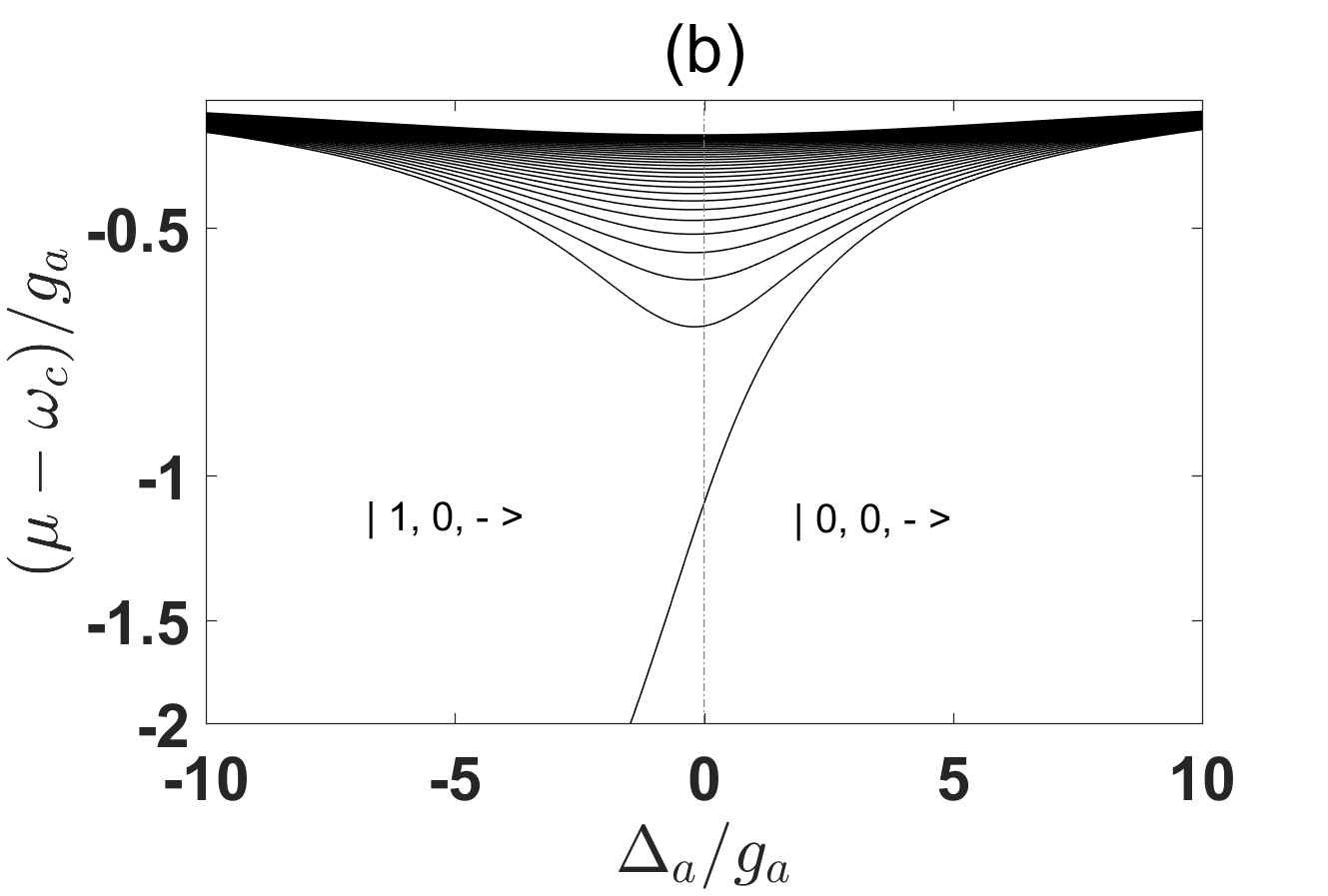}
\includegraphics[width=4.2cm,height=3.8cm]{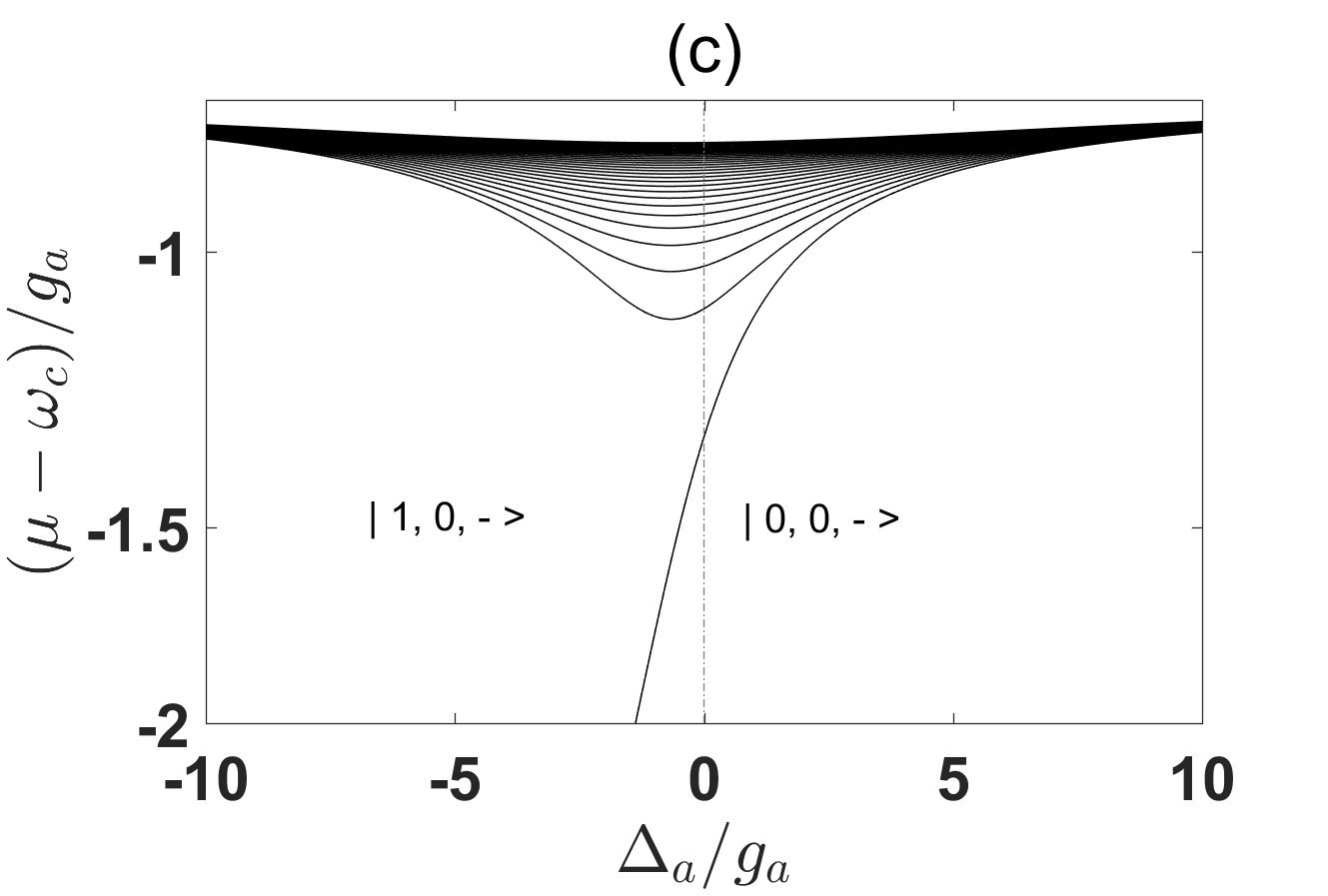}
\includegraphics[width=4.2cm,height=3.8cm]{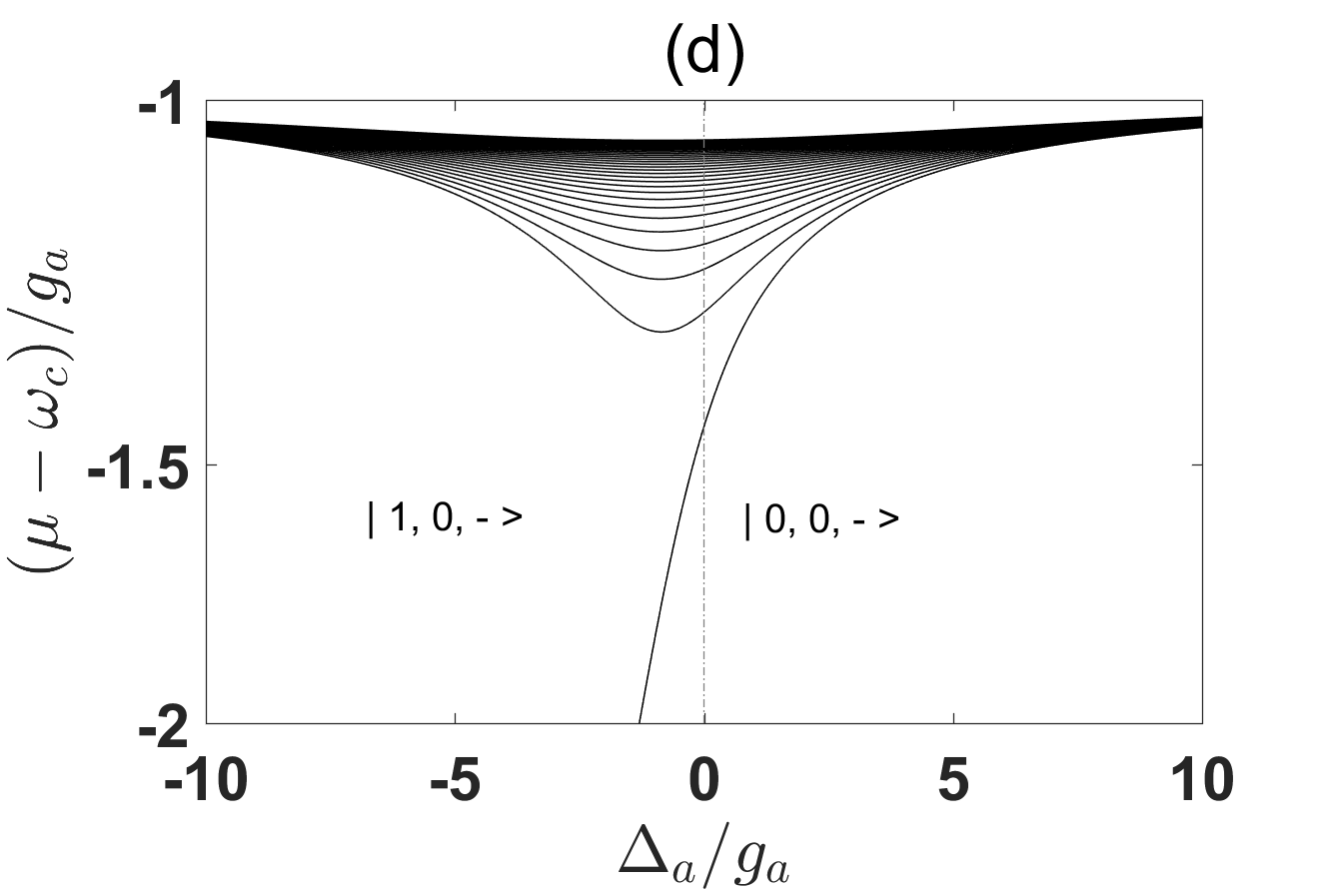}
\includegraphics[width=4.2cm,height=3.8cm]{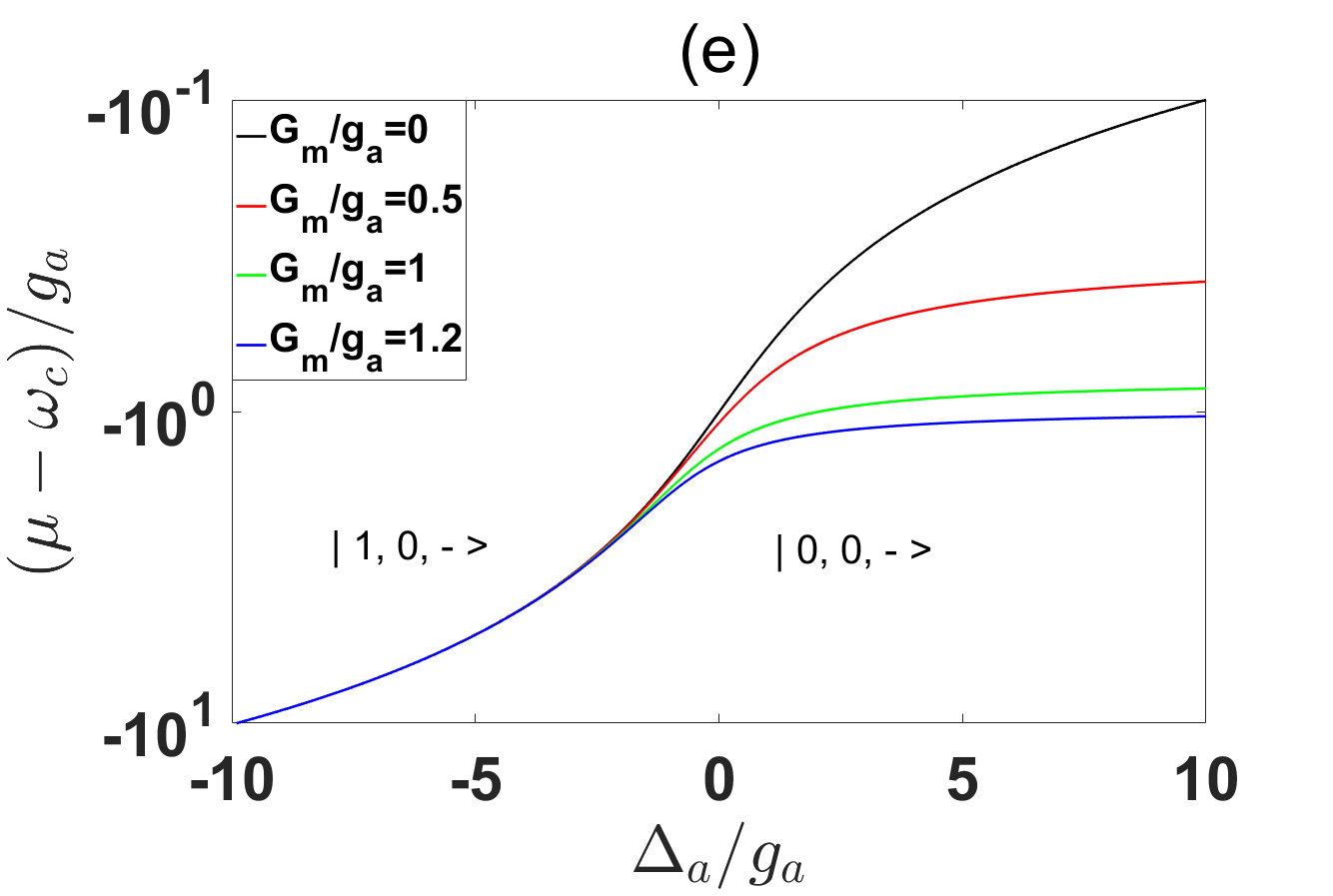}
\caption{Boundaries between Mott lobes as a function of $\mu$ and $\Delta_a$ when the hopping rate approaches zero for different $G_m$ with $\Delta_m=0.5$. (a) $G_m/g_a=0$. (b) $G_m/g_a=0.5$, (c) $G_m/g_a=1$, and (d) $G_m/g_a=1.2$, and (f) exhibits the boundary between states $|0, 0, -\rangle$ and $|1, 0, -\rangle$ with different $ G_m/g_a$.}
\label{fig3}
\end{figure}
Figure \ref{fig3} show the critical chemical potential as a function of $\Delta_a$ for different coupling strength $G_m/g_a$. As shown in Figs. \ref{fig3}(a)-\ref{fig3}(d), the Mott lobes get smaller and closer together with increasing $|\Delta_a/g_a|$, which means that the stable area decreases with $|\Delta_a/g_a|$ increasing. At the same time, the stable area decreases with the increase of the total excitation number, which can be observed in Figs. \ref{fig3}(a)-\ref{fig3}(d). However, it is easy to see that all Mott lobes are symmetric with respect to the detuning $\Delta_a$, except for the states between $|0, 0, -\rangle$ and $|1, 0, -\rangle$ according to Fig. \ref{fig3}(a), which means that the phase boundary at higher excitation number is symmetric only when $G_m/g_a=0$. Furthermore, the stable region of state $|0, 0, -\rangle$ decreases with increasing of the coupling strength $G_m/g_a$ as shown in Fig. \ref{fig3}(e).

\begin{figure}[!htbp]
\includegraphics[width=4.2cm,height=3.0cm]{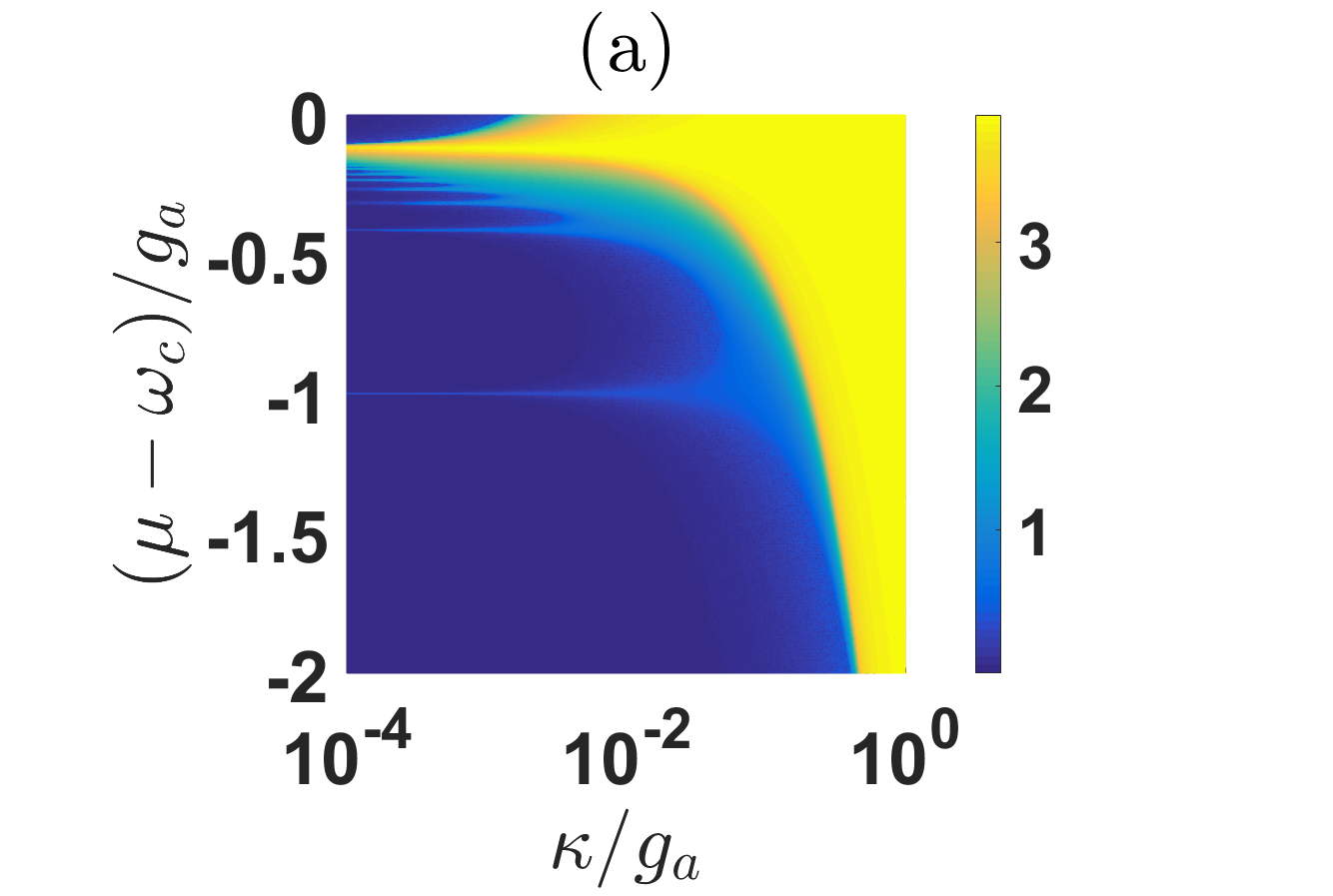}
\includegraphics[width=4.2cm,height=3.0cm]{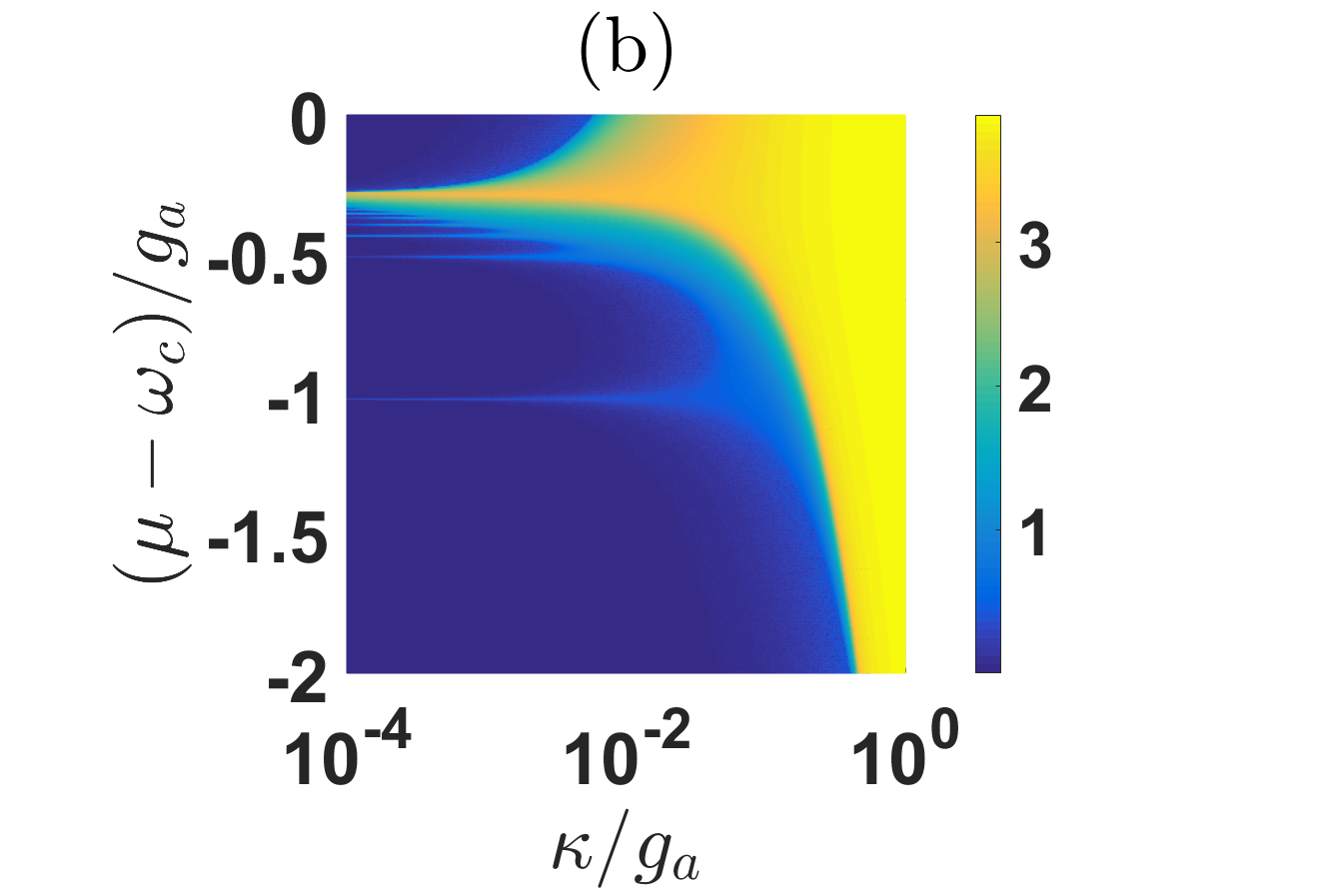}
\includegraphics[width=4.2cm,height=3.0cm]{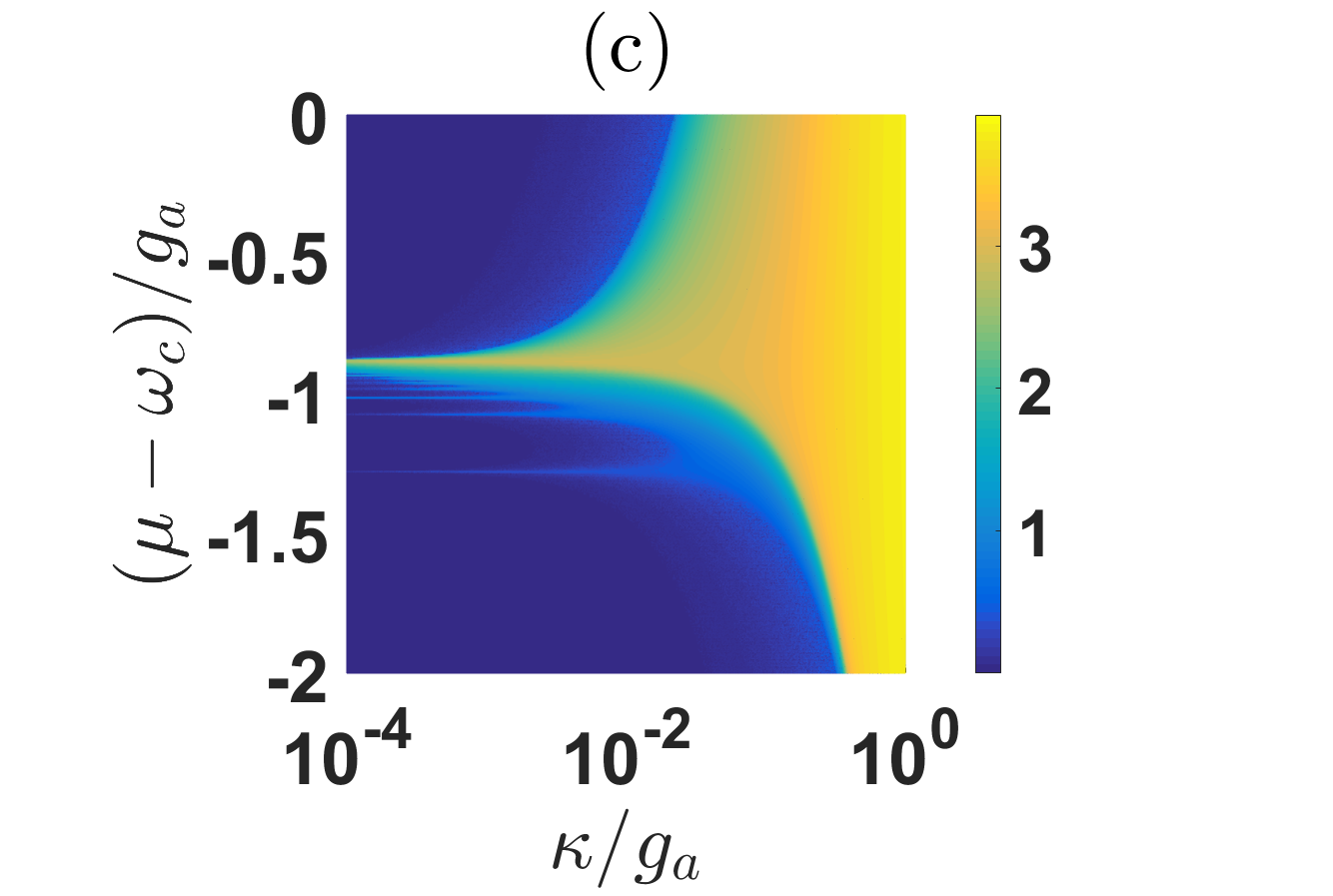}
\includegraphics[width=4.2cm,height=3.0cm]{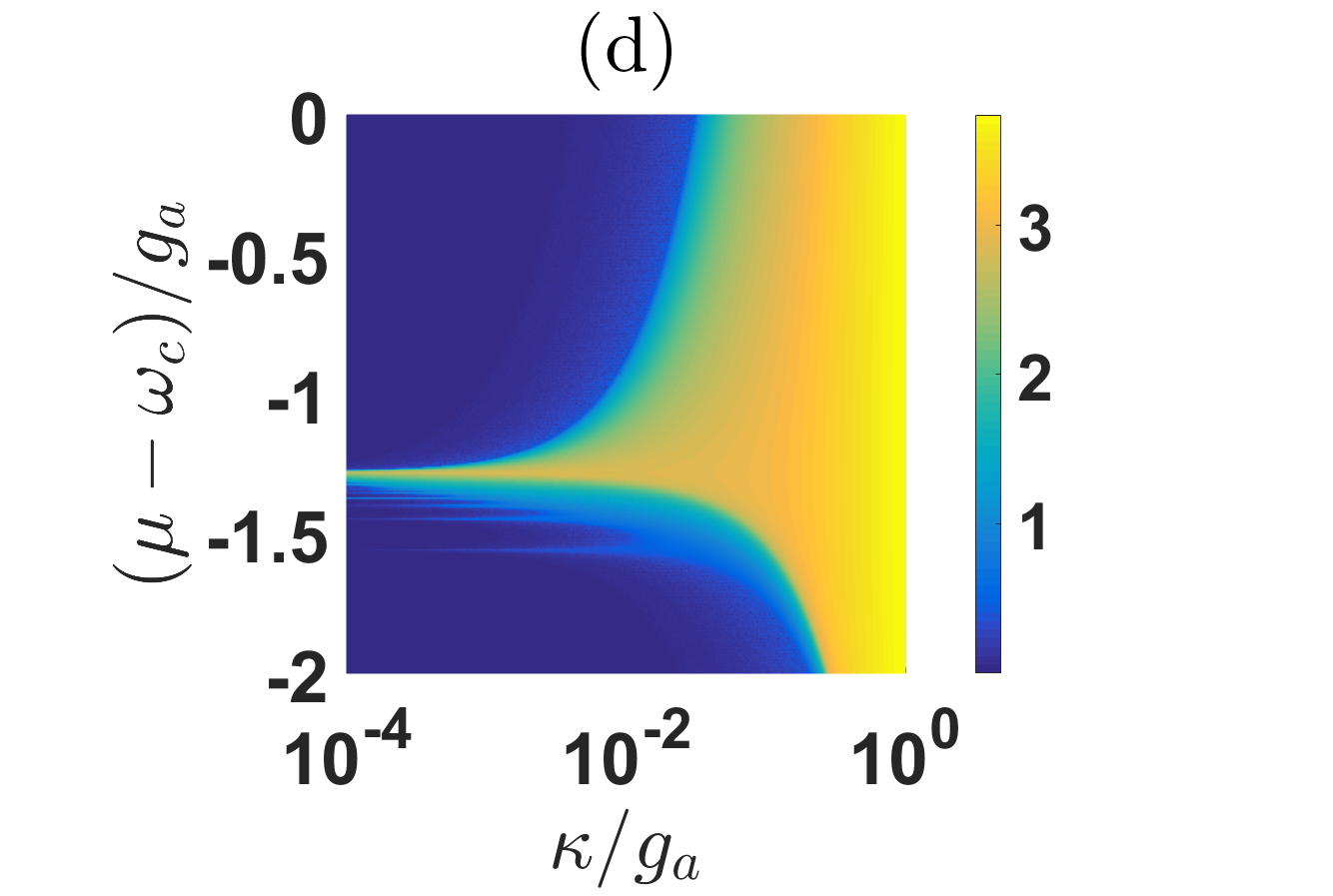}
\includegraphics[width=3.8cm,height=3.0cm]{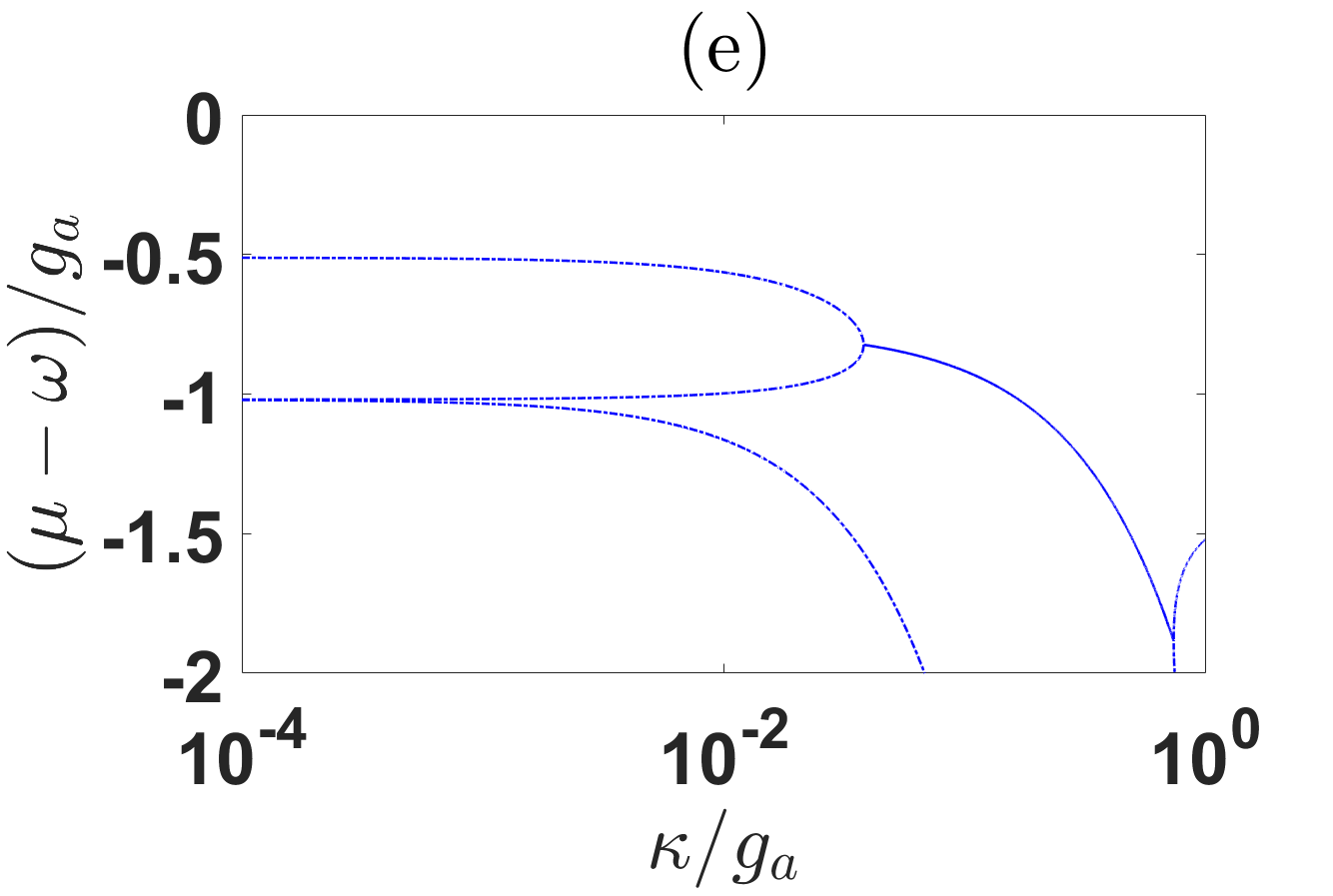}
\includegraphics[width=3.8cm,height=3.0cm]{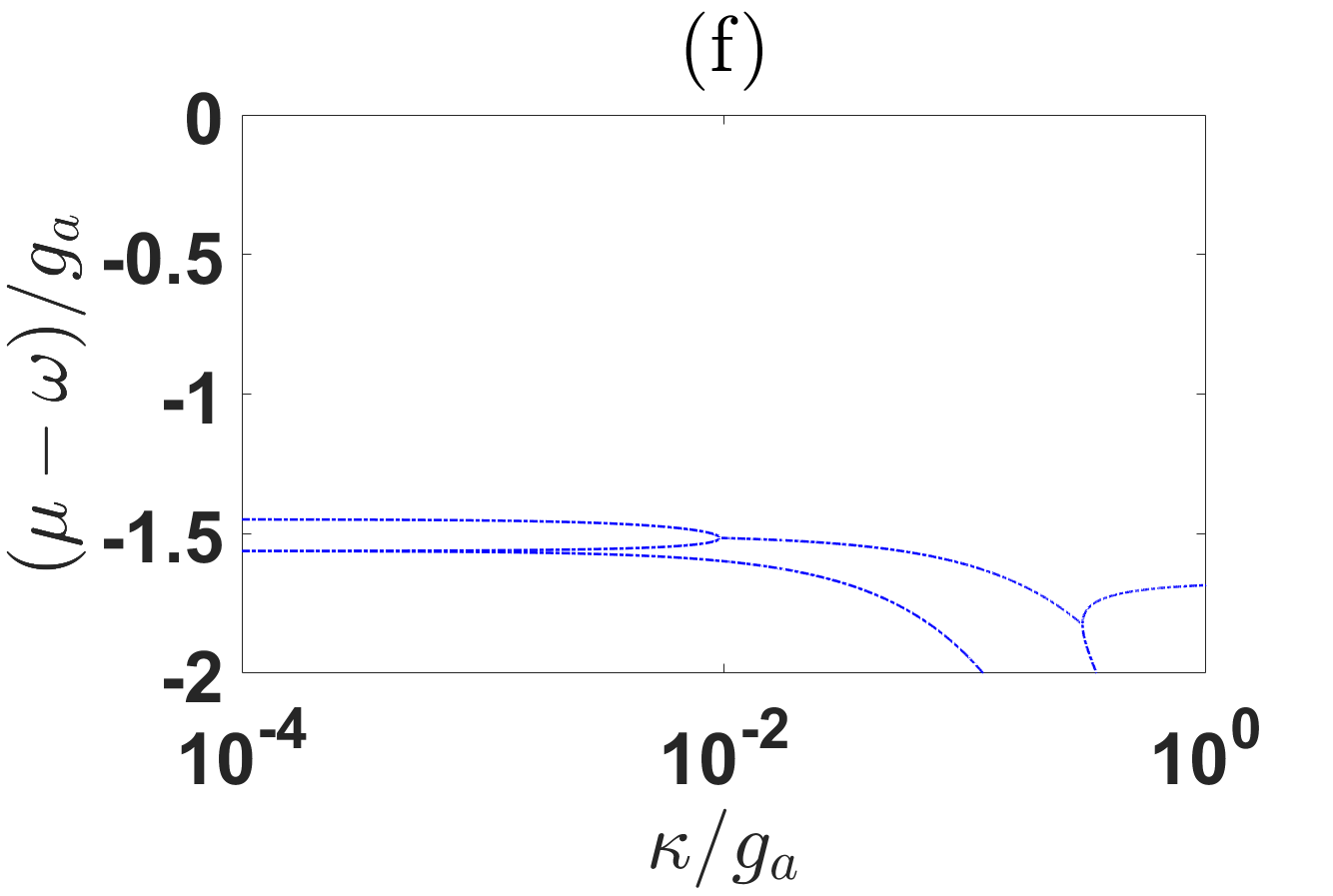}
\caption{The superfluid order parameter as a function of the photon hopping rate $\kappa$ and the chemical potential $\mu$ for different coupling strengths $G_m$ with $\Delta_m=\Delta_a=0$. (a) $G_m/g_a=0$, (b) $G_m/g_a=0.2$, (c) $G_m/g_a=0.8$, (d) $G_m/g_a=1.2$. (e)-(f) Analytical boundaries between different phases as a function of $\mu-\omega_c$ and $\kappa/g_a$ with $\Delta_a=\Delta_m=0$. (e) $G_m=0.2$, (f) $G_m=1.2$.}
\label{fig5}
\end{figure}
After investigating the boundaries between different Mott lobes within the dressed-state formalism in our system, we will calculate the phase diagram by applying the mean-field theory. The well-known feature of the superfluid-Mott insulator quantum  phase transition is the Mott lobe which is exhibited in Fig. \ref{fig5} for different values of $G_m$ with a cutoff $N=20$ to the total excitation number basis so that $0<m<N$. Note that, without photon-magnon interaction in Eqs.\eqref{2.1}-\eqref{2.2}, i.e., $G_m=0$, the hybrid system considered here can be reduced to the usual JCH model. In order to make a comparison with the cases discussed below, the phase diagram for the JCH model is also plotted in Fig. \ref{fig5}(a). One can easily find the superfluid-Mott insulator quantum phase transition that the parameter space is separated into two distinct phases as shown in Figs. \ref{fig5}(b)- \ref{fig5}(d). There are clearly the superfluid phase corresponding to the regions where $\psi\neq0$ for large hopping rate $\kappa$, and the stable ground state of each site is a coherent state. The Mott insulator phase corresponds to the case of $\psi=0$ for a small hopping rate $\kappa$. Each Mott lobe corresponds to a state with an integer number of the total excitations per site. Compared Figs. \ref{fig5}(b)- \ref{fig5}(d) with Fig. \ref{fig5}(a), one can find that a small hopping rate is needed to delocalize the photons and make them enter the superfluid phase. Additionally, the enhanced photon-magnon coupling strength $G_m$ causes the region of each Mott lobe to decrease and the superfluid phase area is increased with the increasing of $G_m$ correspondingly. These results mean that the coupling of the photon-magnon favors the superfluid phase, which is consistent with the results shown in Fig. \ref{fig3}. Furthermore, the analytical results based on Eq. \eqref{12} are also shown in Figs. \ref{fig5}(e)-\ref{fig5}(f), which are described by the blue dashed contours. As expected, we find well agreement between this analytical calculation and the full mean-field calculation in determining the boundary between the superfluid and the Mott insulator phase when the hopping rate is weak. By comparing Fig. \ref{fig5}(e) (\ref{fig5}(f)) with Fig. \ref{fig5}(b) (\ref{fig5}(d)), the analytical results obtained from the second-order perturbation theory are no longer applicable at large hopping rate.

\begin{figure}[!htbp]
\includegraphics[width=4.2cm,height=3.0cm]{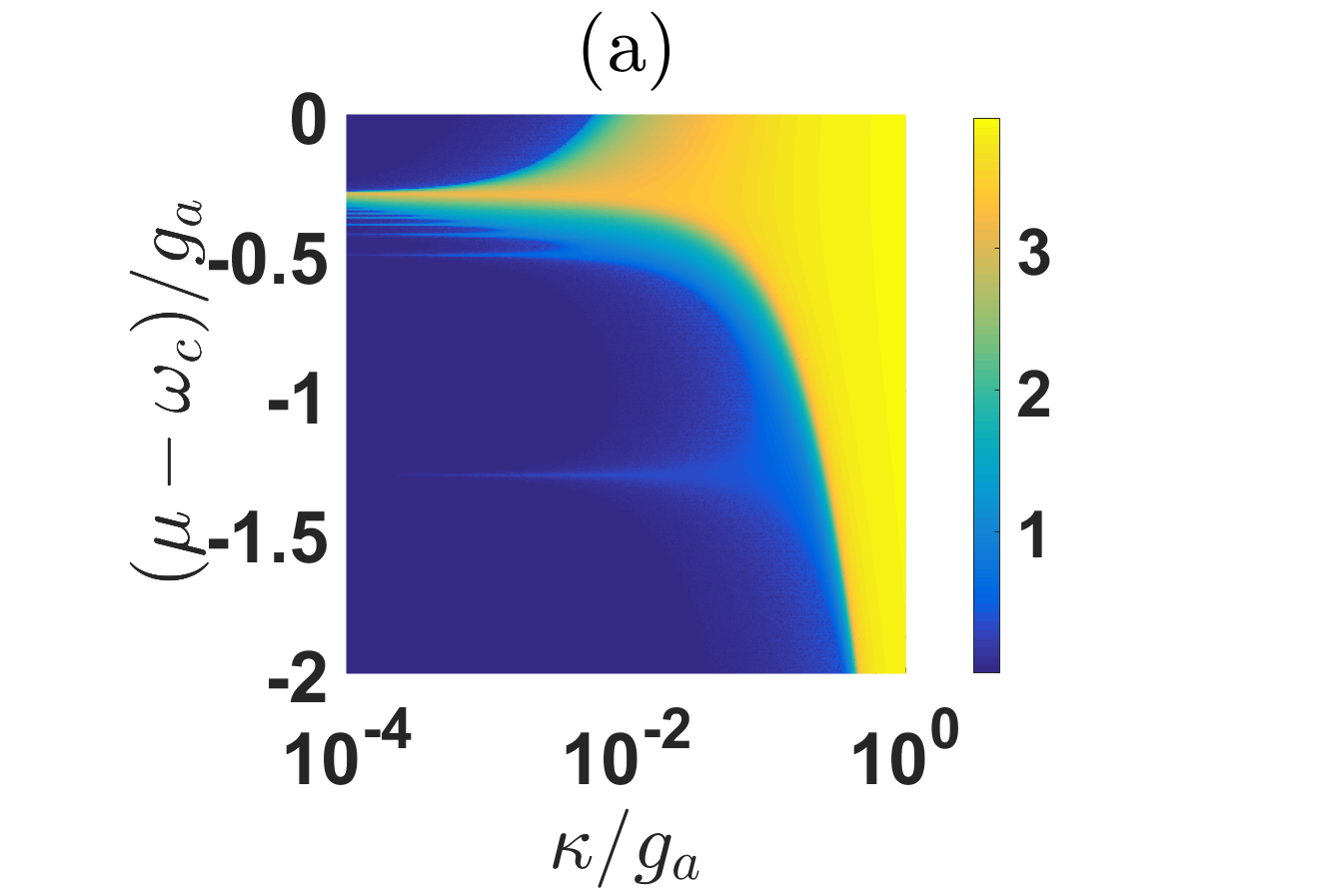}

\includegraphics[width=4.2cm,height=3.0cm]{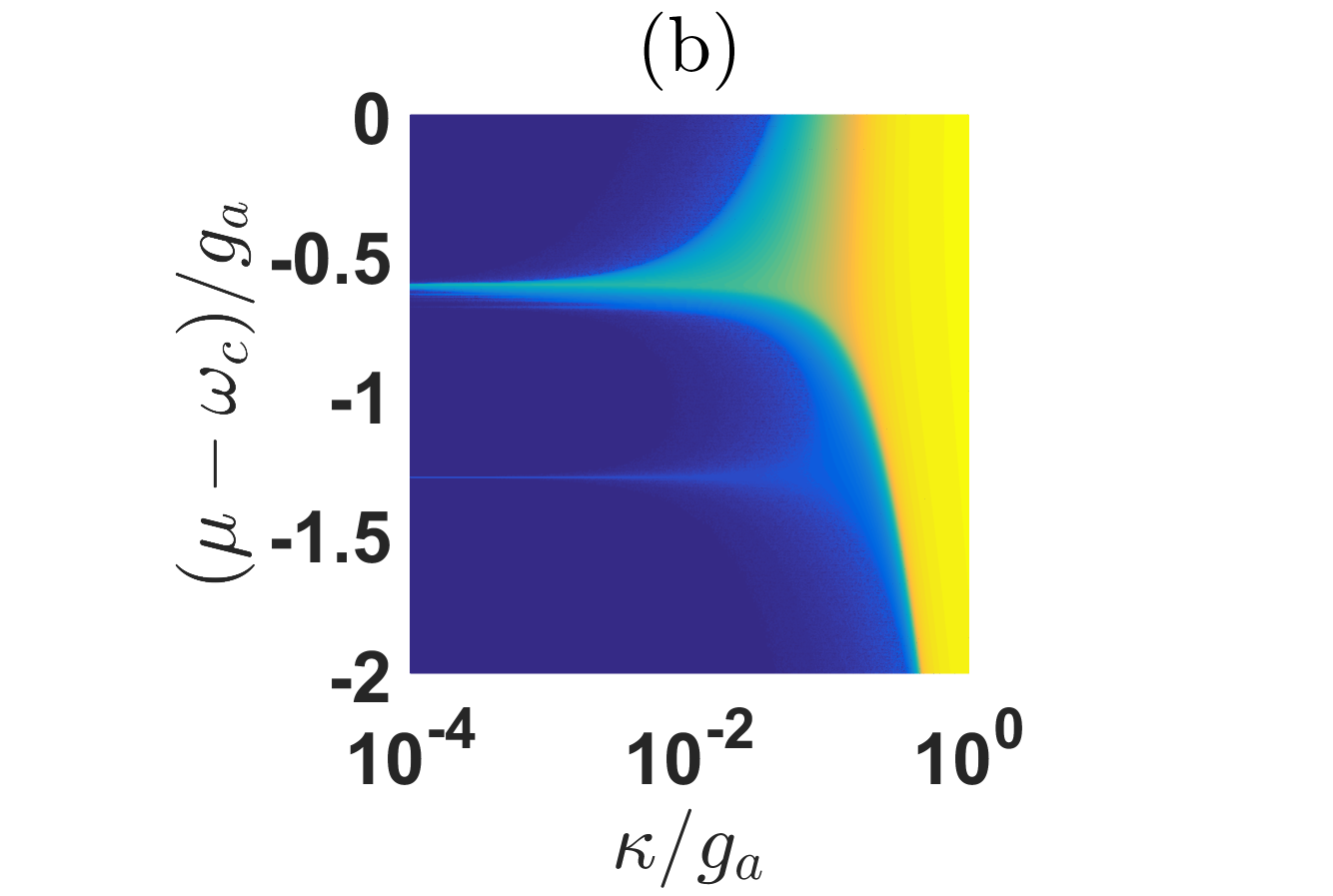}
\includegraphics[width=4.2cm,height=3.0cm]{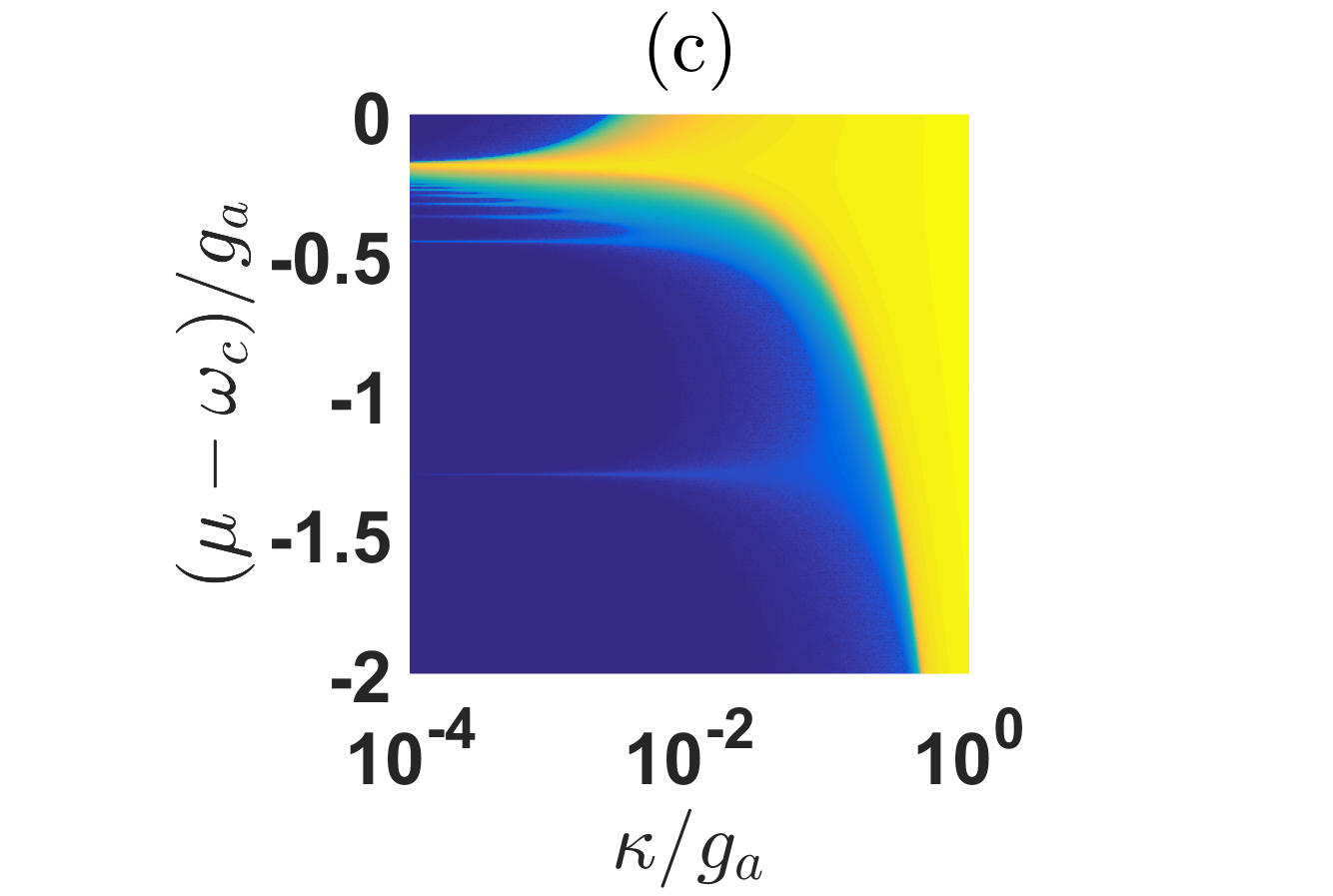}
\includegraphics[width=4.2cm,height=3.0cm]{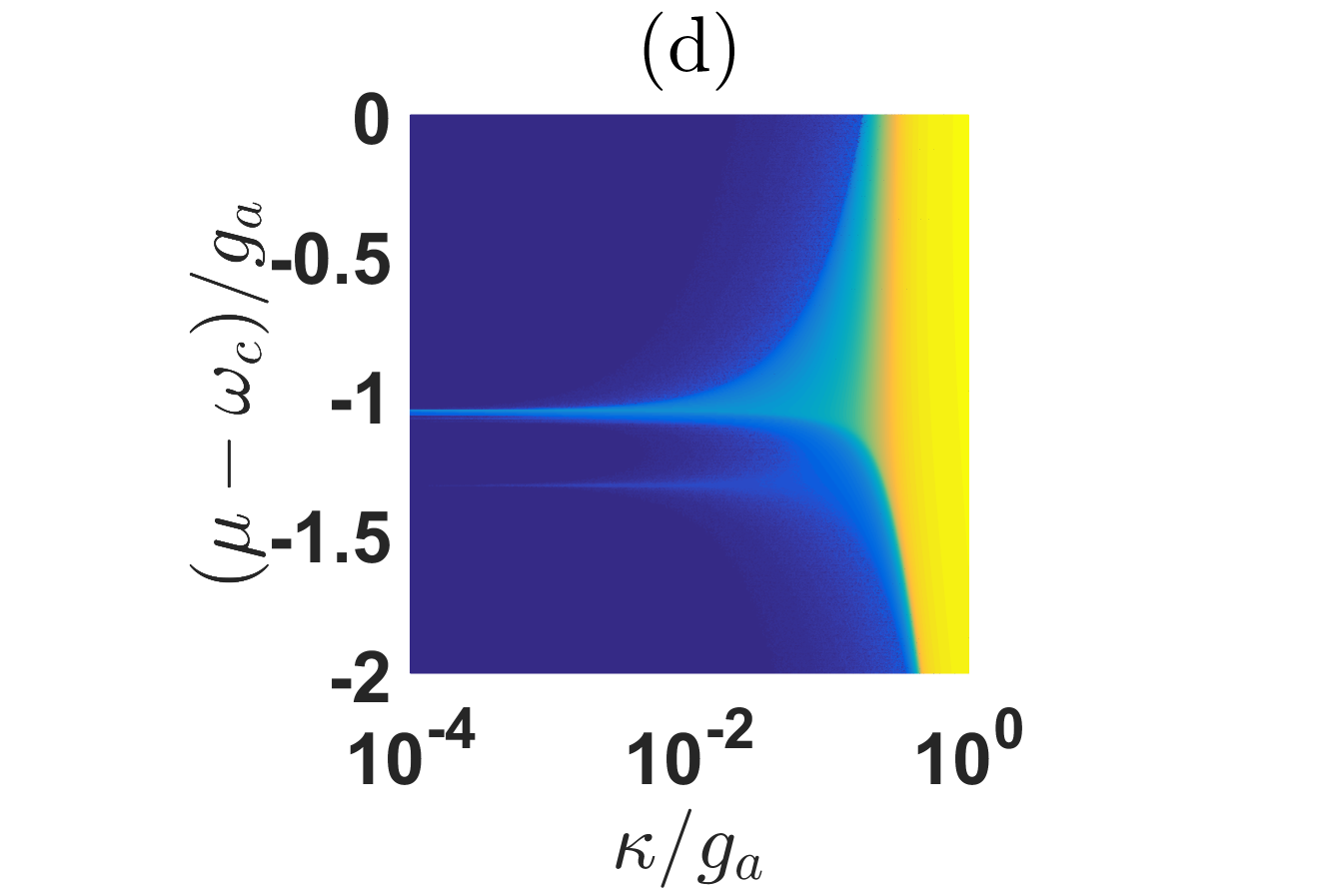}
\includegraphics[width=4.2cm,height=3.0cm]{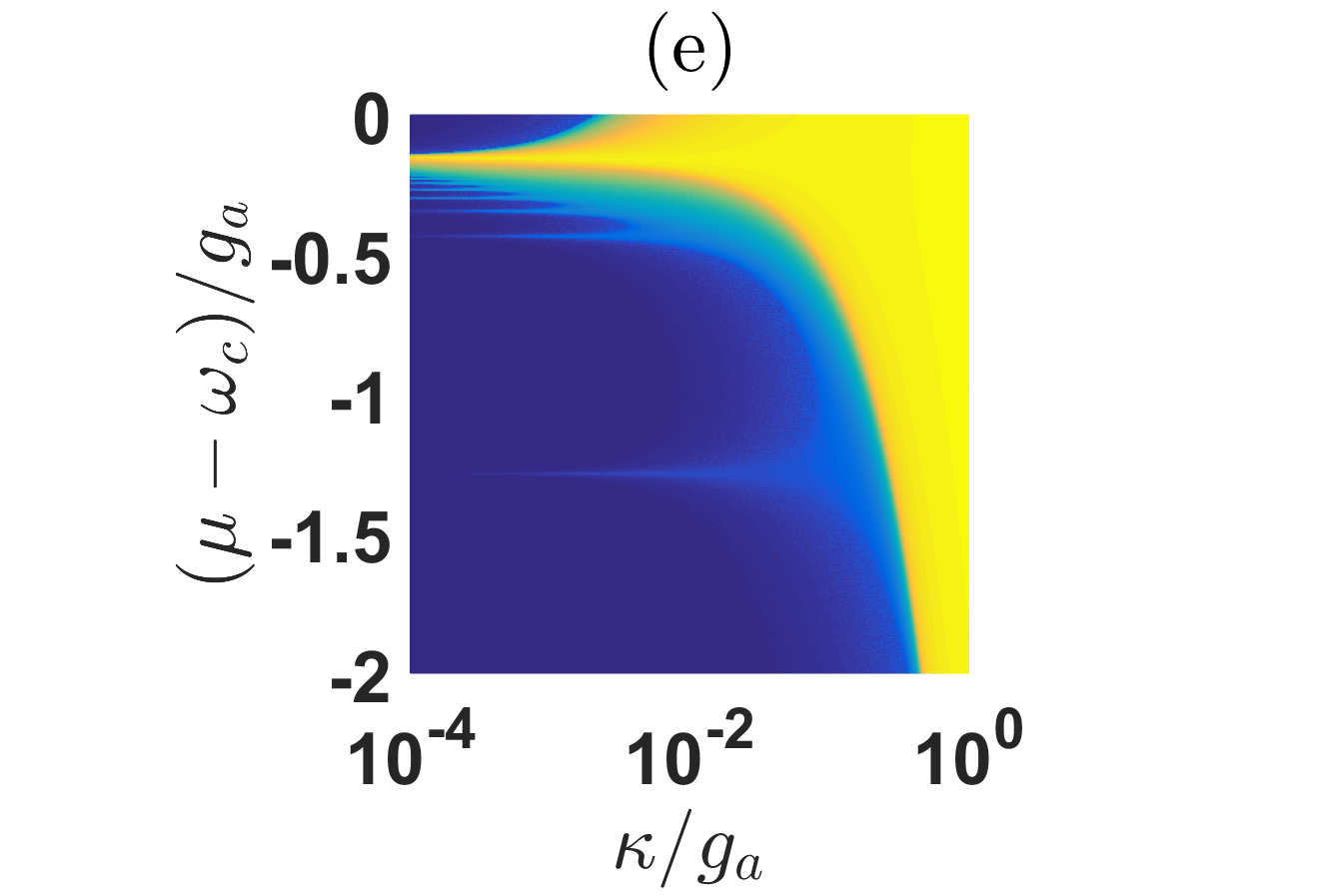}
\caption{The superfluid order parameter as a function of the photon hopping rate $\kappa$ and the chemical potential $\mu$ for different detuning of magnon and cavity photon $\Delta_m$ with $G_m=0.2, \Delta_a=0.5$.  (a) $\Delta_m/g_a=0$, (b) $\Delta_m/g_a=-0.5$, (c) $\Delta_m/g_a=0.5$, (d) $\Delta_m/g_a=-1$, (e) $\Delta_m/g_a=1$.}
\label{fig6}
\end{figure}
According to the H-P transformation, the magnon frequency can be regulated by a bias magnetic field[\onlinecite{Dany}, \onlinecite{Holstein}, \onlinecite{Huebl}]. Therefore, the detuning $\Delta_m$ between the cavity photon and the magnon can be used as an experimentally feasible parameter for adjusting the superfluid-Mott insulator quantum phase transition. Figure \ref{fig6} exhibits the change of the phase diagram for different detuning $\Delta_m$. Figures \ref{fig6}(a), \ref{fig6}(c) and \ref{fig6}(e) show that the region of each Mott lobe does not change significantly but tends to increase slightly for the detuning $\Delta_m$ postive increasing. While for a negative detuning, i.e., $\Delta_m<0$, the effect on the transition of the superfluid-Mott insulator can be enhanced as shown in Figs. \ref{fig6}(b) and \ref{fig6}(d). It can be found that the detunings decreases not only reduce the area of the Mott lobe, but also have a tendency to diminish the Mott lobe with large total excitation number $N$. Then it is more favorable for the generation of highly excited Mott lobes when $\Delta_m$ is positive.

\begin{figure}[!htbp]
\includegraphics[width=4.2cm,height=3.0cm]{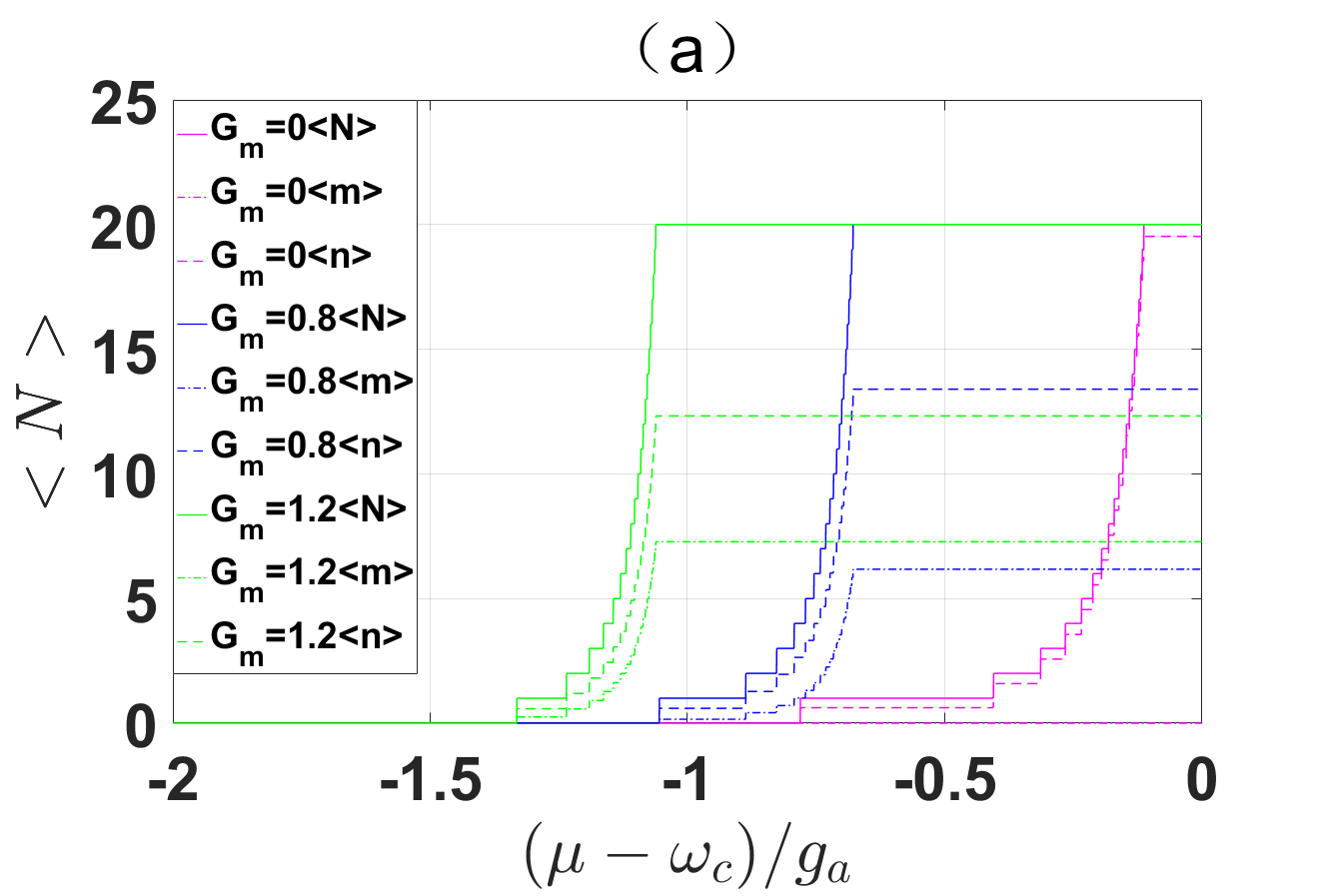}
\includegraphics[width=4.2cm,height=3.0cm]{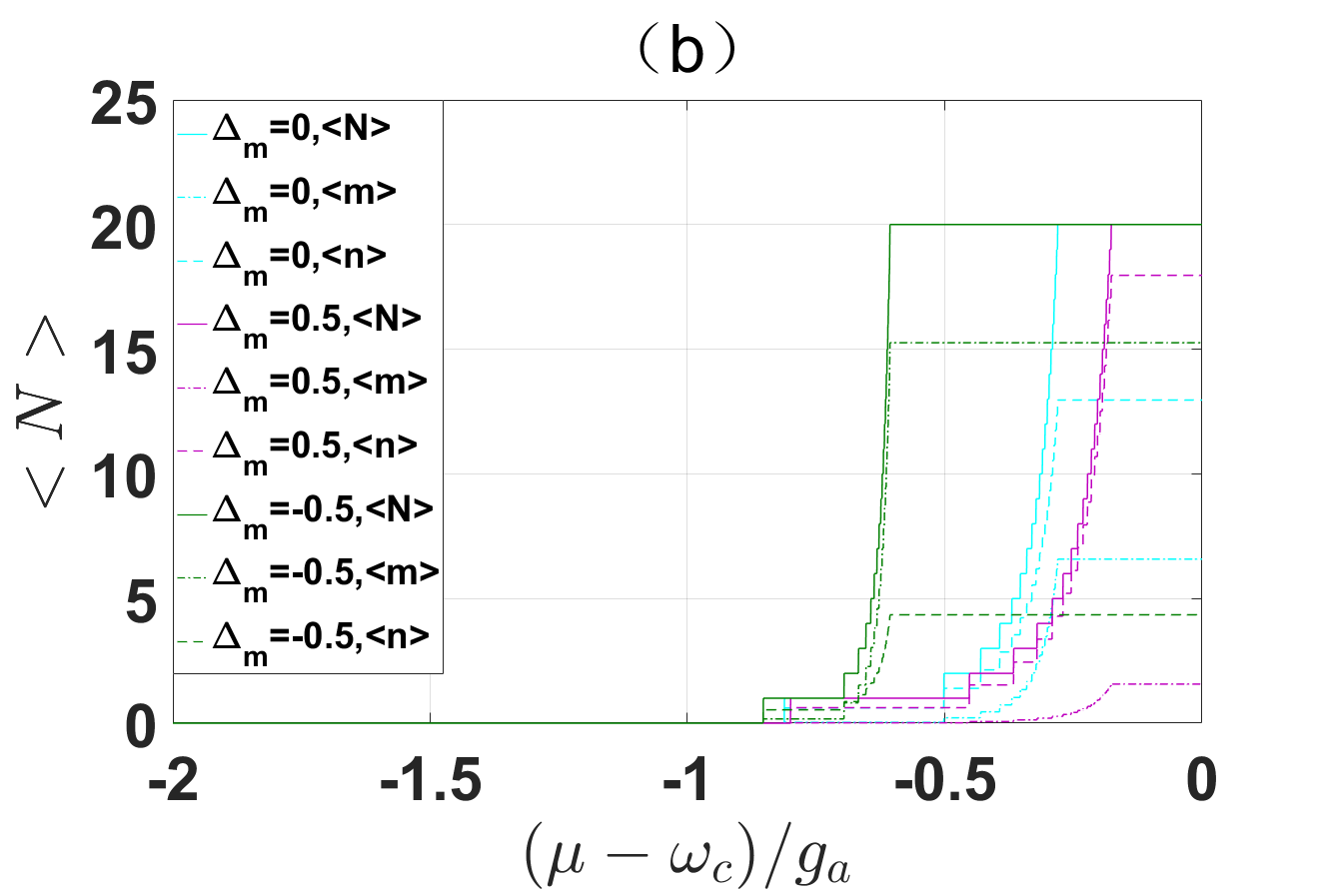}
\caption{The average excitations number $\langle N\rangle$ (photon number $\langle n\rangle$ and magnon number $\langle m\rangle$  ) as a function of the normalized chemical potential $(\mu-\omega_c)/g_a$ for different detunings and photon-magnon coupling strength. (a) $\Delta_a/g_a=\Delta_{m}/g_a=0.5$. (b) $\Delta_a/g_a=0.5,G_m/g_a=0.2$.}
\label{fig7}
\end{figure}

In order to determine the excitation numbers corresponding to each Mott lobe in the phase diagram. We plot the average excitations number $\langle N\rangle$ and the average photon (magnon) number $\langle n\rangle$  $(\langle m\rangle)$ per site for the normalized chemical potential $(\mu-\omega_c)/g_a$ in Fig. \ref{fig7} for $N=20$. It is easy to see that the evolutions of $\langle N\rangle$, $\langle n\rangle$, $\langle m\rangle$ reflect a conspicuous staircase due to the competition between diverse ground-states, and accordingly, each Mott lobe in the phase diagram is characterized by the corresponding plateaus. Figure \ref{fig7} also exhibits $\langle N\rangle$, $\langle n\rangle$ and $\langle m\rangle$ for different coupling strengths $G_m$ and detunings $\Delta_m$. It is easy to notice that the enhanced photon-magnon coupling strength $G_m$ leads to a increase of the average magnon number $\langle m\rangle$ per site and a decrease of the average photon number $\langle n\rangle$ per site correspondingly. It is not difficult to understand that the large detuning $\Delta_m$ can cause the photon-magnon coupling to become weaker and thus reduces the excitation of the magnons. Therefore, with increasing $\Delta_m$, the average number of magnons tends to decrease as shown in Fig. \ref{fig7}(b).

\begin{figure}[!htbp]
\includegraphics[width=4.2cm,height=3.0cm]{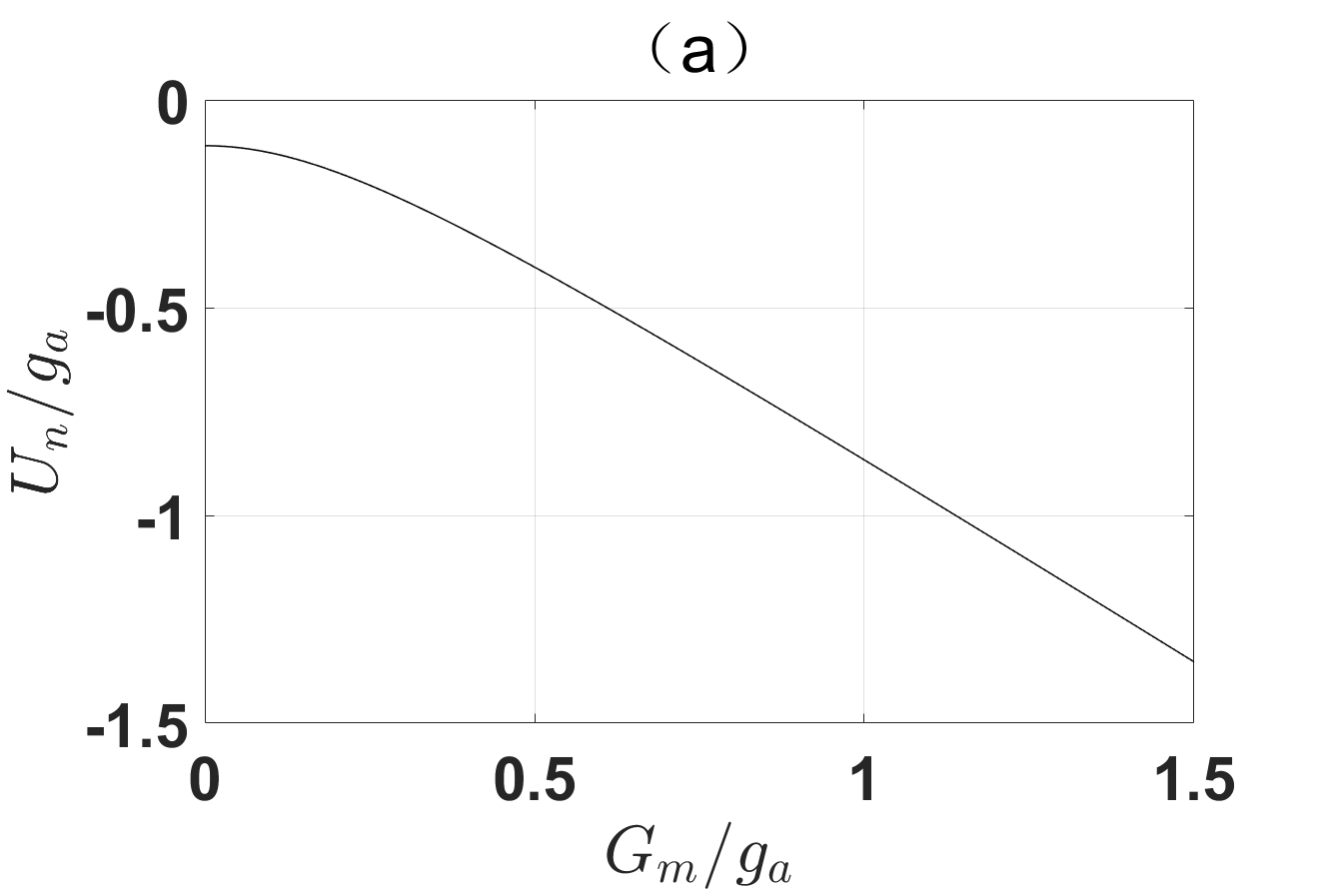}
\includegraphics[width=4.2cm,height=3.0cm]{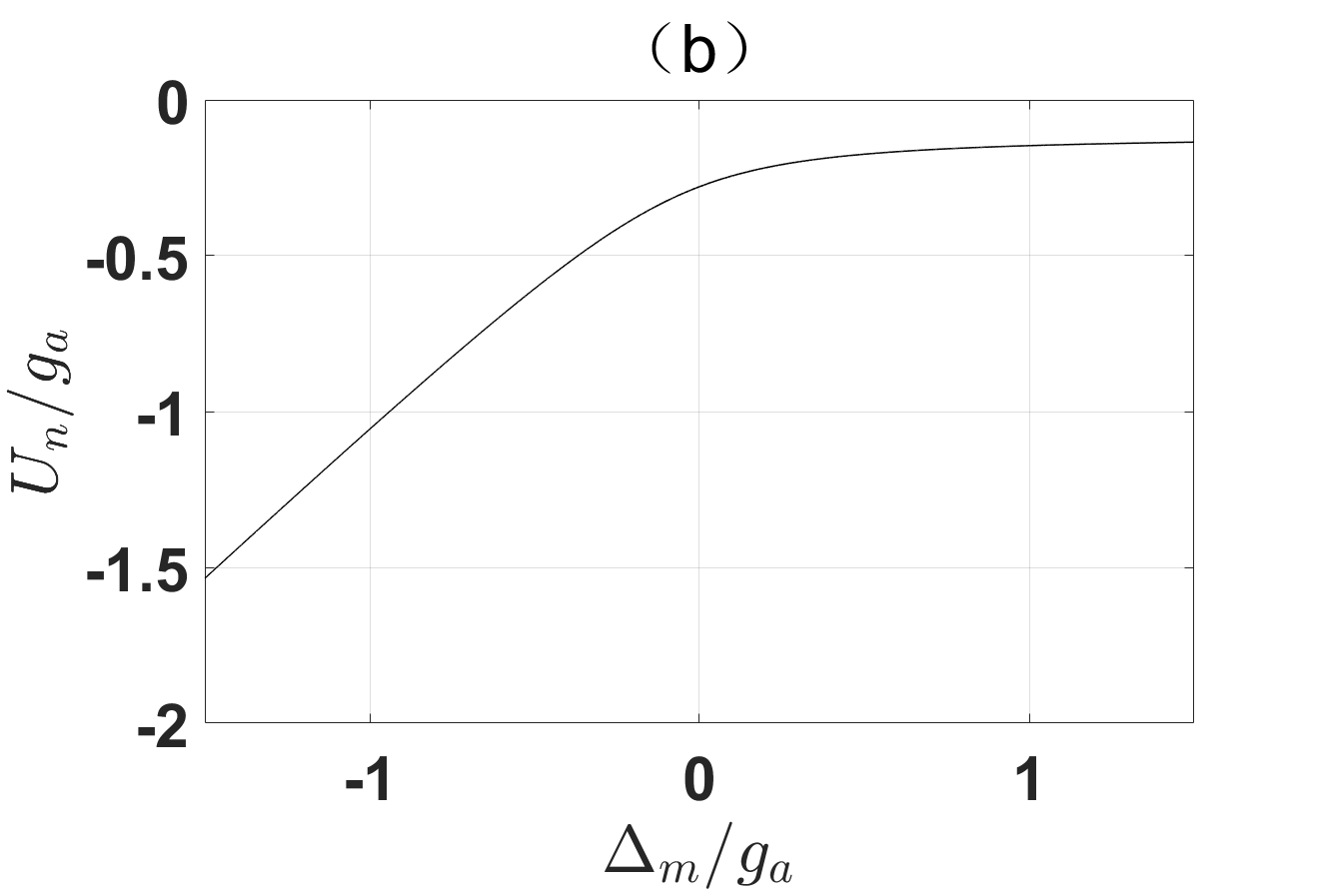}
\caption{(a) Show the normalized effective repulsive potential $U_n$ versus the cavity-magnon coupling strength $G_m/g_a$. (b) Show the normalized effective repulsive potential $U_n$ as a function of the detuning $\Delta_m/g_a$. Other parameters are the same as in Fig. \ref{fig5} and Fig. \ref{fig6}. }
\label{fig8}
\end{figure}
In general, the superfluid-Mott insulator quantum phase transition can partly be understood by the competition arising from the effective on-site repulsive potential and the photon hopping. As the effective on-site repulsive potential dominates the hopping rate, the system should be in a Mott insulator phase; on the contrary, the system is in a superfluid state. The effective on-site repulsive potential $U_n$ can be defined as $U_n=E_{N+1,-}-E_{N,-}-\omega _c$. Figure. \ref{fig8}(a) shows the effective on-site repulsive potential $U_n$ for the coupling strength $G_m$. Obviously, $U_n$ decreases with the increase of $G_m$. Thus, for a strong photon-magnon coupling, the diminished effective on-site repulsive potential leads the system tend to be more delocalized for a certain hopping rate. Then, the superfluid region increases accordingly as shown in Fig. \ref{fig5}. One can also notice that $U_n$ increases with the increase of detuning $\Delta_m$, which leads the Mott insulator phase area increase corresponding to Fig. \ref{fig6}.
\section{CONCLUSION AND DISCUSSION}\label{V}
In conclusion, we have investigated the superfluid-Mott insulator quantum phase transition of light in a two-dimensional cavity optomagnonic array system. Firstly, the critical hopping rate of lower excitations is obtained by the second perturbation theory and Landau second-order phase transition theory. In addition, the phase boundaries between the Mott insulator phase and the superfluid phase are given accordingly the critical hopping rate, and these results are consistent with the numerical ones when the hopping rate is weak. The coupling strength is favorable to the superfluid phase, and the stable region of the Mott lobe decreases with the increase of the photon-magnon coupling strength. Finally, the complete and stable phase diagram is exhibited on the positive photon-magnon detuning, and the highly excited Mott lobe tends to disappear when the detuning is negative. The effective on-site repulsive potential can explain these results. Additionally, our work may extend the studies based on the cavity optomagnonic system and offer a novel idea to explore the superfluid-Mott insulator quantum phase transition of light.

Experimentally, it is a mature technology that a strong coupling magnon-photon system can be engineered in experiments[\onlinecite{Tabuchi}, \onlinecite{Dengke}, \onlinecite{Huebl},  \onlinecite{GuoQiang},\onlinecite{YiPu}]. In addition, the system of a two-level superconducting flux qubit, playing the role of an artificial atom, coupled with the cavity mode has also been realized in experiments[\onlinecite{Buks}]. And the linear array of 3D cavities and qubits for experiments has been developed[\onlinecite{Dmytro}]. Then, the theoretical model proposed here may be experimentally realized if one integrates the processes of the three points mentioned above. For simplicity, the parameters are scaled by $g_a$ for numerical results. According to the theoretical results obtained here, the parameters values for superfluid-Mott insulator quantum phase transition of light will vary for different conditions. To observe these phenomena, the range of parameters is as follows: the coupling strength between photon and magnon $G_m/2\pi$ is 0 to 180MHz, the detuning of photon-magnon $\Delta_m$ is $-0.942$GHz to $0.942$GHz, and the hopping rate $\kappa$ is $94.2$kHz to $28\times 10^3$kHz[\onlinecite{Buks}, \onlinecite{Chang}], which can be achievable in cavity optomagnonic system experimentally. Furthermore, the disorder of this system induces some interesting effects for the quantum phase transition. Taking the JCH model as an example, the disorder of the light-matter interaction and the disorser of detuning between light-matter both induce the transition superfluid phase to the Mott insulator phase, and the disorder in the hopping induces a glassy fluid phase[\onlinecite{Eduardo}]. On the other hand, the effects of the tiny disorder and the weak fluctuations due to temperature can be suppressed by the excitation-hole gap in the Mott-insulator region, then the Mott phase are robustness and should be possible in the considering system[\onlinecite{Tahan}].

\section{ACKNOWLEDGMENTS}
This work was supported by National Natural Science Foundation of China (Grants No. 11874190, No. 61835013 and No. 12047501), and National Key R$\&$D Program of China under grants No. 2016YFA0301500. Support was also provided by Supercomputing Center of Lanzhou University.

\bibliography{paper}

\begin{thebibliography}{89}%
\makeatletter
\providecommand \@ifxundefined [1]{%
 \@ifx{#1\undefined}
}%
\providecommand \@ifnum [1]{%
 \ifnum #1\expandafter \@firstoftwo
 \else \expandafter \@secondoftwo
 \fi
}%
\providecommand \@ifx [1]{%
 \ifx #1\expandafter \@firstoftwo
 \else \expandafter \@secondoftwo
 \fi
}%
\providecommand \natexlab [1]{#1}%
\providecommand \enquote  [1]{``#1''}%
\providecommand \bibnamefont  [1]{#1}%
\providecommand \bibfnamefont [1]{#1}%
\providecommand \citenamefont [1]{#1}%
\providecommand \href@noop [0]{\@secondoftwo}%
\providecommand \href [0]{\begingroup \@sanitize@url \@href}%
\providecommand \@href[1]{\@@startlink{#1}\@@href}%
\providecommand \@@href[1]{\endgroup#1\@@endlink}%
\providecommand \@sanitize@url [0]{\catcode `\\12\catcode `\$12\catcode
  `\&12\catcode `\#12\catcode `\^12\catcode `\_12\catcode `\%12\relax}%
\providecommand \@@startlink[1]{}%
\providecommand \@@endlink[0]{}%
\providecommand \url  [0]{\begingroup\@sanitize@url \@url }%
\providecommand \@url [1]{\endgroup\@href {#1}{\urlprefix }}%
\providecommand \urlprefix  [0]{URL }%
\providecommand \Eprint [0]{\href }%
\providecommand \doibase [0]{http://dx.doi.org/}%
\providecommand \selectlanguage [0]{\@gobble}%
\providecommand \bibinfo  [0]{\@secondoftwo}%
\providecommand \bibfield  [0]{\@secondoftwo}%
\providecommand \translation [1]{[#1]}%
\providecommand \BibitemOpen [0]{}%
\providecommand \bibitemStop [0]{}%
\providecommand \bibitemNoStop [0]{.\EOS\space}%
\providecommand \EOS [0]{\spacefactor3000\relax}%
\providecommand \BibitemShut  [1]{\csname bibitem#1\endcsname}%
\let\auto@bib@innerbib\@empty
\bibitem [{\citenamefont {Georgescu}\ \emph {et~al.}(2014)\citenamefont
  {Georgescu}, \citenamefont {Ashhab},\ and\ \citenamefont {Nori}}]{Georgescu}%
  \BibitemOpen
  \bibfield  {author} {\bibinfo {author} {\bibfnamefont {I.~M.}\ \bibnamefont
  {Georgescu}}, \bibinfo {author} {\bibfnamefont {S.}~\bibnamefont {Ashhab}}, \
  and\ \bibinfo {author} {\bibfnamefont {F.}~\bibnamefont {Nori}},\ }\href
  {\doibase 10.1103/RevModPhys.86.153} {\bibfield  {journal} {\bibinfo
  {journal} {Rev. Mod. Phys.}\ }\textbf {\bibinfo {volume} {86}},\ \bibinfo
  {pages} {153} (\bibinfo {year} {2014})}\BibitemShut {NoStop}%
\bibitem [{\citenamefont {L\"ohneysen}\ \emph {et~al.}(2007)\citenamefont
  {L\"ohneysen}, \citenamefont {Rosch}, \citenamefont {Vojta},\ and\
  \citenamefont {W\"olfle}}]{ohneysen}%
  \BibitemOpen
  \bibfield  {author} {\bibinfo {author} {\bibfnamefont {H.~v.}\ \bibnamefont
  {L\"ohneysen}}, \bibinfo {author} {\bibfnamefont {A.}~\bibnamefont {Rosch}},
  \bibinfo {author} {\bibfnamefont {M.}~\bibnamefont {Vojta}}, \ and\ \bibinfo
  {author} {\bibfnamefont {P.}~\bibnamefont {W\"olfle}},\ }\href {\doibase
  10.1103/RevModPhys.79.1015} {\bibfield  {journal} {\bibinfo  {journal} {Rev.
  Mod. Phys.}\ }\textbf {\bibinfo {volume} {79}},\ \bibinfo {pages} {1015}
  (\bibinfo {year} {2007})}\BibitemShut {NoStop}%
\bibitem [{\citenamefont {Altman}\ \emph {et~al.}(2003)\citenamefont {Altman},
  \citenamefont {Hofstetter}, \citenamefont {Demler},\ and\ \citenamefont
  {Lukin}}]{EhudAltman}%
  \BibitemOpen
  \bibfield  {author} {\bibinfo {author} {\bibfnamefont {E.}~\bibnamefont
  {Altman}}, \bibinfo {author} {\bibfnamefont {W.}~\bibnamefont {Hofstetter}},
  \bibinfo {author} {\bibfnamefont {E.}~\bibnamefont {Demler}}, \ and\ \bibinfo
  {author} {\bibfnamefont {M.~D.}\ \bibnamefont {Lukin}},\ }\href {\doibase
  10.1088/1367-2630/5/1/113} {\bibfield  {journal} {\bibinfo  {journal} {New
  Journal of Physics}\ }\textbf {\bibinfo {volume} {5}},\ \bibinfo {pages}
  {113} (\bibinfo {year} {2003})}\BibitemShut {NoStop}%
\bibitem [{\citenamefont {Han}\ \emph {et~al.}(2004)\citenamefont {Han},
  \citenamefont {Zhang}, \citenamefont {Wang},\ and\ \citenamefont
  {Liu}}]{Jiu-RongHan}%
  \BibitemOpen
  \bibfield  {author} {\bibinfo {author} {\bibfnamefont {J.-R.}\ \bibnamefont
  {Han}}, \bibinfo {author} {\bibfnamefont {T.}~\bibnamefont {Zhang}}, \bibinfo
  {author} {\bibfnamefont {Y.-Z.}\ \bibnamefont {Wang}}, \ and\ \bibinfo
  {author} {\bibfnamefont {W.}~\bibnamefont {Liu}},\ }\href {\doibase
  https://doi.org/10.1016/j.physleta.2004.09.059} {\bibfield  {journal}
  {\bibinfo  {journal} {Physics Letters A}\ }\textbf {\bibinfo {volume}
  {332}},\ \bibinfo {pages} {131} (\bibinfo {year} {2004})}\BibitemShut
  {NoStop}%
\bibitem [{\citenamefont {Orth}\ \emph {et~al.}(2008)\citenamefont {Orth},
  \citenamefont {Stanic},\ and\ \citenamefont {Le~Hur}}]{Orth}%
  \BibitemOpen
  \bibfield  {author} {\bibinfo {author} {\bibfnamefont {P.~P.}\ \bibnamefont
  {Orth}}, \bibinfo {author} {\bibfnamefont {I.}~\bibnamefont {Stanic}}, \ and\
  \bibinfo {author} {\bibfnamefont {K.}~\bibnamefont {Le~Hur}},\ }\href
  {\doibase 10.1103/PhysRevA.77.051601} {\bibfield  {journal} {\bibinfo
  {journal} {Phys. Rev. A}\ }\textbf {\bibinfo {volume} {77}},\ \bibinfo
  {pages} {051601} (\bibinfo {year} {2008})}\BibitemShut {NoStop}%
\bibitem [{\citenamefont {Dicke}(1954)}]{Dicke}%
  \BibitemOpen
  \bibfield  {author} {\bibinfo {author} {\bibfnamefont {R.~H.}\ \bibnamefont
  {Dicke}},\ }\href {\doibase 10.1103/PhysRev.93.99} {\bibfield  {journal}
  {\bibinfo  {journal} {Phys. Rev.}\ }\textbf {\bibinfo {volume} {93}},\
  \bibinfo {pages} {99} (\bibinfo {year} {1954})}\BibitemShut {NoStop}%
\bibitem [{\citenamefont {Fisher}\ \emph {et~al.}(1989)\citenamefont {Fisher},
  \citenamefont {Weichman}, \citenamefont {Grinstein},\ and\ \citenamefont
  {Fisher}}]{Fisher}%
  \BibitemOpen
  \bibfield  {author} {\bibinfo {author} {\bibfnamefont {M.~P.~A.}\
  \bibnamefont {Fisher}}, \bibinfo {author} {\bibfnamefont {P.~B.}\
  \bibnamefont {Weichman}}, \bibinfo {author} {\bibfnamefont {G.}~\bibnamefont
  {Grinstein}}, \ and\ \bibinfo {author} {\bibfnamefont {D.~S.}\ \bibnamefont
  {Fisher}},\ }\href {\doibase 10.1103/PhysRevB.40.546} {\bibfield  {journal}
  {\bibinfo  {journal} {Phys. Rev. B}\ }\textbf {\bibinfo {volume} {40}},\
  \bibinfo {pages} {546} (\bibinfo {year} {1989})}\BibitemShut {NoStop}%
\bibitem [{\citenamefont {Jaksch}\ \emph {et~al.}(1998)\citenamefont {Jaksch},
  \citenamefont {Bruder}, \citenamefont {Cirac}, \citenamefont {Gardiner},\
  and\ \citenamefont {Zoller}}]{Jaksch}%
  \BibitemOpen
  \bibfield  {author} {\bibinfo {author} {\bibfnamefont {D.}~\bibnamefont
  {Jaksch}}, \bibinfo {author} {\bibfnamefont {C.}~\bibnamefont {Bruder}},
  \bibinfo {author} {\bibfnamefont {J.~I.}\ \bibnamefont {Cirac}}, \bibinfo
  {author} {\bibfnamefont {C.~W.}\ \bibnamefont {Gardiner}}, \ and\ \bibinfo
  {author} {\bibfnamefont {P.}~\bibnamefont {Zoller}},\ }\href {\doibase
  10.1103/PhysRevLett.81.3108} {\bibfield  {journal} {\bibinfo  {journal}
  {Phys. Rev. Lett.}\ }\textbf {\bibinfo {volume} {81}},\ \bibinfo {pages}
  {3108} (\bibinfo {year} {1998})}\BibitemShut {NoStop}%
\bibitem [{\citenamefont {Sachdev}(2011)}]{Sachdev}%
  \BibitemOpen
  \bibfield  {author} {\bibinfo {author} {\bibfnamefont {S.}~\bibnamefont
  {Sachdev}},\ }\href {\doibase 10.1017/CBO9780511622540} {\bibfield  {journal}
  {\bibinfo  {journal} {Quantum Phase Transitions, by Subir Sachdev, Cambridge,
  UK: Cambridge University Press, 2011}\ } (\bibinfo {year} {2011}),\
  10.1017/CBO9780511622540}\BibitemShut {NoStop}%
\bibitem [{\citenamefont {Greiner}()}]{Greiner}%
  \BibitemOpen
  \bibfield  {author} {\bibinfo {author} {\bibfnamefont {M.~O. E. T. B.-T. W.
  H.-I.}\ \bibnamefont {Greiner}, \bibfnamefont {Markus}},\ }\href {\doibase
  10.1038/415039a} {\bibfield  {journal} {\bibinfo  {journal} {Nature}\
  }\textbf {\bibinfo {volume} {415}},\ \bibinfo {pages} {39}}\BibitemShut
  {NoStop}%
\bibitem [{\citenamefont {Sherson}()}]{Sherson}%
  \BibitemOpen
  \bibfield  {author} {\bibinfo {author} {\bibfnamefont {W.~C. E. M.-C. M.
  B.-I. K.~S.}\ \bibnamefont {Sherson}, \bibfnamefont {Jacob~F.}},\ }\href
  {\doibase 10.1038/nature09378} {\bibfield  {journal} {\bibinfo  {journal}
  {Nature}\ }\textbf {\bibinfo {volume} {467}},\ \bibinfo {pages}
  {68}}\BibitemShut {NoStop}%
\bibitem [{\citenamefont {Gheri}\ and\ \citenamefont {Ritsch}(1997)}]{Gheri}%
  \BibitemOpen
  \bibfield  {author} {\bibinfo {author} {\bibfnamefont {K.~M.}\ \bibnamefont
  {Gheri}}\ and\ \bibinfo {author} {\bibfnamefont {H.}~\bibnamefont {Ritsch}},\
  }\href {\doibase 10.1103/PhysRevA.56.3187} {\bibfield  {journal} {\bibinfo
  {journal} {Phys. Rev. A}\ }\textbf {\bibinfo {volume} {56}},\ \bibinfo
  {pages} {3187} (\bibinfo {year} {1997})}\BibitemShut {NoStop}%
\bibitem [{\citenamefont {DiCarlo}()}]{DiCarlo}%
  \BibitemOpen
  \bibfield  {author} {\bibinfo {author} {\bibfnamefont {C.~J. M. G. J. M. B.
  L. S. S. D. I. M. J. B. A. F. L. G. S. M. S. R.~J.}\ \bibnamefont {DiCarlo},
  \bibfnamefont {L.}},\ }\href {\doibase 10.1038/nature08121} {\bibfield
  {journal} {\bibinfo  {journal} {Nature}\ }\textbf {\bibinfo {volume} {460}},\
  \bibinfo {pages} {240}}\BibitemShut {NoStop}%
\bibitem [{\citenamefont {Olmschenk}\ \emph {et~al.}(2007)\citenamefont
  {Olmschenk}, \citenamefont {Younge}, \citenamefont {Moehring}, \citenamefont
  {Matsukevich}, \citenamefont {Maunz},\ and\ \citenamefont
  {Monroe}}]{Olmschenk}%
  \BibitemOpen
  \bibfield  {author} {\bibinfo {author} {\bibfnamefont {S.}~\bibnamefont
  {Olmschenk}}, \bibinfo {author} {\bibfnamefont {K.~C.}\ \bibnamefont
  {Younge}}, \bibinfo {author} {\bibfnamefont {D.~L.}\ \bibnamefont
  {Moehring}}, \bibinfo {author} {\bibfnamefont {D.~N.}\ \bibnamefont
  {Matsukevich}}, \bibinfo {author} {\bibfnamefont {P.}~\bibnamefont {Maunz}},
  \ and\ \bibinfo {author} {\bibfnamefont {C.}~\bibnamefont {Monroe}},\ }\href
  {\doibase 10.1103/PhysRevA.76.052314} {\bibfield  {journal} {\bibinfo
  {journal} {Phys. Rev. A}\ }\textbf {\bibinfo {volume} {76}},\ \bibinfo
  {pages} {052314} (\bibinfo {year} {2007})}\BibitemShut {NoStop}%
\bibitem [{\citenamefont {Tahan}\ and\ \citenamefont
  {Hollenberg}(2006)}]{Tahan}%
  \BibitemOpen
  \bibfield  {author} {\bibinfo {author} {\bibfnamefont {C.~J. H. G. A.~D.}\
  \bibnamefont {Tahan}, \bibfnamefont {Charles}}\ and\ \bibinfo {author}
  {\bibfnamefont {L.~C.~L.}\ \bibnamefont {Hollenberg}},\ }\href {\doibase
  10.1038/nphys466} {\bibfield  {journal} {\bibinfo  {journal} {Nature
  Physics}\ }\textbf {\bibinfo {volume} {2}},\ \bibinfo {pages} {856} (\bibinfo
  {year} {2006})}\BibitemShut {NoStop}%
\bibitem [{\citenamefont {Hartmann}\ and\ \citenamefont
  {Hollenberg}(2006)}]{Hartmann}%
  \BibitemOpen
  \bibfield  {author} {\bibinfo {author} {\bibfnamefont {B.~a. P. M. B. C. J.
  H. G. A.~D.}\ \bibnamefont {Hartmann}, \bibfnamefont {Michael~J.}}\ and\
  \bibinfo {author} {\bibfnamefont {L.~C.~L.}\ \bibnamefont {Hollenberg}},\
  }\href {\doibase 10.1038/nphys462} {\bibfield  {journal} {\bibinfo  {journal}
  {Nature Physics}\ }\textbf {\bibinfo {volume} {2}},\ \bibinfo {pages} {849}
  (\bibinfo {year} {2006})}\BibitemShut {NoStop}%
\bibitem [{\citenamefont {Angelakis}\ \emph {et~al.}(2007)\citenamefont
  {Angelakis}, \citenamefont {Santos},\ and\ \citenamefont {Bose}}]{Angelakis}%
  \BibitemOpen
  \bibfield  {author} {\bibinfo {author} {\bibfnamefont {D.~G.}\ \bibnamefont
  {Angelakis}}, \bibinfo {author} {\bibfnamefont {M.~F.}\ \bibnamefont
  {Santos}}, \ and\ \bibinfo {author} {\bibfnamefont {S.}~\bibnamefont
  {Bose}},\ }\href {\doibase 10.1103/PhysRevA.76.031805} {\bibfield  {journal}
  {\bibinfo  {journal} {Phys. Rev. A}\ }\textbf {\bibinfo {volume} {76}},\
  \bibinfo {pages} {031805} (\bibinfo {year} {2007})}\BibitemShut {NoStop}%
\bibitem [{\citenamefont {Lin}\ \emph {et~al.}(2021)\citenamefont {Lin},
  \citenamefont {Georges}, \citenamefont {Klinder}, \citenamefont {Molignini},
  \citenamefont {B¨¹ttner}, \citenamefont {Lode}, \citenamefont {Chitra},
  \citenamefont {Hemmerich},\ and\ \citenamefont {Ke?ler}}]{lin2021mott}%
  \BibitemOpen
  \bibfield  {author} {\bibinfo {author} {\bibfnamefont {R.}~\bibnamefont
  {Lin}}, \bibinfo {author} {\bibfnamefont {C.}~\bibnamefont {Georges}},
  \bibinfo {author} {\bibfnamefont {J.}~\bibnamefont {Klinder}}, \bibinfo
  {author} {\bibfnamefont {P.}~\bibnamefont {Molignini}}, \bibinfo {author}
  {\bibfnamefont {M.}~\bibnamefont {B¨¹ttner}}, \bibinfo {author}
  {\bibfnamefont {A.~U.~J.}\ \bibnamefont {Lode}}, \bibinfo {author}
  {\bibfnamefont {R.}~\bibnamefont {Chitra}}, \bibinfo {author} {\bibfnamefont
  {A.}~\bibnamefont {Hemmerich}}, \ and\ \bibinfo {author} {\bibfnamefont
  {H.}~\bibnamefont {Ke?ler}},\ }\href@noop {} {\enquote {\bibinfo {title}
  {Mott transition in a cavity-boson system: A quantitative comparison between
  theory and experiment},}\ } (\bibinfo {year} {2021}),\ \Eprint
  {http://arxiv.org/abs/2104.11253} {arXiv:2104.11253 [cond-mat.quant-gas]}
  \BibitemShut {NoStop}%
\bibitem [{\citenamefont {Song}\ \emph {et~al.}(2021)\citenamefont {Song},
  \citenamefont {Dutta}, \citenamefont {Bhave}, \citenamefont {Yu},
  \citenamefont {Carter}, \citenamefont {Cooper},\ and\ \citenamefont
  {Schneider}}]{song2021realizing}%
  \BibitemOpen
  \bibfield  {author} {\bibinfo {author} {\bibfnamefont {B.}~\bibnamefont
  {Song}}, \bibinfo {author} {\bibfnamefont {S.}~\bibnamefont {Dutta}},
  \bibinfo {author} {\bibfnamefont {S.}~\bibnamefont {Bhave}}, \bibinfo
  {author} {\bibfnamefont {J.-C.}\ \bibnamefont {Yu}}, \bibinfo {author}
  {\bibfnamefont {E.}~\bibnamefont {Carter}}, \bibinfo {author} {\bibfnamefont
  {N.}~\bibnamefont {Cooper}}, \ and\ \bibinfo {author} {\bibfnamefont
  {U.}~\bibnamefont {Schneider}},\ }\href@noop {} {\enquote {\bibinfo {title}
  {Realizing discontinuous quantum phase transitions in a strongly-correlated
  driven optical lattice},}\ } (\bibinfo {year} {2021}),\ \Eprint
  {http://arxiv.org/abs/2105.12146} {arXiv:2105.12146 [cond-mat.quant-gas]}
  \BibitemShut {NoStop}%
\bibitem [{\citenamefont {Zhu}\ \emph {et~al.}(2021)\citenamefont {Zhu},
  \citenamefont {Guo}, \citenamefont {Breuckmann}, \citenamefont {Guo},\ and\
  \citenamefont {Feng}}]{zhu2021quantum}%
  \BibitemOpen
  \bibfield  {author} {\bibinfo {author} {\bibfnamefont {X.}~\bibnamefont
  {Zhu}}, \bibinfo {author} {\bibfnamefont {J.}~\bibnamefont {Guo}}, \bibinfo
  {author} {\bibfnamefont {N.~P.}\ \bibnamefont {Breuckmann}}, \bibinfo
  {author} {\bibfnamefont {H.}~\bibnamefont {Guo}}, \ and\ \bibinfo {author}
  {\bibfnamefont {S.}~\bibnamefont {Feng}},\ }\href@noop {} {\enquote {\bibinfo
  {title} {Quantum phase transitions of interacting bosons on hyperbolic
  lattices},}\ } (\bibinfo {year} {2021}),\ \Eprint
  {http://arxiv.org/abs/2103.15274} {arXiv:2103.15274 [cond-mat.str-el]}
  \BibitemShut {NoStop}%
\bibitem [{\citenamefont {Chen}\ and\ \citenamefont {Xie}(2021)}]{Chen_2021}%
  \BibitemOpen
  \bibfield  {author} {\bibinfo {author} {\bibfnamefont {H.}~\bibnamefont
  {Chen}}\ and\ \bibinfo {author} {\bibfnamefont {X.~C.}\ \bibnamefont {Xie}},\
  }\href {\doibase 10.1103/physrevb.103.205144} {\bibfield  {journal} {\bibinfo
   {journal} {Physical Review B}\ }\textbf {\bibinfo {volume} {103}} (\bibinfo
  {year} {2021}),\ 10.1103/physrevb.103.205144}\BibitemShut {NoStop}%
\bibitem [{\citenamefont {Huang}\ \emph {et~al.}(2020)\citenamefont {Huang},
  \citenamefont {Yao}, \citenamefont {Liang}, \citenamefont {Wang},
  \citenamefont {Zheng}, \citenamefont {Li}, \citenamefont {Xiong},
  \citenamefont {Zhou}, \citenamefont {Chen}, \citenamefont {Chen},\ and\
  \citenamefont {Hu}}]{huang2020observation}%
  \BibitemOpen
  \bibfield  {author} {\bibinfo {author} {\bibfnamefont {Q.}~\bibnamefont
  {Huang}}, \bibinfo {author} {\bibfnamefont {R.}~\bibnamefont {Yao}}, \bibinfo
  {author} {\bibfnamefont {L.}~\bibnamefont {Liang}}, \bibinfo {author}
  {\bibfnamefont {S.}~\bibnamefont {Wang}}, \bibinfo {author} {\bibfnamefont
  {Q.}~\bibnamefont {Zheng}}, \bibinfo {author} {\bibfnamefont
  {D.}~\bibnamefont {Li}}, \bibinfo {author} {\bibfnamefont {W.}~\bibnamefont
  {Xiong}}, \bibinfo {author} {\bibfnamefont {X.}~\bibnamefont {Zhou}},
  \bibinfo {author} {\bibfnamefont {W.}~\bibnamefont {Chen}}, \bibinfo {author}
  {\bibfnamefont {X.}~\bibnamefont {Chen}}, \ and\ \bibinfo {author}
  {\bibfnamefont {J.}~\bibnamefont {Hu}},\ }\href@noop {} {\enquote {\bibinfo
  {title} {Observation of many-body quantum phase transitions beyond the
  kibble-zurek mechanism},}\ } (\bibinfo {year} {2020}),\ \Eprint
  {http://arxiv.org/abs/2012.01734} {arXiv:2012.01734 [quant-ph]} \BibitemShut
  {NoStop}%
\bibitem [{\citenamefont {Kamide}\ \emph {et~al.}(2013)\citenamefont {Kamide},
  \citenamefont {Yamaguchi}, \citenamefont {Kimura},\ and\ \citenamefont
  {Ogawa}}]{Kamide}%
  \BibitemOpen
  \bibfield  {author} {\bibinfo {author} {\bibfnamefont {K.}~\bibnamefont
  {Kamide}}, \bibinfo {author} {\bibfnamefont {M.}~\bibnamefont {Yamaguchi}},
  \bibinfo {author} {\bibfnamefont {T.}~\bibnamefont {Kimura}}, \ and\ \bibinfo
  {author} {\bibfnamefont {T.}~\bibnamefont {Ogawa}},\ }\href {\doibase
  10.1103/PhysRevA.87.053842} {\bibfield  {journal} {\bibinfo  {journal} {Phys.
  Rev. A}\ }\textbf {\bibinfo {volume} {87}},\ \bibinfo {pages} {053842}
  (\bibinfo {year} {2013})}\BibitemShut {NoStop}%
\bibitem [{\citenamefont {Gomes}\ \emph {et~al.}(2012)\citenamefont {Gomes},
  \citenamefont {Souza},\ and\ \citenamefont {Almeida}}]{gomes2012mott}%
  \BibitemOpen
  \bibfield  {author} {\bibinfo {author} {\bibfnamefont {C.~B.}\ \bibnamefont
  {Gomes}}, \bibinfo {author} {\bibfnamefont {A.~M.~C.}\ \bibnamefont {Souza}},
  \ and\ \bibinfo {author} {\bibfnamefont {F.~A.~G.}\ \bibnamefont {Almeida}},\
  }\href@noop {} {\enquote {\bibinfo {title} {Mott insulator to superfluid
  phase transition in bravais lattices via the jaynes-cummings-hubbard
  model},}\ } (\bibinfo {year} {2012}),\ \Eprint
  {http://arxiv.org/abs/1211.5515} {arXiv:1211.5515 [cond-mat.str-el]}
  \BibitemShut {NoStop}%
\bibitem [{\citenamefont {Soykal}\ and\ \citenamefont {Tahan}(2013)}]{Soykal}%
  \BibitemOpen
  \bibfield  {author} {\bibinfo {author} {\bibfnamefont {O.~O.}\ \bibnamefont
  {Soykal}}\ and\ \bibinfo {author} {\bibfnamefont {C.}~\bibnamefont {Tahan}},\
  }\href {\doibase 10.1103/PhysRevB.88.134511} {\bibfield  {journal} {\bibinfo
  {journal} {Phys. Rev. B}\ }\textbf {\bibinfo {volume} {88}},\ \bibinfo
  {pages} {134511} (\bibinfo {year} {2013})}\BibitemShut {NoStop}%
\bibitem [{\citenamefont {Zheng}\ \emph {et~al.}(2017)\citenamefont {Zheng},
  \citenamefont {Li}, \citenamefont {L\"u},\ and\ \citenamefont {Wu}}]{Zheng}%
  \BibitemOpen
  \bibfield  {author} {\bibinfo {author} {\bibfnamefont {L.-L.}\ \bibnamefont
  {Zheng}}, \bibinfo {author} {\bibfnamefont {K.-M.}\ \bibnamefont {Li}},
  \bibinfo {author} {\bibfnamefont {X.-Y.}\ \bibnamefont {L\"u}}, \ and\
  \bibinfo {author} {\bibfnamefont {Y.}~\bibnamefont {Wu}},\ }\href {\doibase
  10.1103/PhysRevA.96.053809} {\bibfield  {journal} {\bibinfo  {journal} {Phys.
  Rev. A}\ }\textbf {\bibinfo {volume} {96}},\ \bibinfo {pages} {053809}
  (\bibinfo {year} {2017})}\BibitemShut {NoStop}%
\bibitem [{\citenamefont {Noh}\ and\ \citenamefont
  {Angelakis}(2016)}]{Noh_2016}%
  \BibitemOpen
  \bibfield  {author} {\bibinfo {author} {\bibfnamefont {C.}~\bibnamefont
  {Noh}}\ and\ \bibinfo {author} {\bibfnamefont {D.~G.}\ \bibnamefont
  {Angelakis}},\ }\href {\doibase 10.1088/0034-4885/80/1/016401} {\bibfield
  {journal} {\bibinfo  {journal} {Reports on Progress in Physics}\ }\textbf
  {\bibinfo {volume} {80}},\ \bibinfo {pages} {016401} (\bibinfo {year}
  {2016})}\BibitemShut {NoStop}%
\bibitem [{\citenamefont {Schmidt}\ and\ \citenamefont {Koch}(2013)}]{Schmidt}%
  \BibitemOpen
  \bibfield  {author} {\bibinfo {author} {\bibfnamefont {S.}~\bibnamefont
  {Schmidt}}\ and\ \bibinfo {author} {\bibfnamefont {J.}~\bibnamefont {Koch}},\
  }\href {\doibase https://doi.org/10.1002/andp.201200261} {\bibfield
  {journal} {\bibinfo  {journal} {Annalen der Physik}\ }\textbf {\bibinfo
  {volume} {525}},\ \bibinfo {pages} {395} (\bibinfo {year}
  {2013})}\BibitemShut {NoStop}%
\bibitem [{\citenamefont {Houck}\ \emph {et~al.}(2012)\citenamefont {Houck},
  \citenamefont {T\"{u}reci},\ and\ \citenamefont {Koch}}]{Houck}%
  \BibitemOpen
  \bibfield  {author} {\bibinfo {author} {\bibfnamefont {A.~A.}\ \bibnamefont
  {Houck}}, \bibinfo {author} {\bibfnamefont {H.~E.}\ \bibnamefont
  {T\"{u}reci}}, \ and\ \bibinfo {author} {\bibfnamefont {J.}~\bibnamefont
  {Koch}},\ }\href {\doibase 10.1103/PhysRevLett.123.246801} {\bibfield
  {journal} {\bibinfo  {journal} {Nature Physics}\ }\textbf {\bibinfo {volume}
  {8}} (\bibinfo {year} {2012}),\ 10.1103/PhysRevLett.123.246801}\BibitemShut
  {NoStop}%
\bibitem [{\citenamefont {Hartmann}\ \emph {et~al.}(2008)\citenamefont
  {Hartmann}, \citenamefont {Brand\~{a}o},\ and\ \citenamefont
  {Plenio}}]{Hartmannn}%
  \BibitemOpen
  \bibfield  {author} {\bibinfo {author} {\bibfnamefont {M.}~\bibnamefont
  {Hartmann}}, \bibinfo {author} {\bibfnamefont {F.}~\bibnamefont
  {Brand\~{a}o}}, \ and\ \bibinfo {author} {\bibfnamefont {M.}~\bibnamefont
  {Plenio}},\ }\href {\doibase https://doi.org/10.1002/lpor.200810046}
  {\bibfield  {journal} {\bibinfo  {journal} {Laser \& Photonics Reviews}\
  }\textbf {\bibinfo {volume} {2}},\ \bibinfo {pages} {527} (\bibinfo {year}
  {2008})}\BibitemShut {NoStop}%
\bibitem [{\citenamefont {Mari}\ \emph {et~al.}(2013)\citenamefont {Mari},
  \citenamefont {Farace}, \citenamefont {Didier}, \citenamefont {Giovannetti},\
  and\ \citenamefont {Fazio}}]{Mari}%
  \BibitemOpen
  \bibfield  {author} {\bibinfo {author} {\bibfnamefont {A.}~\bibnamefont
  {Mari}}, \bibinfo {author} {\bibfnamefont {A.}~\bibnamefont {Farace}},
  \bibinfo {author} {\bibfnamefont {N.}~\bibnamefont {Didier}}, \bibinfo
  {author} {\bibfnamefont {V.}~\bibnamefont {Giovannetti}}, \ and\ \bibinfo
  {author} {\bibfnamefont {R.}~\bibnamefont {Fazio}},\ }\href {\doibase
  10.1103/PhysRevLett.111.103605} {\bibfield  {journal} {\bibinfo  {journal}
  {Phys. Rev. Lett.}\ }\textbf {\bibinfo {volume} {111}},\ \bibinfo {pages}
  {103605} (\bibinfo {year} {2013})}\BibitemShut {NoStop}%
\bibitem [{\citenamefont {Zhang}\ \emph {et~al.}(2015)\citenamefont {Zhang},
  \citenamefont {Shah}, \citenamefont {Cardenas},\ and\ \citenamefont
  {Lipson}}]{Mian}%
  \BibitemOpen
  \bibfield  {author} {\bibinfo {author} {\bibfnamefont {M.}~\bibnamefont
  {Zhang}}, \bibinfo {author} {\bibfnamefont {S.}~\bibnamefont {Shah}},
  \bibinfo {author} {\bibfnamefont {J.}~\bibnamefont {Cardenas}}, \ and\
  \bibinfo {author} {\bibfnamefont {M.}~\bibnamefont {Lipson}},\ }\href
  {\doibase 10.1103/PhysRevLett.115.163902} {\bibfield  {journal} {\bibinfo
  {journal} {Phys. Rev. Lett.}\ }\textbf {\bibinfo {volume} {115}},\ \bibinfo
  {pages} {163902} (\bibinfo {year} {2015})}\BibitemShut {NoStop}%
\bibitem [{\citenamefont {Peano}\ \emph {et~al.}(2015)\citenamefont {Peano},
  \citenamefont {Brendel}, \citenamefont {Schmidt},\ and\ \citenamefont
  {Marquardt}}]{Peano}%
  \BibitemOpen
  \bibfield  {author} {\bibinfo {author} {\bibfnamefont {V.}~\bibnamefont
  {Peano}}, \bibinfo {author} {\bibfnamefont {C.}~\bibnamefont {Brendel}},
  \bibinfo {author} {\bibfnamefont {M.}~\bibnamefont {Schmidt}}, \ and\
  \bibinfo {author} {\bibfnamefont {F.}~\bibnamefont {Marquardt}},\ }\href
  {\doibase 10.1103/PhysRevX.5.031011} {\bibfield  {journal} {\bibinfo
  {journal} {Phys. Rev. X}\ }\textbf {\bibinfo {volume} {5}},\ \bibinfo {pages}
  {031011} (\bibinfo {year} {2015})}\BibitemShut {NoStop}%
\bibitem [{\citenamefont {Jin-Lou~Ma}()}]{JinLouMa}%
  \BibitemOpen
  \bibfield  {author} {\bibinfo {author} {\bibfnamefont {Q.~L. H.-Q. G.
  W.-M.~L.}\ \bibnamefont {Jin-Lou~Ma}, \bibfnamefont {Lei~Tan}},\ }\href
  {https://doi.org/10.1088/1361-6455/aba582} {\bibfield  {journal} {\bibinfo
  {journal} {J. Phys. B}\ }\textbf {\bibinfo {volume} {53}},\ \bibinfo {pages}
  {195402}}\BibitemShut {NoStop}%
\bibitem [{\citenamefont {Lachance-Quirion}\ \emph
  {et~al.}(2019{\natexlab{a}})\citenamefont {Lachance-Quirion}, \citenamefont
  {Tabuchi}, \citenamefont {Gloppe}, \citenamefont {Usami},\ and\ \citenamefont
  {Nakamura}}]{Dany}%
  \BibitemOpen
  \bibfield  {author} {\bibinfo {author} {\bibfnamefont {D.}~\bibnamefont
  {Lachance-Quirion}}, \bibinfo {author} {\bibfnamefont {Y.}~\bibnamefont
  {Tabuchi}}, \bibinfo {author} {\bibfnamefont {A.}~\bibnamefont {Gloppe}},
  \bibinfo {author} {\bibfnamefont {K.}~\bibnamefont {Usami}}, \ and\ \bibinfo
  {author} {\bibfnamefont {Y.}~\bibnamefont {Nakamura}},\ }\href {\doibase
  10.7567/1882-0786/ab248d} {\bibfield  {journal} {\bibinfo  {journal} {Applied
  Physics Express}\ }\textbf {\bibinfo {volume} {12}},\ \bibinfo {pages}
  {070101} (\bibinfo {year} {2019}{\natexlab{a}})}\BibitemShut {NoStop}%
\bibitem [{\citenamefont {Zhao}\ \emph
  {et~al.}(2020{\natexlab{a}})\citenamefont {Zhao}, \citenamefont {Li},
  \citenamefont {Chao}, \citenamefont {Peng}, \citenamefont {Li},\ and\
  \citenamefont {Zhou}}]{Zhao}%
  \BibitemOpen
  \bibfield  {author} {\bibinfo {author} {\bibfnamefont {C.}~\bibnamefont
  {Zhao}}, \bibinfo {author} {\bibfnamefont {X.}~\bibnamefont {Li}}, \bibinfo
  {author} {\bibfnamefont {S.}~\bibnamefont {Chao}}, \bibinfo {author}
  {\bibfnamefont {R.}~\bibnamefont {Peng}}, \bibinfo {author} {\bibfnamefont
  {C.}~\bibnamefont {Li}}, \ and\ \bibinfo {author} {\bibfnamefont
  {L.}~\bibnamefont {Zhou}},\ }\href {\doibase 10.1103/PhysRevA.101.063838}
  {\bibfield  {journal} {\bibinfo  {journal} {Phys. Rev. A}\ }\textbf {\bibinfo
  {volume} {101}},\ \bibinfo {pages} {063838} (\bibinfo {year}
  {2020}{\natexlab{a}})}\BibitemShut {NoStop}%
\bibitem [{\citenamefont {Osada}\ \emph {et~al.}(2016)\citenamefont {Osada},
  \citenamefont {Hisatomi}, \citenamefont {Noguchi}, \citenamefont {Tabuchi},
  \citenamefont {Yamazaki}, \citenamefont {Usami}, \citenamefont {Sadgrove},
  \citenamefont {Yalla}, \citenamefont {Nomura},\ and\ \citenamefont
  {Nakamura}}]{Osada}%
  \BibitemOpen
  \bibfield  {author} {\bibinfo {author} {\bibfnamefont {A.}~\bibnamefont
  {Osada}}, \bibinfo {author} {\bibfnamefont {R.}~\bibnamefont {Hisatomi}},
  \bibinfo {author} {\bibfnamefont {A.}~\bibnamefont {Noguchi}}, \bibinfo
  {author} {\bibfnamefont {Y.}~\bibnamefont {Tabuchi}}, \bibinfo {author}
  {\bibfnamefont {R.}~\bibnamefont {Yamazaki}}, \bibinfo {author}
  {\bibfnamefont {K.}~\bibnamefont {Usami}}, \bibinfo {author} {\bibfnamefont
  {M.}~\bibnamefont {Sadgrove}}, \bibinfo {author} {\bibfnamefont
  {R.}~\bibnamefont {Yalla}}, \bibinfo {author} {\bibfnamefont
  {M.}~\bibnamefont {Nomura}}, \ and\ \bibinfo {author} {\bibfnamefont
  {Y.}~\bibnamefont {Nakamura}},\ }\href {\doibase
  10.1103/PhysRevLett.116.223601} {\bibfield  {journal} {\bibinfo  {journal}
  {Phys. Rev. Lett.}\ }\textbf {\bibinfo {volume} {116}},\ \bibinfo {pages}
  {223601} (\bibinfo {year} {2016})}\BibitemShut {NoStop}%
\bibitem [{\citenamefont {Gao}\ \emph {et~al.}(2019{\natexlab{a}})\citenamefont
  {Gao}, \citenamefont {Liu}, \citenamefont {Wang}, \citenamefont {Cao},\ and\
  \citenamefont {Wang}}]{Gao}%
  \BibitemOpen
  \bibfield  {author} {\bibinfo {author} {\bibfnamefont {Y.-P.}\ \bibnamefont
  {Gao}}, \bibinfo {author} {\bibfnamefont {X.-F.}\ \bibnamefont {Liu}},
  \bibinfo {author} {\bibfnamefont {T.-J.}\ \bibnamefont {Wang}}, \bibinfo
  {author} {\bibfnamefont {C.}~\bibnamefont {Cao}}, \ and\ \bibinfo {author}
  {\bibfnamefont {C.}~\bibnamefont {Wang}},\ }\href {\doibase
  10.1103/PhysRevA.100.043831} {\bibfield  {journal} {\bibinfo  {journal}
  {Phys. Rev. A}\ }\textbf {\bibinfo {volume} {100}},\ \bibinfo {pages}
  {043831} (\bibinfo {year} {2019}{\natexlab{a}})}\BibitemShut {NoStop}%
\bibitem [{\citenamefont {Haigh}\ \emph {et~al.}(2016)\citenamefont {Haigh},
  \citenamefont {Nunnenkamp}, \citenamefont {Ramsay},\ and\ \citenamefont
  {Ferguson}}]{Haigh}%
  \BibitemOpen
  \bibfield  {author} {\bibinfo {author} {\bibfnamefont {J.~A.}\ \bibnamefont
  {Haigh}}, \bibinfo {author} {\bibfnamefont {A.}~\bibnamefont {Nunnenkamp}},
  \bibinfo {author} {\bibfnamefont {A.~J.}\ \bibnamefont {Ramsay}}, \ and\
  \bibinfo {author} {\bibfnamefont {A.~J.}\ \bibnamefont {Ferguson}},\ }\href
  {\doibase 10.1103/PhysRevLett.117.133602} {\bibfield  {journal} {\bibinfo
  {journal} {Phys. Rev. Lett.}\ }\textbf {\bibinfo {volume} {117}},\ \bibinfo
  {pages} {133602} (\bibinfo {year} {2016})}\BibitemShut {NoStop}%
\bibitem [{\citenamefont {Zhang}\ \emph
  {et~al.}(2014{\natexlab{a}})\citenamefont {Zhang}, \citenamefont {Zou},
  \citenamefont {Jiang},\ and\ \citenamefont {Tang}}]{Xufeng}%
  \BibitemOpen
  \bibfield  {author} {\bibinfo {author} {\bibfnamefont {X.}~\bibnamefont
  {Zhang}}, \bibinfo {author} {\bibfnamefont {C.-L.}\ \bibnamefont {Zou}},
  \bibinfo {author} {\bibfnamefont {L.}~\bibnamefont {Jiang}}, \ and\ \bibinfo
  {author} {\bibfnamefont {H.~X.}\ \bibnamefont {Tang}},\ }\href {\doibase
  10.1103/PhysRevLett.113.156401} {\bibfield  {journal} {\bibinfo  {journal}
  {Phys. Rev. Lett.}\ }\textbf {\bibinfo {volume} {113}},\ \bibinfo {pages}
  {156401} (\bibinfo {year} {2014}{\natexlab{a}})}\BibitemShut {NoStop}%
\bibitem [{\citenamefont {Zhang}\ \emph {et~al.}(2016)\citenamefont {Zhang},
  \citenamefont {Zhu}, \citenamefont {Zou},\ and\ \citenamefont
  {Tang}}]{Zhang}%
  \BibitemOpen
  \bibfield  {author} {\bibinfo {author} {\bibfnamefont {X.}~\bibnamefont
  {Zhang}}, \bibinfo {author} {\bibfnamefont {N.}~\bibnamefont {Zhu}}, \bibinfo
  {author} {\bibfnamefont {C.-L.}\ \bibnamefont {Zou}}, \ and\ \bibinfo
  {author} {\bibfnamefont {H.~X.}\ \bibnamefont {Tang}},\ }\href {\doibase
  10.1103/PhysRevLett.117.123605} {\bibfield  {journal} {\bibinfo  {journal}
  {Phys. Rev. Lett.}\ }\textbf {\bibinfo {volume} {117}},\ \bibinfo {pages}
  {123605} (\bibinfo {year} {2016})}\BibitemShut {NoStop}%
\bibitem [{\citenamefont {Tabuchi}\ \emph {et~al.}(2014)\citenamefont
  {Tabuchi}, \citenamefont {Ishino}, \citenamefont {Ishikawa}, \citenamefont
  {Yamazaki}, \citenamefont {Usami},\ and\ \citenamefont {Nakamura}}]{Tabuchi}%
  \BibitemOpen
  \bibfield  {author} {\bibinfo {author} {\bibfnamefont {Y.}~\bibnamefont
  {Tabuchi}}, \bibinfo {author} {\bibfnamefont {S.}~\bibnamefont {Ishino}},
  \bibinfo {author} {\bibfnamefont {T.}~\bibnamefont {Ishikawa}}, \bibinfo
  {author} {\bibfnamefont {R.}~\bibnamefont {Yamazaki}}, \bibinfo {author}
  {\bibfnamefont {K.}~\bibnamefont {Usami}}, \ and\ \bibinfo {author}
  {\bibfnamefont {Y.}~\bibnamefont {Nakamura}},\ }\href {\doibase
  10.1103/PhysRevLett.113.083603} {\bibfield  {journal} {\bibinfo  {journal}
  {Phys. Rev. Lett.}\ }\textbf {\bibinfo {volume} {113}},\ \bibinfo {pages}
  {083603} (\bibinfo {year} {2014})}\BibitemShut {NoStop}%
\bibitem [{\citenamefont {Goryachev}\ \emph {et~al.}(2014)\citenamefont
  {Goryachev}, \citenamefont {Farr}, \citenamefont {Creedon}, \citenamefont
  {Fan}, \citenamefont {Kostylev},\ and\ \citenamefont {Tobar}}]{Goryachev}%
  \BibitemOpen
  \bibfield  {author} {\bibinfo {author} {\bibfnamefont {M.}~\bibnamefont
  {Goryachev}}, \bibinfo {author} {\bibfnamefont {W.~G.}\ \bibnamefont {Farr}},
  \bibinfo {author} {\bibfnamefont {D.~L.}\ \bibnamefont {Creedon}}, \bibinfo
  {author} {\bibfnamefont {Y.}~\bibnamefont {Fan}}, \bibinfo {author}
  {\bibfnamefont {M.}~\bibnamefont {Kostylev}}, \ and\ \bibinfo {author}
  {\bibfnamefont {M.~E.}\ \bibnamefont {Tobar}},\ }\href {\doibase
  10.1103/PhysRevApplied.2.054002} {\bibfield  {journal} {\bibinfo  {journal}
  {Phys. Rev. Applied}\ }\textbf {\bibinfo {volume} {2}},\ \bibinfo {pages}
  {054002} (\bibinfo {year} {2014})}\BibitemShut {NoStop}%
\bibitem [{\citenamefont {Kittel}(1958)}]{Kittel}%
  \BibitemOpen
  \bibfield  {author} {\bibinfo {author} {\bibfnamefont {C.}~\bibnamefont
  {Kittel}},\ }\href {\doibase 10.1103/PhysRev.110.836} {\bibfield  {journal}
  {\bibinfo  {journal} {Phys. Rev.}\ }\textbf {\bibinfo {volume} {110}},\
  \bibinfo {pages} {836} (\bibinfo {year} {1958})}\BibitemShut {NoStop}%
\bibitem [{\citenamefont {Sinha}\ and\ \citenamefont
  {Upadhyaya}(1962)}]{Sinha}%
  \BibitemOpen
  \bibfield  {author} {\bibinfo {author} {\bibfnamefont {K.~P.}\ \bibnamefont
  {Sinha}}\ and\ \bibinfo {author} {\bibfnamefont {U.~N.}\ \bibnamefont
  {Upadhyaya}},\ }\href {\doibase 10.1103/PhysRev.127.432} {\bibfield
  {journal} {\bibinfo  {journal} {Phys. Rev.}\ }\textbf {\bibinfo {volume}
  {127}},\ \bibinfo {pages} {432} (\bibinfo {year} {1962})}\BibitemShut
  {NoStop}%
\bibitem [{\citenamefont {Shen}\ and\ \citenamefont
  {Bloembergen}(1966)}]{Shen}%
  \BibitemOpen
  \bibfield  {author} {\bibinfo {author} {\bibfnamefont {Y.~R.}\ \bibnamefont
  {Shen}}\ and\ \bibinfo {author} {\bibfnamefont {N.}~\bibnamefont
  {Bloembergen}},\ }\href {\doibase 10.1103/PhysRev.143.372} {\bibfield
  {journal} {\bibinfo  {journal} {Phys. Rev.}\ }\textbf {\bibinfo {volume}
  {143}},\ \bibinfo {pages} {372} (\bibinfo {year} {1966})}\BibitemShut
  {NoStop}%
\bibitem [{\citenamefont {Demokritov}\ \emph {et~al.}(2001)\citenamefont
  {Demokritov}, \citenamefont {Hillebrands},\ and\ \citenamefont
  {Slavin}}]{Demokritov}%
  \BibitemOpen
  \bibfield  {author} {\bibinfo {author} {\bibfnamefont {S.}~\bibnamefont
  {Demokritov}}, \bibinfo {author} {\bibfnamefont {B.}~\bibnamefont
  {Hillebrands}}, \ and\ \bibinfo {author} {\bibfnamefont {A.}~\bibnamefont
  {Slavin}},\ }\href {\doibase https://doi.org/10.1016/S0370-1573(00)00116-2}
  {\bibfield  {journal} {\bibinfo  {journal} {Physics Reports}\ }\textbf
  {\bibinfo {volume} {348}},\ \bibinfo {pages} {441} (\bibinfo {year}
  {2001})}\BibitemShut {NoStop}%
\bibitem [{\citenamefont {Hillebrands}()}]{Hillebrands}%
  \BibitemOpen
  \bibfield  {author} {\bibinfo {author} {\bibfnamefont {C.~A.~V.}\
  \bibnamefont {Hillebrands}, \bibfnamefont {B.}},\ }\href {\doibase
  10.1038/nphys3347} {\bibfield  {journal} {\bibinfo  {journal} {Nature
  Physics}\ }\textbf {\bibinfo {volume} {11}},\ \bibinfo {pages}
  {453}}\BibitemShut {NoStop}%
\bibitem [{\citenamefont {Tan}\ and\ \citenamefont
  {Li}(2021{\natexlab{a}})}]{Tan}%
  \BibitemOpen
  \bibfield  {author} {\bibinfo {author} {\bibfnamefont {H.}~\bibnamefont
  {Tan}}\ and\ \bibinfo {author} {\bibfnamefont {J.}~\bibnamefont {Li}},\
  }\href {\doibase 10.1103/PhysRevResearch.3.013192} {\bibfield  {journal}
  {\bibinfo  {journal} {Phys. Rev. Research}\ }\textbf {\bibinfo {volume}
  {3}},\ \bibinfo {pages} {013192} (\bibinfo {year}
  {2021}{\natexlab{a}})}\BibitemShut {NoStop}%
\bibitem [{\citenamefont {Xiao}\ \emph {et~al.}(2019)\citenamefont {Xiao},
  \citenamefont {Yan}, \citenamefont {Zhang}, \citenamefont {Grigoryan},
  \citenamefont {Hu}, \citenamefont {Guo},\ and\ \citenamefont {Xia}}]{Xiao}%
  \BibitemOpen
  \bibfield  {author} {\bibinfo {author} {\bibfnamefont {Y.}~\bibnamefont
  {Xiao}}, \bibinfo {author} {\bibfnamefont {X.~H.}\ \bibnamefont {Yan}},
  \bibinfo {author} {\bibfnamefont {Y.}~\bibnamefont {Zhang}}, \bibinfo
  {author} {\bibfnamefont {V.~L.}\ \bibnamefont {Grigoryan}}, \bibinfo {author}
  {\bibfnamefont {C.~M.}\ \bibnamefont {Hu}}, \bibinfo {author} {\bibfnamefont
  {H.}~\bibnamefont {Guo}}, \ and\ \bibinfo {author} {\bibfnamefont
  {K.}~\bibnamefont {Xia}},\ }\href {\doibase 10.1103/PhysRevB.99.094407}
  {\bibfield  {journal} {\bibinfo  {journal} {Phys. Rev. B}\ }\textbf {\bibinfo
  {volume} {99}},\ \bibinfo {pages} {094407} (\bibinfo {year}
  {2019})}\BibitemShut {NoStop}%
\bibitem [{\citenamefont {Zhang}\ and\ \citenamefont
  {Hollenberg}(2015)}]{ZhangXufeng}%
  \BibitemOpen
  \bibfield  {author} {\bibinfo {author} {\bibfnamefont {Z.~C.-L. Z. N. J.-L.
  T. H. X. C. J. H. G. A.~D.}\ \bibnamefont {Zhang}, \bibfnamefont {Xufeng}}\
  and\ \bibinfo {author} {\bibfnamefont {L.~C.~L.}\ \bibnamefont
  {Hollenberg}},\ }\href {\doibase 10.1038/ncomms9914} {\bibfield  {journal}
  {\bibinfo  {journal} {Nature Communications}\ }\textbf {\bibinfo {volume}
  {6}},\ \bibinfo {pages} {8914} (\bibinfo {year} {2015})}\BibitemShut
  {NoStop}%
\bibitem [{\citenamefont {Lachance-Quirion}\ \emph
  {et~al.}(2019{\natexlab{b}})\citenamefont {Lachance-Quirion}, \citenamefont
  {Tabuchi}, \citenamefont {Gloppe}, \citenamefont {Usami},\ and\ \citenamefont
  {Nakamura}}]{Lachance}%
  \BibitemOpen
  \bibfield  {author} {\bibinfo {author} {\bibfnamefont {D.}~\bibnamefont
  {Lachance-Quirion}}, \bibinfo {author} {\bibfnamefont {Y.}~\bibnamefont
  {Tabuchi}}, \bibinfo {author} {\bibfnamefont {A.}~\bibnamefont {Gloppe}},
  \bibinfo {author} {\bibfnamefont {K.}~\bibnamefont {Usami}}, \ and\ \bibinfo
  {author} {\bibfnamefont {Y.}~\bibnamefont {Nakamura}},\ }\href {\doibase
  10.7567/1882-0786/ab248d} {\bibfield  {journal} {\bibinfo  {journal} {Applied
  Physics Express}\ }\textbf {\bibinfo {volume} {12}},\ \bibinfo {pages}
  {070101} (\bibinfo {year} {2019}{\natexlab{b}})}\BibitemShut {NoStop}%
\bibitem [{\citenamefont {Hisatomi}\ \emph {et~al.}(2016)\citenamefont
  {Hisatomi}, \citenamefont {Osada}, \citenamefont {Tabuchi}, \citenamefont
  {Ishikawa}, \citenamefont {Noguchi}, \citenamefont {Yamazaki}, \citenamefont
  {Usami},\ and\ \citenamefont {Nakamura}}]{Hisatomi}%
  \BibitemOpen
  \bibfield  {author} {\bibinfo {author} {\bibfnamefont {R.}~\bibnamefont
  {Hisatomi}}, \bibinfo {author} {\bibfnamefont {A.}~\bibnamefont {Osada}},
  \bibinfo {author} {\bibfnamefont {Y.}~\bibnamefont {Tabuchi}}, \bibinfo
  {author} {\bibfnamefont {T.}~\bibnamefont {Ishikawa}}, \bibinfo {author}
  {\bibfnamefont {A.}~\bibnamefont {Noguchi}}, \bibinfo {author} {\bibfnamefont
  {R.}~\bibnamefont {Yamazaki}}, \bibinfo {author} {\bibfnamefont
  {K.}~\bibnamefont {Usami}}, \ and\ \bibinfo {author} {\bibfnamefont
  {Y.}~\bibnamefont {Nakamura}},\ }\href {\doibase 10.1103/PhysRevB.93.174427}
  {\bibfield  {journal} {\bibinfo  {journal} {Phys. Rev. B}\ }\textbf {\bibinfo
  {volume} {93}},\ \bibinfo {pages} {174427} (\bibinfo {year}
  {2016})}\BibitemShut {NoStop}%
\bibitem [{\citenamefont {Viola~Kusminskiy}\ \emph
  {et~al.}(2016{\natexlab{a}})\citenamefont {Viola~Kusminskiy}, \citenamefont
  {Tang},\ and\ \citenamefont {Marquardt}}]{Viola}%
  \BibitemOpen
  \bibfield  {author} {\bibinfo {author} {\bibfnamefont {S.}~\bibnamefont
  {Viola~Kusminskiy}}, \bibinfo {author} {\bibfnamefont {H.~X.}\ \bibnamefont
  {Tang}}, \ and\ \bibinfo {author} {\bibfnamefont {F.}~\bibnamefont
  {Marquardt}},\ }\href {\doibase 10.1103/PhysRevA.94.033821} {\bibfield
  {journal} {\bibinfo  {journal} {Phys. Rev. A}\ }\textbf {\bibinfo {volume}
  {94}},\ \bibinfo {pages} {033821} (\bibinfo {year}
  {2016}{\natexlab{a}})}\BibitemShut {NoStop}%
\bibitem [{\citenamefont {Harder}\ \emph {et~al.}(2018)\citenamefont {Harder},
  \citenamefont {Yang}, \citenamefont {Yao}, \citenamefont {Yu}, \citenamefont
  {Rao}, \citenamefont {Gui}, \citenamefont {Stamps},\ and\ \citenamefont
  {Hu}}]{Harder}%
  \BibitemOpen
  \bibfield  {author} {\bibinfo {author} {\bibfnamefont {M.}~\bibnamefont
  {Harder}}, \bibinfo {author} {\bibfnamefont {Y.}~\bibnamefont {Yang}},
  \bibinfo {author} {\bibfnamefont {B.~M.}\ \bibnamefont {Yao}}, \bibinfo
  {author} {\bibfnamefont {C.~H.}\ \bibnamefont {Yu}}, \bibinfo {author}
  {\bibfnamefont {J.~W.}\ \bibnamefont {Rao}}, \bibinfo {author} {\bibfnamefont
  {Y.~S.}\ \bibnamefont {Gui}}, \bibinfo {author} {\bibfnamefont {R.~L.}\
  \bibnamefont {Stamps}}, \ and\ \bibinfo {author} {\bibfnamefont {C.-M.}\
  \bibnamefont {Hu}},\ }\href {\doibase 10.1103/PhysRevLett.121.137203}
  {\bibfield  {journal} {\bibinfo  {journal} {Phys. Rev. Lett.}\ }\textbf
  {\bibinfo {volume} {121}},\ \bibinfo {pages} {137203} (\bibinfo {year}
  {2018})}\BibitemShut {NoStop}%
\bibitem [{\citenamefont {Grigoryan}\ \emph {et~al.}(2018)\citenamefont
  {Grigoryan}, \citenamefont {Shen},\ and\ \citenamefont {Xia}}]{Grigoryan}%
  \BibitemOpen
  \bibfield  {author} {\bibinfo {author} {\bibfnamefont {V.~L.}\ \bibnamefont
  {Grigoryan}}, \bibinfo {author} {\bibfnamefont {K.}~\bibnamefont {Shen}}, \
  and\ \bibinfo {author} {\bibfnamefont {K.}~\bibnamefont {Xia}},\ }\href
  {\doibase 10.1103/PhysRevB.98.024406} {\bibfield  {journal} {\bibinfo
  {journal} {Phys. Rev. B}\ }\textbf {\bibinfo {volume} {98}},\ \bibinfo
  {pages} {024406} (\bibinfo {year} {2018})}\BibitemShut {NoStop}%
\bibitem [{\citenamefont {Wang}\ \emph
  {et~al.}(2019{\natexlab{a}})\citenamefont {Wang}, \citenamefont {Rao},
  \citenamefont {Yang}, \citenamefont {Xu}, \citenamefont {Gui}, \citenamefont
  {Yao}, \citenamefont {You},\ and\ \citenamefont {Hu}}]{Wang}%
  \BibitemOpen
  \bibfield  {author} {\bibinfo {author} {\bibfnamefont {Y.-P.}\ \bibnamefont
  {Wang}}, \bibinfo {author} {\bibfnamefont {J.~W.}\ \bibnamefont {Rao}},
  \bibinfo {author} {\bibfnamefont {Y.}~\bibnamefont {Yang}}, \bibinfo {author}
  {\bibfnamefont {P.-C.}\ \bibnamefont {Xu}}, \bibinfo {author} {\bibfnamefont
  {Y.~S.}\ \bibnamefont {Gui}}, \bibinfo {author} {\bibfnamefont {B.~M.}\
  \bibnamefont {Yao}}, \bibinfo {author} {\bibfnamefont {J.~Q.}\ \bibnamefont
  {You}}, \ and\ \bibinfo {author} {\bibfnamefont {C.-M.}\ \bibnamefont {Hu}},\
  }\href {\doibase 10.1103/PhysRevLett.123.127202} {\bibfield  {journal}
  {\bibinfo  {journal} {Phys. Rev. Lett.}\ }\textbf {\bibinfo {volume} {123}},\
  \bibinfo {pages} {127202} (\bibinfo {year} {2019}{\natexlab{a}})}\BibitemShut
  {NoStop}%
\bibitem [{\citenamefont {Yu}\ \emph {et~al.}(2019)\citenamefont {Yu},
  \citenamefont {Wang}, \citenamefont {Yuan},\ and\ \citenamefont {Xiao}}]{Yu}%
  \BibitemOpen
  \bibfield  {author} {\bibinfo {author} {\bibfnamefont {W.}~\bibnamefont
  {Yu}}, \bibinfo {author} {\bibfnamefont {J.}~\bibnamefont {Wang}}, \bibinfo
  {author} {\bibfnamefont {H.~Y.}\ \bibnamefont {Yuan}}, \ and\ \bibinfo
  {author} {\bibfnamefont {J.}~\bibnamefont {Xiao}},\ }\href {\doibase
  10.1103/PhysRevLett.123.227201} {\bibfield  {journal} {\bibinfo  {journal}
  {Phys. Rev. Lett.}\ }\textbf {\bibinfo {volume} {123}},\ \bibinfo {pages}
  {227201} (\bibinfo {year} {2019})}\BibitemShut {NoStop}%
\bibitem [{\citenamefont {Wang}\ \emph
  {et~al.}(2019{\natexlab{b}})\citenamefont {Wang}, \citenamefont {Rao},
  \citenamefont {Yang}, \citenamefont {Xu}, \citenamefont {Gui}, \citenamefont
  {Yao}, \citenamefont {You},\ and\ \citenamefont {Hu}}]{Yi-Pu}%
  \BibitemOpen
  \bibfield  {author} {\bibinfo {author} {\bibfnamefont {Y.-P.}\ \bibnamefont
  {Wang}}, \bibinfo {author} {\bibfnamefont {J.~W.}\ \bibnamefont {Rao}},
  \bibinfo {author} {\bibfnamefont {Y.}~\bibnamefont {Yang}}, \bibinfo {author}
  {\bibfnamefont {P.-C.}\ \bibnamefont {Xu}}, \bibinfo {author} {\bibfnamefont
  {Y.~S.}\ \bibnamefont {Gui}}, \bibinfo {author} {\bibfnamefont {B.~M.}\
  \bibnamefont {Yao}}, \bibinfo {author} {\bibfnamefont {J.~Q.}\ \bibnamefont
  {You}}, \ and\ \bibinfo {author} {\bibfnamefont {C.-M.}\ \bibnamefont {Hu}},\
  }\href {\doibase 10.1103/PhysRevLett.123.127202} {\bibfield  {journal}
  {\bibinfo  {journal} {Phys. Rev. Lett.}\ }\textbf {\bibinfo {volume} {123}},\
  \bibinfo {pages} {127202} (\bibinfo {year} {2019}{\natexlab{b}})}\BibitemShut
  {NoStop}%
\bibitem [{\citenamefont {Xu}\ \emph {et~al.}(2020{\natexlab{a}})\citenamefont
  {Xu}, \citenamefont {Gao}, \citenamefont {Wang},\ and\ \citenamefont
  {Wang}}]{wen}%
  \BibitemOpen
  \bibfield  {author} {\bibinfo {author} {\bibfnamefont {W.-L.}\ \bibnamefont
  {Xu}}, \bibinfo {author} {\bibfnamefont {Y.-P.}\ \bibnamefont {Gao}},
  \bibinfo {author} {\bibfnamefont {T.-J.}\ \bibnamefont {Wang}}, \ and\
  \bibinfo {author} {\bibfnamefont {C.}~\bibnamefont {Wang}},\ }\href {\doibase
  10.1364/OE.394488} {\bibfield  {journal} {\bibinfo  {journal} {Opt. Express}\
  }\textbf {\bibinfo {volume} {28}},\ \bibinfo {pages} {22334} (\bibinfo {year}
  {2020}{\natexlab{a}})}\BibitemShut {NoStop}%
\bibitem [{\citenamefont {Viola~Kusminskiy}\ \emph
  {et~al.}(2016{\natexlab{b}})\citenamefont {Viola~Kusminskiy}, \citenamefont
  {Tang},\ and\ \citenamefont {Marquardt}}]{Kusminskiy}%
  \BibitemOpen
  \bibfield  {author} {\bibinfo {author} {\bibfnamefont {S.}~\bibnamefont
  {Viola~Kusminskiy}}, \bibinfo {author} {\bibfnamefont {H.~X.}\ \bibnamefont
  {Tang}}, \ and\ \bibinfo {author} {\bibfnamefont {F.}~\bibnamefont
  {Marquardt}},\ }\href {\doibase 10.1103/PhysRevA.94.033821} {\bibfield
  {journal} {\bibinfo  {journal} {Phys. Rev. A}\ }\textbf {\bibinfo {volume}
  {94}},\ \bibinfo {pages} {033821} (\bibinfo {year}
  {2016}{\natexlab{b}})}\BibitemShut {NoStop}%
\bibitem [{\citenamefont {Gao}\ \emph {et~al.}(2020)\citenamefont {Gao},
  \citenamefont {Cao}, \citenamefont {Duan}, \citenamefont {Liu}, \citenamefont
  {Pang}, \citenamefont {Wang},\ and\ \citenamefont {Wang}}]{Yong}%
  \BibitemOpen
  \bibfield  {author} {\bibinfo {author} {\bibfnamefont {Y.-P.}\ \bibnamefont
  {Gao}}, \bibinfo {author} {\bibfnamefont {C.}~\bibnamefont {Cao}}, \bibinfo
  {author} {\bibfnamefont {Y.-W.}\ \bibnamefont {Duan}}, \bibinfo {author}
  {\bibfnamefont {X.-F.}\ \bibnamefont {Liu}}, \bibinfo {author} {\bibfnamefont
  {T.-T.}\ \bibnamefont {Pang}}, \bibinfo {author} {\bibfnamefont {T.-J.}\
  \bibnamefont {Wang}}, \ and\ \bibinfo {author} {\bibfnamefont
  {C.}~\bibnamefont {Wang}},\ }\href {\doibase doi:10.1515/nanoph-2019-0441}
  {\bibfield  {journal} {\bibinfo  {journal} {Nanophotonics}\ }\textbf
  {\bibinfo {volume} {9}},\ \bibinfo {pages} {1953} (\bibinfo {year}
  {2020})}\BibitemShut {NoStop}%
\bibitem [{\citenamefont {Xu}\ \emph {et~al.}(2020{\natexlab{b}})\citenamefont
  {Xu}, \citenamefont {Liu}, \citenamefont {Sun}, \citenamefont {Gao},
  \citenamefont {Wang},\ and\ \citenamefont {Wang}}]{Xu}%
  \BibitemOpen
  \bibfield  {author} {\bibinfo {author} {\bibfnamefont {W.-L.}\ \bibnamefont
  {Xu}}, \bibinfo {author} {\bibfnamefont {X.-F.}\ \bibnamefont {Liu}},
  \bibinfo {author} {\bibfnamefont {Y.}~\bibnamefont {Sun}}, \bibinfo {author}
  {\bibfnamefont {Y.-P.}\ \bibnamefont {Gao}}, \bibinfo {author} {\bibfnamefont
  {T.-J.}\ \bibnamefont {Wang}}, \ and\ \bibinfo {author} {\bibfnamefont
  {C.}~\bibnamefont {Wang}},\ }\href {\doibase 10.1103/PhysRevE.101.012205}
  {\bibfield  {journal} {\bibinfo  {journal} {Phys. Rev. E}\ }\textbf {\bibinfo
  {volume} {101}},\ \bibinfo {pages} {012205} (\bibinfo {year}
  {2020}{\natexlab{b}})}\BibitemShut {NoStop}%
\bibitem [{\citenamefont {Zhang}(2017)}]{Dengke}%
  \BibitemOpen
  \bibfield  {author} {\bibinfo {author} {\bibfnamefont {L.~X.-Q. W. Y.-P.-L.
  T.-F. Y. Y. J.~Q.}\ \bibnamefont {Zhang}, \bibfnamefont {Dengke}},\ }\href
  {\doibase 10.1038/s41467-017-01634-w} {\bibfield  {journal} {\bibinfo
  {journal} {Nature Communications}\ }\textbf {\bibinfo {volume} {8}},\
  \bibinfo {pages} {1368} (\bibinfo {year} {2017})}\BibitemShut {NoStop}%
\bibitem [{\citenamefont {Zhang}\ and\ \citenamefont {You}(2019)}]{Qiang}%
  \BibitemOpen
  \bibfield  {author} {\bibinfo {author} {\bibfnamefont {G.-Q.}\ \bibnamefont
  {Zhang}}\ and\ \bibinfo {author} {\bibfnamefont {J.~Q.}\ \bibnamefont
  {You}},\ }\href {\doibase 10.1103/PhysRevB.99.054404} {\bibfield  {journal}
  {\bibinfo  {journal} {Phys. Rev. B}\ }\textbf {\bibinfo {volume} {99}},\
  \bibinfo {pages} {054404} (\bibinfo {year} {2019})}\BibitemShut {NoStop}%
\bibitem [{\citenamefont {Kong}\ \emph {et~al.}(2021)\citenamefont {Kong},
  \citenamefont {Hu}, \citenamefont {Hu},\ and\ \citenamefont {Xu}}]{Deyi}%
  \BibitemOpen
  \bibfield  {author} {\bibinfo {author} {\bibfnamefont {D.}~\bibnamefont
  {Kong}}, \bibinfo {author} {\bibfnamefont {X.}~\bibnamefont {Hu}}, \bibinfo
  {author} {\bibfnamefont {L.}~\bibnamefont {Hu}}, \ and\ \bibinfo {author}
  {\bibfnamefont {J.}~\bibnamefont {Xu}},\ }\href {\doibase
  10.1103/PhysRevB.103.224416} {\bibfield  {journal} {\bibinfo  {journal}
  {Phys. Rev. B}\ }\textbf {\bibinfo {volume} {103}},\ \bibinfo {pages}
  {224416} (\bibinfo {year} {2021})}\BibitemShut {NoStop}%
\bibitem [{\citenamefont {Li}\ \emph {et~al.}(2018)\citenamefont {Li},
  \citenamefont {Zhu},\ and\ \citenamefont {Agarwal}}]{ShiYao}%
  \BibitemOpen
  \bibfield  {author} {\bibinfo {author} {\bibfnamefont {J.}~\bibnamefont
  {Li}}, \bibinfo {author} {\bibfnamefont {S.-Y.}\ \bibnamefont {Zhu}}, \ and\
  \bibinfo {author} {\bibfnamefont {G.~S.}\ \bibnamefont {Agarwal}},\ }\href
  {\doibase 10.1103/PhysRevLett.121.203601} {\bibfield  {journal} {\bibinfo
  {journal} {Phys. Rev. Lett.}\ }\textbf {\bibinfo {volume} {121}},\ \bibinfo
  {pages} {203601} (\bibinfo {year} {2018})}\BibitemShut {NoStop}%
\bibitem [{\citenamefont {Yang}\ \emph {et~al.}(2021)\citenamefont {Yang},
  \citenamefont {Jin}, \citenamefont {Jin}, \citenamefont {Liu}, \citenamefont
  {Liu},\ and\ \citenamefont {Yang}}]{Yang}%
  \BibitemOpen
  \bibfield  {author} {\bibinfo {author} {\bibfnamefont {Z.-B.}\ \bibnamefont
  {Yang}}, \bibinfo {author} {\bibfnamefont {H.}~\bibnamefont {Jin}}, \bibinfo
  {author} {\bibfnamefont {J.-W.}\ \bibnamefont {Jin}}, \bibinfo {author}
  {\bibfnamefont {J.-Y.}\ \bibnamefont {Liu}}, \bibinfo {author} {\bibfnamefont
  {H.-Y.}\ \bibnamefont {Liu}}, \ and\ \bibinfo {author} {\bibfnamefont
  {R.-C.}\ \bibnamefont {Yang}},\ }\href {\doibase
  10.1103/PhysRevResearch.3.023126} {\bibfield  {journal} {\bibinfo  {journal}
  {Phys. Rev. Research}\ }\textbf {\bibinfo {volume} {3}},\ \bibinfo {pages}
  {023126} (\bibinfo {year} {2021})}\BibitemShut {NoStop}%
\bibitem [{\citenamefont {Tan}\ and\ \citenamefont
  {Li}(2021{\natexlab{b}})}]{Huatang}%
  \BibitemOpen
  \bibfield  {author} {\bibinfo {author} {\bibfnamefont {H.}~\bibnamefont
  {Tan}}\ and\ \bibinfo {author} {\bibfnamefont {J.}~\bibnamefont {Li}},\
  }\href {\doibase 10.1103/PhysRevResearch.3.013192} {\bibfield  {journal}
  {\bibinfo  {journal} {Phys. Rev. Research}\ }\textbf {\bibinfo {volume}
  {3}},\ \bibinfo {pages} {013192} (\bibinfo {year}
  {2021}{\natexlab{b}})}\BibitemShut {NoStop}%
\bibitem [{\citenamefont {Li}\ and\ \citenamefont {Zhu}(2019)}]{Li_2019}%
  \BibitemOpen
  \bibfield  {author} {\bibinfo {author} {\bibfnamefont {J.}~\bibnamefont
  {Li}}\ and\ \bibinfo {author} {\bibfnamefont {S.-Y.}\ \bibnamefont {Zhu}},\
  }\href {\doibase 10.1088/1367-2630/ab3508} {\bibfield  {journal} {\bibinfo
  {journal} {New Journal of Physics}\ }\textbf {\bibinfo {volume} {21}},\
  \bibinfo {pages} {085001} (\bibinfo {year} {2019})}\BibitemShut {NoStop}%
\bibitem [{\citenamefont {Rao}(2021)}]{Rao}%
  \BibitemOpen
  \bibfield  {author} {\bibinfo {author} {\bibfnamefont {X.-P. C. G.-Y. S.
  W.-Y. P. Y. Y. Y. B. D.~J.}\ \bibnamefont {Rao}, \bibfnamefont {J.~W.}},\
  }\href {\doibase 10.1038/s41467-021-22171-7} {\bibfield  {journal} {\bibinfo
  {journal} {Nature Communications}\ }\textbf {\bibinfo {volume} {12}},\
  \bibinfo {pages} {1933} (\bibinfo {year} {2021})}\BibitemShut {NoStop}%
\bibitem [{\citenamefont {Bittencourt}\ \emph {et~al.}(2019)\citenamefont
  {Bittencourt}, \citenamefont {Feulner},\ and\ \citenamefont
  {Kusminskiy}}]{Bittencourt}%
  \BibitemOpen
  \bibfield  {author} {\bibinfo {author} {\bibfnamefont {V.~A. S.~V.}\
  \bibnamefont {Bittencourt}}, \bibinfo {author} {\bibfnamefont
  {V.}~\bibnamefont {Feulner}}, \ and\ \bibinfo {author} {\bibfnamefont
  {S.~V.}\ \bibnamefont {Kusminskiy}},\ }\href {\doibase
  10.1103/PhysRevA.100.013810} {\bibfield  {journal} {\bibinfo  {journal}
  {Phys. Rev. A}\ }\textbf {\bibinfo {volume} {100}},\ \bibinfo {pages}
  {013810} (\bibinfo {year} {2019})}\BibitemShut {NoStop}%
\bibitem [{\citenamefont {Li}\ \emph {et~al.}(2019)\citenamefont {Li},
  \citenamefont {Zhu},\ and\ \citenamefont {Agarwal}}]{Shi-Yao}%
  \BibitemOpen
  \bibfield  {author} {\bibinfo {author} {\bibfnamefont {J.}~\bibnamefont
  {Li}}, \bibinfo {author} {\bibfnamefont {S.-Y.}\ \bibnamefont {Zhu}}, \ and\
  \bibinfo {author} {\bibfnamefont {G.~S.}\ \bibnamefont {Agarwal}},\ }\href
  {\doibase 10.1103/PhysRevA.99.021801} {\bibfield  {journal} {\bibinfo
  {journal} {Phys. Rev. A}\ }\textbf {\bibinfo {volume} {99}},\ \bibinfo
  {pages} {021801} (\bibinfo {year} {2019})}\BibitemShut {NoStop}%
\bibitem [{\citenamefont {Zhao}\ \emph
  {et~al.}(2020{\natexlab{b}})\citenamefont {Zhao}, \citenamefont {Li},
  \citenamefont {Chao}, \citenamefont {Peng}, \citenamefont {Li},\ and\
  \citenamefont {Zhou}}]{Chengsong}%
  \BibitemOpen
  \bibfield  {author} {\bibinfo {author} {\bibfnamefont {C.}~\bibnamefont
  {Zhao}}, \bibinfo {author} {\bibfnamefont {X.}~\bibnamefont {Li}}, \bibinfo
  {author} {\bibfnamefont {S.}~\bibnamefont {Chao}}, \bibinfo {author}
  {\bibfnamefont {R.}~\bibnamefont {Peng}}, \bibinfo {author} {\bibfnamefont
  {C.}~\bibnamefont {Li}}, \ and\ \bibinfo {author} {\bibfnamefont
  {L.}~\bibnamefont {Zhou}},\ }\href {\doibase 10.1103/PhysRevA.101.063838}
  {\bibfield  {journal} {\bibinfo  {journal} {Phys. Rev. A}\ }\textbf {\bibinfo
  {volume} {101}},\ \bibinfo {pages} {063838} (\bibinfo {year}
  {2020}{\natexlab{b}})}\BibitemShut {NoStop}%
\bibitem [{\citenamefont {Gao}\ \emph {et~al.}(2019{\natexlab{b}})\citenamefont
  {Gao}, \citenamefont {Liu}, \citenamefont {Wang}, \citenamefont {Cao},\ and\
  \citenamefont {Wang}}]{Fei}%
  \BibitemOpen
  \bibfield  {author} {\bibinfo {author} {\bibfnamefont {Y.-P.}\ \bibnamefont
  {Gao}}, \bibinfo {author} {\bibfnamefont {X.-F.}\ \bibnamefont {Liu}},
  \bibinfo {author} {\bibfnamefont {T.-J.}\ \bibnamefont {Wang}}, \bibinfo
  {author} {\bibfnamefont {C.}~\bibnamefont {Cao}}, \ and\ \bibinfo {author}
  {\bibfnamefont {C.}~\bibnamefont {Wang}},\ }\href {\doibase
  10.1103/PhysRevA.100.043831} {\bibfield  {journal} {\bibinfo  {journal}
  {Phys. Rev. A}\ }\textbf {\bibinfo {volume} {100}},\ \bibinfo {pages}
  {043831} (\bibinfo {year} {2019}{\natexlab{b}})}\BibitemShut {NoStop}%
\bibitem [{\citenamefont {Zou}\ \emph {et~al.}(2020)\citenamefont {Zou},
  \citenamefont {Liao},\ and\ \citenamefont {Liao}}]{Fen}%
  \BibitemOpen
  \bibfield  {author} {\bibinfo {author} {\bibfnamefont {F.}~\bibnamefont
  {Zou}}, \bibinfo {author} {\bibfnamefont {D.-G.}\ \bibnamefont {Liao}}, \
  and\ \bibinfo {author} {\bibfnamefont {J.-Q.}\ \bibnamefont {Liao}},\
  }\href@noop {} {\bibfield  {journal} {\bibinfo  {journal} {Optics express}\
  }\textbf {\bibinfo {volume} {28}},\ \bibinfo {pages} {16175} (\bibinfo {year}
  {2020})}\BibitemShut {NoStop}%
\bibitem [{\citenamefont {Nietner}(2010)}]{0Quantum}%
  \BibitemOpen
  \bibfield  {author} {\bibinfo {author} {\bibfnamefont {C.}~\bibnamefont
  {Nietner}},\ }\href@noop {} {\  (\bibinfo {year} {2010})}\BibitemShut
  {NoStop}%
\bibitem [{\citenamefont {Hohenadler}\ \emph {et~al.}(2012)\citenamefont
  {Hohenadler}, \citenamefont {Aichhorn}, \citenamefont {Pollet},\ and\
  \citenamefont {Schmidt}}]{Hohenadler}%
  \BibitemOpen
  \bibfield  {author} {\bibinfo {author} {\bibfnamefont {M.}~\bibnamefont
  {Hohenadler}}, \bibinfo {author} {\bibfnamefont {M.}~\bibnamefont
  {Aichhorn}}, \bibinfo {author} {\bibfnamefont {L.}~\bibnamefont {Pollet}}, \
  and\ \bibinfo {author} {\bibfnamefont {S.}~\bibnamefont {Schmidt}},\ }\href
  {\doibase 10.1103/PhysRevA.85.013810} {\bibfield  {journal} {\bibinfo
  {journal} {Phys. Rev. A}\ }\textbf {\bibinfo {volume} {85}},\ \bibinfo
  {pages} {013810} (\bibinfo {year} {2012})}\BibitemShut {NoStop}%
\bibitem [{\citenamefont {Holstein}\ and\ \citenamefont
  {Primakoff}(1940)}]{Holstein}%
  \BibitemOpen
  \bibfield  {author} {\bibinfo {author} {\bibfnamefont {T.}~\bibnamefont
  {Holstein}}\ and\ \bibinfo {author} {\bibfnamefont {H.}~\bibnamefont
  {Primakoff}},\ }\href {\doibase 10.1103/PhysRev.58.1098} {\bibfield
  {journal} {\bibinfo  {journal} {Phys. Rev.}\ }\textbf {\bibinfo {volume}
  {58}},\ \bibinfo {pages} {1098} (\bibinfo {year} {1940})}\BibitemShut
  {NoStop}%
\bibitem [{\citenamefont {Huebl}\ \emph {et~al.}(2013)\citenamefont {Huebl},
  \citenamefont {Zollitsch}, \citenamefont {Lotze}, \citenamefont {Hocke},
  \citenamefont {Greifenstein}, \citenamefont {Marx}, \citenamefont {Gross},\
  and\ \citenamefont {Goennenwein}}]{Huebl}%
  \BibitemOpen
  \bibfield  {author} {\bibinfo {author} {\bibfnamefont {H.}~\bibnamefont
  {Huebl}}, \bibinfo {author} {\bibfnamefont {C.~W.}\ \bibnamefont
  {Zollitsch}}, \bibinfo {author} {\bibfnamefont {J.}~\bibnamefont {Lotze}},
  \bibinfo {author} {\bibfnamefont {F.}~\bibnamefont {Hocke}}, \bibinfo
  {author} {\bibfnamefont {M.}~\bibnamefont {Greifenstein}}, \bibinfo {author}
  {\bibfnamefont {A.}~\bibnamefont {Marx}}, \bibinfo {author} {\bibfnamefont
  {R.}~\bibnamefont {Gross}}, \ and\ \bibinfo {author} {\bibfnamefont
  {S.~T.~B.}\ \bibnamefont {Goennenwein}},\ }\href {\doibase
  10.1103/PhysRevLett.111.127003} {\bibfield  {journal} {\bibinfo  {journal}
  {Phys. Rev. Lett.}\ }\textbf {\bibinfo {volume} {111}},\ \bibinfo {pages}
  {127003} (\bibinfo {year} {2013})}\BibitemShut {NoStop}%
\bibitem [{\citenamefont {van Oosten}\ \emph {et~al.}(2001)\citenamefont {van
  Oosten}, \citenamefont {van~der Straten},\ and\ \citenamefont
  {Stoof}}]{Oosten}%
  \BibitemOpen
  \bibfield  {author} {\bibinfo {author} {\bibfnamefont {D.}~\bibnamefont {van
  Oosten}}, \bibinfo {author} {\bibfnamefont {P.}~\bibnamefont {van~der
  Straten}}, \ and\ \bibinfo {author} {\bibfnamefont {H.~T.~C.}\ \bibnamefont
  {Stoof}},\ }\href {\doibase 10.1103/PhysRevA.63.053601} {\bibfield  {journal}
  {\bibinfo  {journal} {Phys. Rev. A}\ }\textbf {\bibinfo {volume} {63}},\
  \bibinfo {pages} {053601} (\bibinfo {year} {2001})}\BibitemShut {NoStop}%
\bibitem [{\citenamefont {Sethna}(2011)}]{Sethna}%
  \BibitemOpen
  \bibfield  {author} {\bibinfo {author} {\bibfnamefont {J.~P.}\ \bibnamefont
  {Sethna}},\ }\href@noop {} {\bibfield  {journal} {\bibinfo  {journal}
  {Statistical Mechanics}\ }\textbf {\bibinfo {volume} {126}},\ \bibinfo
  {pages} {429} (\bibinfo {year} {2011})}\BibitemShut {NoStop}%
\bibitem [{\citenamefont {Huang}\ and\ \citenamefont {Holbrow}(1963)}]{huang}%
  \BibitemOpen
  \bibfield  {author} {\bibinfo {author} {\bibfnamefont {K.}~\bibnamefont
  {Huang}}\ and\ \bibinfo {author} {\bibfnamefont {C.~H.}\ \bibnamefont
  {Holbrow}},\ }\href@noop {} {\bibfield  {journal} {\bibinfo  {journal}
  {Physics Today}\ }\textbf {\bibinfo {volume} {18}},\ \bibinfo {pages} {92}
  (\bibinfo {year} {1963})}\BibitemShut {NoStop}%
\bibitem [{\citenamefont {Wang}\ \emph {et~al.}(2016)\citenamefont {Wang},
  \citenamefont {Zhang}, \citenamefont {Zhang}, \citenamefont {Luo},
  \citenamefont {Xiong}, \citenamefont {Wang}, \citenamefont {Li},
  \citenamefont {Hu},\ and\ \citenamefont {You}}]{GuoQiang}%
  \BibitemOpen
  \bibfield  {author} {\bibinfo {author} {\bibfnamefont {Y.-P.}\ \bibnamefont
  {Wang}}, \bibinfo {author} {\bibfnamefont {G.-Q.}\ \bibnamefont {Zhang}},
  \bibinfo {author} {\bibfnamefont {D.}~\bibnamefont {Zhang}}, \bibinfo
  {author} {\bibfnamefont {X.-Q.}\ \bibnamefont {Luo}}, \bibinfo {author}
  {\bibfnamefont {W.}~\bibnamefont {Xiong}}, \bibinfo {author} {\bibfnamefont
  {S.-P.}\ \bibnamefont {Wang}}, \bibinfo {author} {\bibfnamefont {T.-F.}\
  \bibnamefont {Li}}, \bibinfo {author} {\bibfnamefont {C.-M.}\ \bibnamefont
  {Hu}}, \ and\ \bibinfo {author} {\bibfnamefont {J.~Q.}\ \bibnamefont {You}},\
  }\href {\doibase 10.1103/PhysRevB.94.224410} {\bibfield  {journal} {\bibinfo
  {journal} {Phys. Rev. B}\ }\textbf {\bibinfo {volume} {94}},\ \bibinfo
  {pages} {224410} (\bibinfo {year} {2016})}\BibitemShut {NoStop}%
\bibitem [{\citenamefont {Wang}\ \emph {et~al.}()\citenamefont {Wang},
  \citenamefont {Zhang}, \citenamefont {Zhang}, \citenamefont {Li},
  \citenamefont {Hu},\ and\ \citenamefont {You}}]{YiPu}%
  \BibitemOpen
  \bibfield  {author} {\bibinfo {author} {\bibfnamefont {Y.-P.}\ \bibnamefont
  {Wang}}, \bibinfo {author} {\bibfnamefont {G.-Q.}\ \bibnamefont {Zhang}},
  \bibinfo {author} {\bibfnamefont {D.}~\bibnamefont {Zhang}}, \bibinfo
  {author} {\bibfnamefont {T.-F.}\ \bibnamefont {Li}}, \bibinfo {author}
  {\bibfnamefont {C.-M.}\ \bibnamefont {Hu}}, \ and\ \bibinfo {author}
  {\bibfnamefont {J.~Q.}\ \bibnamefont {You}},\ }\href@noop {} {\bibfield
  {journal} {\bibinfo  {journal} {Phys. Rev. Lett.}\ }\textbf {\bibinfo
  {volume} {120}}}\BibitemShut {NoStop}%
\bibitem [{\citenamefont {Buks}\ \emph {et~al.}(2020)\citenamefont {Buks},
  \citenamefont {Brookes}, \citenamefont {Ginossar}, \citenamefont {Deng},
  \citenamefont {Orgiazzi}, \citenamefont {Otto},\ and\ \citenamefont
  {Lupascu}}]{Buks}%
  \BibitemOpen
  \bibfield  {author} {\bibinfo {author} {\bibfnamefont {E.}~\bibnamefont
  {Buks}}, \bibinfo {author} {\bibfnamefont {P.}~\bibnamefont {Brookes}},
  \bibinfo {author} {\bibfnamefont {E.}~\bibnamefont {Ginossar}}, \bibinfo
  {author} {\bibfnamefont {C.}~\bibnamefont {Deng}}, \bibinfo {author}
  {\bibfnamefont {J.-L. F.~X.}\ \bibnamefont {Orgiazzi}}, \bibinfo {author}
  {\bibfnamefont {M.}~\bibnamefont {Otto}}, \ and\ \bibinfo {author}
  {\bibfnamefont {A.}~\bibnamefont {Lupascu}},\ }\href {\doibase
  10.1103/PhysRevA.102.043716} {\bibfield  {journal} {\bibinfo  {journal}
  {Phys. Rev. A}\ }\textbf {\bibinfo {volume} {102}},\ \bibinfo {pages}
  {043716} (\bibinfo {year} {2020})}\BibitemShut {NoStop}%
\bibitem [{\citenamefont {Dubyna}\ and\ \citenamefont {Kuo}(2020)}]{Dmytro}%
  \BibitemOpen
  \bibfield  {author} {\bibinfo {author} {\bibfnamefont {D.}~\bibnamefont
  {Dubyna}}\ and\ \bibinfo {author} {\bibfnamefont {W.}~\bibnamefont {Kuo}},\
  }\href {\doibase 10.1088/2058-9565/ab8114} {\bibfield  {journal} {\bibinfo
  {journal} {Quantum Science and Technology}\ }\textbf {\bibinfo {volume}
  {5}},\ \bibinfo {pages} {035002} (\bibinfo {year} {2020})}\BibitemShut
  {NoStop}%
\bibitem [{\citenamefont {Zhang}\ \emph
  {et~al.}(2014{\natexlab{b}})\citenamefont {Zhang}, \citenamefont {Zou},
  \citenamefont {Jiang},\ and\ \citenamefont {Tang}}]{Chang}%
  \BibitemOpen
  \bibfield  {author} {\bibinfo {author} {\bibfnamefont {X.}~\bibnamefont
  {Zhang}}, \bibinfo {author} {\bibfnamefont {C.-L.}\ \bibnamefont {Zou}},
  \bibinfo {author} {\bibfnamefont {L.}~\bibnamefont {Jiang}}, \ and\ \bibinfo
  {author} {\bibfnamefont {H.~X.}\ \bibnamefont {Tang}},\ }\href {\doibase
  10.1103/PhysRevLett.113.156401} {\bibfield  {journal} {\bibinfo  {journal}
  {Phys. Rev. Lett.}\ }\textbf {\bibinfo {volume} {113}},\ \bibinfo {pages}
  {156401} (\bibinfo {year} {2014}{\natexlab{b}})}\BibitemShut {NoStop}%
\bibitem [{\citenamefont {Mascarenhas}\ \emph {et~al.}(2012)\citenamefont
  {Mascarenhas}, \citenamefont {Heaney}, \citenamefont {Aguiar},\ and\
  \citenamefont {Santos}}]{Eduardo}%
  \BibitemOpen
  \bibfield  {author} {\bibinfo {author} {\bibfnamefont {E.}~\bibnamefont
  {Mascarenhas}}, \bibinfo {author} {\bibfnamefont {L.}~\bibnamefont {Heaney}},
  \bibinfo {author} {\bibfnamefont {M.~C.~O.}\ \bibnamefont {Aguiar}}, \ and\
  \bibinfo {author} {\bibfnamefont {M.~F.}\ \bibnamefont {Santos}},\ }\href
  {\doibase 10.1088/1367-2630/14/4/043033} {\ \textbf {\bibinfo {volume}
  {14}},\ \bibinfo {pages} {043033} (\bibinfo {year} {2012})}\BibitemShut
  {NoStop}%
\end{thebibliography}%

\section{APPENDIX A}\label{A}
\begin{widetext}
\begin{tiny}
\begin{flalign}
H_1=\left(
\begin{array}{ccccccc}
 \text{N$\omega $}_c-\text{N$\mu $} & \sqrt{N} G_m & 0 &    &    & 0 & 0 \\
 \sqrt{N} G_m & (N-1) \omega _c+\omega _m-\text{N$\mu $} & \sqrt{2 (N-1)} G_m &    &    & 0 & 0 \\
 0 & \sqrt{2 (N-1)} G_m & (N-2) \omega _c+2 \omega _m-\text{N$\mu $} &    &    & 0 & 0 \\
 0 & 0 & \sqrt{3 (N-2)} G_m &    &    & 0 & 0 \\
    &    &    & \ddots & \ddots & 0 & 0 \\
    &    &    & \ddots & \ddots & 0 & 0 \\
 0 & 0 & 0 &    &    & \omega _c+(N-1) \omega _m-\text{N$\mu $} & \sqrt{N} G_m \\
 0 & 0 & 0 &    &    & \sqrt{N} G_m & \text{N$\omega $}_m-\text{N$\mu $} \\
\end{array}
\right)_{(N+1)\times(N+1)} \tag{A1}
\end{flalign}
\end{tiny}
\begin{tiny}
\begin{flalign}
H_2=\left(
\begin{array}{ccccccc}
 \omega _a+(N-1) \omega _c-\text{N$\mu $} & \sqrt{N-1} G_m & 0 &    &    & 0 & 0 \\
 \sqrt{N-1} G_m & \omega _a+(N-2) \omega _c+\omega _m-\text{N$\mu $} & \sqrt{2 (N-2)} G_m &    &    & 0 & 0 \\
 0 & \sqrt{2 (N-2)} G_m & \omega _a+(N-3) \omega _c+2 \omega _m-\text{N$\mu $} &    &    & 0 & 0 \\
 0 & 0 & \sqrt{3 (N-3)} G_m &    &    & 0 & 0 \\
    &    &    & \ddots & \ddots &    &    \\
    &    &    & \ddots & \ddots &    &    \\
 0 & 0 & 0 &    &    & \omega _a+\omega _c+(N-2) \omega _m-\text{N$\mu $} & \sqrt{N-1} G_m \\
 0 & 0 & 0 &    &    & \sqrt{N-1} G_m & \omega _a+(N-1) \omega _m-\text{N$\mu $} \\
\end{array}
 \right)_{N\times N} \tag{A2}
\end{flalign}
\end{tiny}
\end{widetext}

\section{APPENDIX B}\label{B}
The lower excitions eigenvalues are as follows:
\begin{footnotesize}
 \begin{flalign}
 E_{0,0}=0 \label{1.1} \tag{B1}
 \end{flalign}
 \begin{flalign}
 E_{1,0}=\omega -\mu \label{1.2} \tag{B2}
 \end{flalign}
 \begin{flalign}
 E_{1,-}=\omega -\mu-\sqrt{g_a^2+G_m^2} \label{1.3} \tag{B3}
 \end{flalign}
 \begin{flalign}
 E_{1,+}=\omega -\mu+\sqrt{g_a^2+G_m^2} \label{1.4} \tag{B4}
 \end{flalign}
 \begin{flalign}
E_{2,0}=2 (\omega-\mu  ) \label{3.1} \tag{B5}
\end{flalign}
 \begin{flalign}
E_{2,-'}=2 (\omega-\mu)-\frac{\sqrt{3 g_a^2+5 G_m^2-\sqrt{30 g_a^2 G_m^2+g_a^4+9 G_m^4}}}{\sqrt{2}} \label{3.2} \tag{B6}
\end{flalign}
 \begin{flalign}
E_{2,+'}=2 (\omega-\mu)+\frac{\sqrt{3 g_a^2+5 G_m^2-\sqrt{30 g_a^2 G_m^2+g_a^4+9 G_m^4}}}{\sqrt{2}} \label{3.3} \tag{B7}
\end{flalign}
 \begin{flalign}
E_{2,-}=2 (\omega-\mu)-\frac{\sqrt{3 g_a^2+5 G_m^2+\sqrt{30 g_a^2 G_m^2+g_a^4+9 G_m^4}}}{\sqrt{2}} \label{3.4} \tag{B8}
\end{flalign}
 \begin{flalign}
E_{2,+}=2 (\omega-\mu)+\frac{\sqrt{3 g_a^2+5 G_m^2+\sqrt{30 g_a^2 G_m^2+g_a^4+9 G_m^4}}}{\sqrt{2}} \label{3.5} \tag{B9}
\end{flalign}
\end{footnotesize}

The parameters involved in Eq. \eqref{3.8} are following:
\begin{footnotesize}
\begin{equation}
a_1=-\frac{\sqrt{g_a^2+G_m^2}}{G_m},\label{3.0} \tag{B10}
\end{equation}
\begin{equation}
d_1=\frac{g_a}{G_m} \label{3.20} \tag{B11}
\end{equation}
\begin{equation}
B_1=1+a_1^2+d_1^2 \label{3.21} \tag{B12}
\end{equation}
\begin{align}
b=-\frac{\sqrt{30 g_a^2 G_m^2+g_a^4+9 G_m^4}-7 g_a^2+3 G_m^2}{6 \sqrt{2} g_a G_m} \label{3.10} \tag{B13}
\end{align}
\begin{equation}
c=-\frac{\sqrt{30 g_a^2 G_m^2+g_a^4+9 G_m^4}-7 g_a^2+3 G_m^2}{6 \sqrt{2} g_a G_m} \label{3.11} \tag{B14}
\end{equation}
\begin{equation}
d=-\frac{\sqrt{-\sqrt{30 g_a^2 G_m^2+g_a^4+9 G_m^4}+3 g_a^2+5 G_m^2}}{2 G_m} \label{3.12} \tag{B15}
\end{equation}
\begin{equation}
B_{2}=a^2+b^2+c^2+d^2+1 \label{3.13} \tag{B16}
\end{equation}
\end{footnotesize}
\begin{widetext}
\begin{footnotesize}
\begin{equation}
a=\frac{\sqrt{-\sqrt{30 g_a^2 G_m^2+g_a^4+9 G_m^4}+3 g_a^2+5 G_m^2} \left(\sqrt{30 g_a^2 G_m^2+g_a^4+9 G_m^4}-g_a^2+3 G_m^2\right)}{12 g_a G_m^2} \label{3.9} \tag{B17}
\end{equation}
\end{footnotesize}
\end{widetext}

The second-order corrections to the energy and the normalized eigenstaes are following:
\begin{footnotesize}
\begin{equation}
E_{1,-}^{(2)}=(z\kappa \psi)^2(\frac{\left| \left\langle \phi _{2}^{(0)}\left|\hat{a}^{\dagger }\right|\phi _{1}^{(0)}\right\rangle \right|^2}{E_{1,-}^{(0)}-E_{2,-}^{(0)}}+\frac{\left| \left\langle \phi _{0}^{(0)}\left|\hat{a}\right|\phi _{1}^{(0)}\right\rangle \right|^2}{E_{1,-}^{(0)}-E_{0,0}^{(0)}}) \tag{B18}
\label{3.14}
\end{equation}
\end{footnotesize}

\begin{footnotesize}
\begin{equation}
\phi _{1}^{(1)}=-z\kappa \psi \left(\frac{\left\langle \phi _{2}^{(0)}\left|\hat{a}^{\dagger }\right|\phi _{1}^{(0)}\right\rangle }{E_{1,-}^{(0)}-E_{2,-}^{(0)}}\left|\phi _{2}^{(0)}\right\rangle
+\frac{\left\langle \phi _{0}^{(0)}\left|\hat{a}\right|\phi _{1}^{(0)}\right\rangle }{E_{1,-}^{(0)}-E_{0,0}^{(0)}}\left|\phi _{0}^{(0)}\right\rangle\right) \tag{B19}
\label{3.15}
\end{equation}

\begin{equation}
\phi _{0}^{(1)}=-z\kappa \psi \frac{\left\langle \phi _{1}^{(0)}\left|\hat{a}^{\dagger }\right|\phi _{0}^{(0)}\right\rangle }{E_{0,0}^{(0)}-E_{1,-}^{(0)}}\left|\phi _{1}^{(0)}\right\rangle \tag{B20}
\label{3.15}
\end{equation}

\begin{equation}
\left\langle \phi _{2}^{(0)}\left|\hat{a}^{\dagger }\right|\phi _{1}^{(0)}\right\rangle =\frac{1}{\sqrt{B_{1} B_{2}}}\left(d+\sqrt{2} c a_{1}+a d_1\right) \tag{B21}
\label{3.16}
\end{equation}
\begin{equation}
\left\langle \phi _{0}^{(0)}0\left|\hat{a}\right|\phi _{1}^{(0)}\right\rangle =\frac{1}{\sqrt{B_{1}}}a_1.\label{3.17} \tag{B22}
\end{equation}
\end{footnotesize}

\end{document}